\newif\ifaaai
\newif\ifapp
\apptrue
\newif\ifarxiv
\arxivtrue
\ifarxiv
  \aaaifalse
  \apptrue
\else
  \aaaitrue
  \appfalse
\fi

\newcommand{\thetitle}{Properties of Egalitarian Sequences of Committees: Theory and Experiments}
\newcommand{\thertitle}{Properties of Egalitarian Sequences of Committees: Theory and Experiments}
\newcommand{\AFFA}{AFFA~(BR~5207/1 and NI~369/15, Nr.~284041127)}
\newcommand{\PACS}{PACS (FL~1247/1-1, Nr.~522475669)}
\def\HUBaffil{Humboldt-Universität zu Berlin, 
Department of Computer Science, Algorithm Engineering Group, Germany}
\def\TUCaffil{Technische Universität Clausthal, Institut für Informatik, Germany}

\ifaaai{}
  \documentclass{ecai}

  \usepackage{latexsym}
  \usepackage{amssymb}
  \usepackage{amsmath}
  \usepackage{amsthm}
  \usepackage{booktabs}
  \usepackage{graphicx}
  \usepackage{color}
  \usepackage[utf8]{inputenc}
  \usepackage{graphicx}
  \usepackage{amsmath}
  \usepackage{amsthm}
  \usepackage{booktabs}
  \usepackage{algorithm}
  \usepackage{algorithmic}
  \usepackage{nameref}
  \usepackage{amssymb}
  \usepackage{mathtools}
\else{}
  \documentclass[10pt,twocolumn]{article}
  \usepackage[a4paper,includefoot,margin=1.5cm]{geometry}

  \usepackage{latexsym}
  \usepackage{amsmath, amssymb, amsthm}
  
  \usepackage{thm-restate}
  \usepackage{thmtools}
  
  \usepackage{graphicx,scalerel}

  \usepackage[T1]{fontenc}
  \usepackage[%
  breaklinks=true,
  pdfdisplaydoctitle,
  menucolor=orange!40!black,filecolor=magenta!40!black,urlcolor=blue!40!black,linkcolor=red!40!black,citecolor=green!40!black,
  colorlinks
  ]{hyperref} %
  
  \newcommand{\ExternalLink}{%
      \tikz[color=magenta, x=1.2ex, y=1.2ex, baseline=-0.05ex]{%
          \begin{scope}[x=1ex, y=1ex]
              \clip (-0.1,-0.1) --++ (-0, 1.2) --++ (0.6, 0) --++ (0, -0.6) --++ (0.6, 0) --++ (0, -1);
              \path[draw, line width = 0.5, rounded corners=0.5] (0,0) rectangle (1,1);
          \end{scope}
          \path[draw, line width = 0.5] (0.5, 0.5) -- (1, 1);
          \path[draw, line width = 0.5] (0.6, 1) -- (1, 1) -- (1, 0.6);
      }
  }

  \DeclareUnicodeCharacter{1EF3}{\`y}

  \usepackage{natbib}
  \makeatletter
  \def\@maketitle{%
  \newpage\null\vskip 2em%
  \begin{center}%
  \let \footnote \thanks
    {\Large\bf \@title \par}\vskip 1.5em{\large\lineskip .5em\begin{tabular}[t]{c}\@author\end{tabular}\par}\vskip 1em{\large \@date}%
  \end{center}\par\vskip 1.5em
  }
  \def\url@leostyle{\@ifundefined{selectfont}{\def\UrlFont{\sf}}{\def\UrlFont{\small\ttfamily}}}
  \makeatother
  \usepackage[ruled,vlined,linesnumbered]{algorithm2e}
  
  \usepackage{authblk}
  
  \author[1]{Paula Böhm}
  \author[1]{Robert Bredereck}
  \author[2]{Till~Fluschnik}

  \affil[2]{\footnotesize
    \TUCaffil\\
    \texttt{$\{$paula.boehm,robert.bredereck$\}$@tu-clausthal.de}
  }
  \affil[2]{\footnotesize
    \HUBaffil\\
    \texttt{till.fluschnik@hu-berlin.de}
  }

  \usepackage[mathic]{mathtools}
\fi{}

\usepackage{adjustbox}
\usepackage[dvipsnames,table]{xcolor}

\usepackage[textsize=footnotesize,backgroundcolor=green!40!white,linecolor=black!60,disable]{todonotes}

\usepackage{etoolbox}
\ifaaai\else{}
  \AtBeginEnvironment{tabular}{\fontsize{8pt}{10pt}\selectfont}
  \AtBeginEnvironment{tabular}{\setlength{\tabcolsep}{3pt}}
  \AtBeginEnvironment{tblr}{\fontsize{8pt}{10pt}\selectfont}
  \AtBeginEnvironment{tblr}{\setlength{\tabcolsep}{3pt}}
\fi

\usepackage[capitalise]{cleveref}
\crefname{algocf}{Algorithm}{Algorithms}
\Crefname{algocf}{Algorithm}{Algorithms}
\crefname{axiom}{}{}
\creflabelformat{axiom}{(#2I#1#3)}
\providecommand\crefpairconjunction{,}

\usepackage{url}
\usepackage{dsfont, paralist, microtype}
\usepackage{booktabs, tabularx}

\usepackage{tikz}
\usetikzlibrary{arrows, arrows.meta, calc, decorations.pathmorphing, decorations.pathreplacing, calligraphy, backgrounds, math, shapes, shapes.geometric, spy, positioning, quotes}
\usetikzlibrary{external}
\RequirePackage{xargs}     

\usetikzlibrary{decorations.pathreplacing}
\usetikzlibrary{math}
\usetikzlibrary{calc}
\usetikzlibrary{shapes}

\newcommand{\tikzpreamble}{%
  \def\teps{0.075}
  \def\nsc{0.33}
  \def\tanS{1.7320508}
  \def\bsc{0.075}
  \def\colpNPh{red!85!black}
  \def\colXPWh{orange}
  \def\colFPTnoPK{green!80!black}
  \def\colFPTPK{green!40!white}
  \def\colFPT{green}
  \def\colXP{yellow}
  \def\colOpen{brown!20!white}
  \tikzstyle{xnode}=[circle,scale=\nsc,draw,fill=white];
  \tikzstyle{xnodeA}=[circle,scale=\nsc,fill=white,draw=red];
  \tikzstyle{xnodeB}=[circle,scale=\nsc,fill=lightgray,draw=green];
  \tikzstyle{xnodeC}=[circle,scale=\nsc,fill=black,draw=blue];
  \tikzstyle{xnodex}=[circle,fill,scale=\nsc,draw];
  \tikzstyle{xnodey}=[diamond,fill,scale=\nsc,draw];
  \tikzstyle{xstarL}=[star,star points=6,draw,scale=0.4];
  \tikzstyle{xstar}=[star,star points=8,draw,scale=0.4];
  \tikzstyle{xstarB}=[star,star points=10,draw,scale=0.4];
  \tikzstyle{xedge}=[thick,-];
  \tikzstyle{xedgex}=[thick,-,dashed];
  \tikzstyle{xedgedot}=[thick,-,dotted];
  
  \tikzstyle{xpath}=[color=blue,opacity=0.25,line cap=round,line width=6pt];
  \tikzstyle{xpathS}=[color=blue!40!white,opacity=0.3,line cap=round,line join=round,line width=5pt];
  \tikzstyle{xpathSx}=[color=green!40!black,opacity=0.3,line cap=round,line join=round,line width=5pt];
  \tikzstyle{xpathx}=[color=magenta,opacity=0.40,line cap=round,line width=6pt];
  \tikzstyle{xpathy}=[color=green,opacity=0.40,line cap=round,line width=6pt];
  
  \tikzstyle{xhili}=[circle,scale=1.25,opacity=0.25,fill,color=orange,draw];
  \tikzstyle{xhiliS}=[circle,scale=0.625,opacity=0.25,fill,color=orange,draw];
  \tikzstyle{xhiliIS}=[circle,scale=1.25,opacity=0.25,fill,color=magenta,draw];
  \tikzstyle{xxhiliS}=[rectangle,scale=0.85,opacity=0.25,fill,color=cyan,draw];
}

\newcommand{\Grid}[9]{
  \foreach\x in {0,...,#2}{
    \foreach \y in {0,...,#3}{
      \node (#1\x\y) at (#8*\x*\xr+#4*\xr,#9*\y*\yr+#5*\yr)[xnode,#6]{};
    }
  }
  \ifnum#3>0
    \pgfmathsetmacro\yx{int(#3 - 1)}
    \foreach \x in {0,...,#2}
      \foreach \y [count=\yi] in {0,...,\yx}  
        \draw[#7] (#1\x\y)--(#1\x\yi) ;
  \fi
  \ifnum#2>0
  \pgfmathsetmacro\yx{int(#2 - 1)}
  \foreach \x in {0,...,#3}
    \foreach \y [count=\yi] in {0,...,\yx}  
      \draw[#7] (#1\y\x)--(#1\yi\x) ;
  \fi
}

\usepackage{pgfplots}

\usepackage{xcolor-material}
\usepackage{tabularray}
\UseTblrLibrary{booktabs}
\usepackage{siunitx}
\usepackage{stmaryrd}
\makeatletter
\newcommand{\oset}[2]{%
  \mathrel{%
      \mathop{#2}\limits^{%
      \vbox to 0pt{%
        \kern-2\ex@%
        \hbox{\raisebox{0pt}[0pt][0pt]{$\scriptstyle#1$}}\vss%
      }%
    }%
  }%
}
\makeatother

\definecolor{TUCgreen}{cmyk}{1,0,0.9,0.2}
\definecolor{TUCgrey10}{cmyk}{0,0,0,0.1}
\definecolor{TUCgrey50}{cmyk}{0,0,0,0.5}

\ifaaai{}
  \newcommand{\pref}[1]{\textsl{\nameref{#1}~(P\ref{#1})}}
  \newcommand{\pxref}[1]{{\small\textsl{\nameref{#1} (P\ref{#1})}}}
\else{}
  \newcommand{\pref}[1]{\textsl{\nameref{#1}}}
\fi{}
\newcommand{\psref}[1]{P\ref{#1}}
\usepackage{soul}

\usepackage{twemojis}
\AtBeginDocument{
  \setlength{\twemojiDefaultHeight}{0.9em}
}
\usepackage{xargs}
\newcommandx{\mydefenv}[4][3=A]{%
  \newtheorem{#1}{#2}
  \ifstrequal{#3}{A}{\crefname{#1}{#2}{#2s}}{\crefname{#1}{#2}{#3}}%
  \Crefname{#1}{#4{.}}{#4s{.}}
}

\theoremstyle{plain}
\mydefenv{theorem}{Theorem}{Thm}
\mydefenv{lemma}{Lemma}{Lem}
\mydefenv{problem}{Problem}{Prob}
\mydefenv{erule}{Rule}{Rule}
\mydefenv{proposition}{Proposition}{Prop}
\mydefenv{conjecture}{Conjecture}{Conj}
\mydefenv{corollary}{Corollary}[Corollaries]{Cor}
\mydefenv{observation}{Observation}{Obs}
\mydefenv{claim}{Claim}{Claim}
\mydefenv{fact}{Fact}{Fact}
\theoremstyle{definition}
\mydefenv{definition}{Definition}{Def}
\mydefenv{construction}{Construction}{Constr}
\mydefenv{rrule}{Data Reduction Rule}{DRR}
\mydefenv{property}{Property}[Properties]{P}

\mydefenv{propertyN}{Property}[Properties]{\ensuremath{\mathcal{P}}}
\renewcommand*{\thepropertyN}{\Alph{propertyN}}
\theoremstyle{remark}
\mydefenv{example}{Example}{Ex}
\mydefenv{remark}{Remark}{Rem}
\mydefenv{type}{Type}{Type}
\mydefenv{hldesc}{High-Level Description}{High-Level Descr}

\crefname{figure}{Figure}{Figures}
\Crefname{figure}{Fig{.}}{Figs{.}}
\Crefname{section}{S{.}}{S{.}}
\crefname{section}{Section}{Sections}
\ifaaai{}
  \newcommand{\crefrangeconjunction}{--}
  \newcommand{\crefmiddleconjunction}{, }
  \newcommand{\creflastconjunction}{ \& }
  \Crefname{property}{}{}
\fi{}

\newcommand{\cqed}{\hfill$\diamond$}
\newcommand{\rqed}{\hfill$\triangleleft$}

\newcommand{\prob}[1]{\textnormal{\textsc{#1}}}
\newcommand{\wpb}{when parameterized by}
\newcommand{\RD}{$(\Rightarrow)\quad$}
\newcommand{\LD}{$(\Leftarrow)\quad$}
\newcommandx{\compset}[1]{\ensuremath{[#1]}}
\newcommandx{\set}[2][1=1]{\ensuremath{\{#1,\ldots,#2\}}}
\newcommandx{\tlog}[3][1=,3=]{\log_{#1}^{#3}(#2)}
\newcommandx{\ith}[2][1=th]{#2\nobreakdash-#1}
\newcommandx{\decprob}[6][3=Input,5=Question]{\vspace{0.125em}\begin{samepage}\begingroup
\label{DEACTIVATEDprob:#2}{
{\noindent \textsc{#1}}}
  \nopagebreak[4]\nopagebreak[4]\vspace{-0.125em}
  \par\noindent\hangindent=\parindent\textbf{#3}:  #4\nopagebreak[4]
  \par\noindent\hangindent=\parindent\textbf{#5}:  #6
  \par\medskip\endgroup\end{samepage}
}

\newcommandx{\decrule}[6][3=Input,5=Output]{\begin{samepage}\begingroup
\begin{erule}\label{erule:#2}{
{\textsc{#1}}}
  \nopagebreak[4]\end{erule}\nopagebreak[4]\vspace{-0.525em}
  \par\noindent\hangindent=\parindent\textbf{#3}:  #4\nopagebreak[4]
  \par\noindent\hangindent=\parindent\textbf{#5}:  #6
  \par\medskip\endgroup\end{samepage}
}

\DeclarePairedDelimiterX{\abs}[1]{\lvert}{\rvert}{#1}
\DeclarePairedDelimiterX{\norm}[1]{\lVert}{\rVert}{#1}
\DeclarePairedDelimiterX{\ceil}[1]{\lceil}{\rceil}{#1}
\newcommand{\mvert}{\;\middle\vert\;}

\newcommand{\N}{\mathds{N}}
\newcommand{\Nzero}{\mathds{N}_0}
\newcommand{\Z}{\mathds{Z}}
\newcommand{\R}{\mathds{R}}
\newcommand{\Q}{\mathds{Q}}
\newcommand{\F}{\mathds{F}}

\newcommand{\calA}{\mathcal{A}}
\newcommand{\calB}{\mathcal{B}}
\newcommand{\calC}{\mathcal{C}}
\newcommand{\calE}{\mathcal{E}}
\newcommand{\calF}{\mathcal{F}}
\newcommand{\calI}{\mathcal{I}}
\newcommand{\calP}{\mathcal{P}}
\newcommand{\calR}{\mathcal{R}}
\newcommand{\calS}{\mathcal{S}}
\newcommand{\calT}{\mathcal{T}}
\newcommand{\calX}{\mathcal{X}}

\newcommand{\cocl}[1]{\textrm{#1}}
\newcommand{\NP}{\cocl{NP}}
\newcommand{\classP}{\cocl{P}}
\newcommand{\W}[1]{\ensuremath{\cocl{W}[1]}}
\newcommand{\FPT}{\cocl{FPT}}

\newcommand{\bigO}{\mathcal{O}}
\newcommand{\yes}{\emph{yes}}
\newcommand{\no}{\emph{no}}
\newcommand{\ceq}{\coloneqq}
\newcommand{\eqc}{\eqqcolon}
\newcommand{\false}{\bot}
\newcommand{\true}{\top}

\newcommand{\symdiff}{\ensuremath{\bigtriangleup}}
\newcommand{\edit}[1]{\ensuremath{\langle#1\rangle}}
\newcommand{\ol}[1]{\overline{#1}}

\newcommand{\tsps}[1]{\textsuperscript{#1}}
\newcommand{\tsbs}[1]{\textsubscript{#1}}

\newcommand{\tref}[1]{\scriptsize{(\Cref{#1})}}

\DeclareMathOperator{\score}{scr}
\DeclareMathOperator{\poly}{poly}
\DeclareMathOperator{\sgn}{sgn}
\DeclareMathOperator{\pf}{pf}
\DeclareMathOperator{\val}{val}
\DeclareMathOperator{\rep}{rep}
\DeclareMathOperator{\rank}{rank}
\DeclareMathOperator{\argmaxlim}{\arg\,\max}
\DeclareMathOperator{\argminlim}{\arg\,\min}
\newcommand{\argmax}{\mathop{\arg\,\max}\limits}
\newcommand{\argmin}{\mathop{\arg\,\min}\limits}
\newcommand{\repr}{\ensuremath{\subseteq_{\rep}}}

\DeclareMathOperator{\opt}{opt}

\newcommand{\oneto}[1]{[ #1 ]} %

\newcommand{\eps}{\varepsilon}

\newcommand{\wilog}{without loss of generality}
\newcommand{\Wilog}{Without loss of generality}
\newcommand{\ie}{i.\,e.,\ }
\newcommand{\cif}{\text{if~}}

\newcommand{\col}{\ensuremath{c}}

\newcommand{\mybinom}[2]{\Bigl(\begin{array}{@{}c@{}}#1\\#2\end{array}\Bigr)}
\newcommand{\ctwo}[1]{\ensuremath \mybinom{#1}{2}}
\newcommand{\lneg}[1]{\ensuremath{\ol{#1}}}
\newcommand{\setto}{\leftarrow}

\newcommand{\ythenx}{$[y\to x]$}
\newcommand{\xtheny}{$[x\to y]$}
\newcommand{\xgiveny}{$[x\mid y]$}
\newcommand{\ygivenx}{$[y\mid x]$}

\newcommand{\Rule}{\calR}
\newcommand{\Rulebase}{\Rule_{\rm vld}}
\newcommand{\hori}{{\rm A}}
\newcommand{\verti}{{\rm L}}
\newcommand{\xscrTxt}{level}
\newcommand{\yscrTxt}{agent}
\DeclareMathOperator{\yscore}{\score_{\hori}}
\DeclareMathOperator{\yscoremin}{\score_{\hori}^{\min}}
\DeclareMathOperator{\yscoresum}{\score_{\Sigma}}
\DeclareMathOperator{\xscore}{\score_{\verti}}
\DeclareMathOperator{\xscoremin}{\score_{\verti}^{\min}}
\DeclareMathOperator{\xscoresum}{\score_{\Sigma}}
\newcommand{\Elec}{\calE}
\newcommand{\Com}{X}
\newcommand{\ComSeq}{\calX}
\newcommand{\ComSeqSet}{\mathbf{X}}
\newcommand{\tauS}{T}
\newcommand{\Cand}{C}
\newcommand{\CandSeq}{\calC}
\newcommand{\ktop}{k^\top}
\newcommand{\ktopSeq}{\kappa^\top}
\newcommand{\kbot}{k^\bot}
\newcommand{\kbotSeq}{\kappa^\bot}
\newcommand{\kSeq}{\kappa}

\newcommand{\smlaTsc}{\prob{Egalitarian Committee Sequence Election}}
\newcommand{\smlaAcr}{\prob{GCSE}}
\newcommand{\pvcTsc}{\prob{Partial Vertex Cover}}
\newcommand{\pvcAcr}{\prob{PVC}}

\newcommand{\xhorz}{\hori}
\newcommand{\xvert}{\verti}
\newcommand{\ymaxmin}{\hori} %
\newcommand{\xmaxmin}{\verti} %
\newcommand{\maxsum}{\Sigma} %

\newcommand{\Ruleh}{\Rule_{\ymaxmin}}
\newcommand{\Rulev}{\Rule_{\xmaxmin}}
\newcommand{\Rules}{\Rule_{\maxsum}}
\newcommand{\Rulehv}{\Rule_{\ymaxmin,\xmaxmin}}
\newcommand{\Rulevh}{\Rule_{\xmaxmin,\ymaxmin}}
\newcommand{\Rulehs}{\Rule_{\ymaxmin,\maxsum}}
\newcommand{\Rulevs}{\Rule_{\xmaxmin,\maxsum}}
\newcommand{\Rulesh}{\Rule_{\maxsum,\ymaxmin}}
\newcommand{\Rulesv}{\Rule_{\maxsum,\xmaxmin}}
\newcommand{\Rulehvs}{\Rule_{\ymaxmin,\xmaxmin,\maxsum}}
\newcommand{\Rulevhs}{\Rule_{\xmaxmin,\ymaxmin,\maxsum}}
\newcommand{\Rulehsv}{\Rule_{\ymaxmin,\maxsum,\xmaxmin}}
\newcommand{\Rulevsh}{\Rule_{\xmaxmin,\maxsum,\ymaxmin}}
\newcommand{\Ruleshv}{\Rule_{\maxsum,\ymaxmin,\xmaxmin}}
\newcommand{\Rulesvh}{\Rule_{\maxsum,\xmaxmin,\ymaxmin}}

\newcommand{\Ruleapp}{\Rules}
\newcommand{\Rulegreedy}{\Rule_{\rm greedy}}
\newcommand{\Rulelex}{\Rule_{\rm lex}}
\newcommand{\Ruleydom}{\Rule_{\xhorz_{\rm dom}}}

\newcommand{\sat}{\operatorname{sat}}
\newcommand{\ems}{\score^{\rm lex}}%
\newcommand{\lexord}{\operatorname{ord_{\yscore}}}

\newcommand{\hdominate}{dominate}

\newcommand{\subsall}{\subseteq_{\rm all}}
\newcommand{\subslvl}{\subseteq_{\rm lvl}}
\newcommand{\supsall}{\supseteq_{\rm all}}
\newcommand{\supslvl}{\supseteq_{\rm lvl}}

\newcommand{\nB}[2]{($(#1)!0.5!(#2)$)}

\newcommand{\leqElsf}{\preceq_{\rm lsf}}
\newcommand{\leqElex}{\preceq_{\rm lex}}
\newcommand{\leqEsum}{\preceq_{\rm sum}}
\newcommand{\leqEvec}{\preceq_{\rm vec}}

\newcommandx{\tlab}[2][1=]{%
  \ifstrequal{#1}{}{}{\,(#1)}%
}
\newcommand{\tlabtxt}[2]{(#1: #2)}
\newcommand{\dunno}{\cellcolor{gray!33!white}\textcolor{blue}{?!?}}
\newcommand{\tyes}{yes}%
\newcommand{\ttyes}{\tyes\tsps{\tlabA}}
\newcommand{\tno}{\textbf{no}}%
\newcommand{\ttno}{\tno\tsps{\tlabA}}
\ifaaai{}
  \newcommand{\htyes}[1]{\tyes\tsps{\Cref{#1}}}
  \newcommand{\htno}[1]{\tno\tsps{\Cref{#1}}}
  \newcommand{\rref}[2]{#2}
\else{}
  \newcommand{\htyes}[1]{\hyperref[#1]{\tyes}}
  \newcommand{\htno}[1]{\hyperref[#1]{\tno}}
  \newcommand{\rref}[2]{\hyperref[#1]{#2}}
\fi{}
\newcommand{\bigland}{\bigwedge}

\newcounter{acnt}
\newcommandx{\addagent}[2][1=]{%
  \pgfmathsetmacro\val{int(\value{acnt})}
  \node (a\val) at (0,-\yr*\Aysh*\val)[]{};
  \node at ($(a\val)+(-0.05*\xr,0.19*\yr)$)[anchor=base east]{#1$a_{\val}$:};
  \foreach \xA [count=\iA from 1] in {#2}{
    \node (nx\val\iA) at ($(a\val)+(\iA*\Axsh*\xr-0.75*\Axsh*\xr,0.19*\yr)$)[anchor=base]{$\xA$};
  }
  \addtocounter{acnt}{1}
}
\newcommandx{\addagentp}[2][1=]{%
  \pgfmathsetmacro\val{int(\value{acnt})}
  \node (a\val) at (0,-\yr*\Aysh*\val)[]{};
  \node at ($(a\val)+(-0.05*\xr,0.19*\yr)$)[anchor=base east]{#1$a_{\val}'$:};
  \foreach \xA [count=\iA from 1] in {#2}{
    \node (nx\val\iA) at ($(a\val)+(\iA*\Axsh*\xr-0.75*\Axsh*\xr,0.19*\yr)$)[anchor=base]{$\xA$};
  }
  \addtocounter{acnt}{1}
}
\newcommandx{\addagentx}[2][1=]{%
  \pgfmathsetmacro\val{int(\value{acnt})}
  \node (a\val) at (0,-\yr*\Aysh*\val)[]{};
  \node at ($(a\val)+(-0.05*\xr,0.19*\yr)$)[anchor=base east]{#1:};
  \foreach \xA [count=\iA from 1] in {#2}{
    \node (nx\val\iA) at ($(a\val)+(\iA*\Axsh*\xr-0.75*\Axsh*\xr,0.19*\yr)$)[anchor=base]{$\xA$};
  }
  \addtocounter{acnt}{1}
}

\newcommandx{\addagentf}[1]{%
  \pgfmathsetmacro\val{int(\value{acnt})}
  \def\frnd{B}
  \ifnum\val=2 \def\frnd{D} \fi
  \ifnum\val=3 \def\frnd{E} \fi
  \ifnum\val=4 \def\frnd{F} \fi
  \ifnum\val=5 \def\frnd{G} \fi
  \ifnum\val=6 \def\frnd{H} \fi
  \node (a\val) at (0,-\yr*\Aysh*\val)[]{};
  \node at ($(a\val)+(-0.05*\xr,0.19*\yr)$)[anchor=base east]{\frnd:};
  \foreach \xA [count=\iA from 1] in {#1}{
    \node (nx\val\iA) at ($(a\val)+(\iA*\Axsh*\xr-0.75*\Axsh*\xr,0.19*\yr)$)[anchor=base]{$\xA$};
  }
  \addtocounter{acnt}{1}
}

\newcommand{\agentInit}[2]{%
  \def\Aysh{#2}
  \def\Axsh{#1}
  \setcounter{acnt}{1}
}

\newcommandx{\hiliA}[3][2=blue!50!white]{%
  \node at ($(a#1)+(#3*\Axsh*\xr-0.75*\Axsh*\xr,0.075*\yr)$)[anchor=south,opacity=0.3,circle,fill=#2,scale=1.2]{};
}

\newcommandx{\hiliB}[3][2=green!50!black]{%
  \node at ($(a#1)+(#3*\Axsh*\xr-0.75*\Axsh*\xr,0.075*\yr)$)[anchor=south,opacity=0.4,circle,draw=#2,scale=1.2,very thick]{};
}

\newcommand{\FEhiliA}[1]{%
  \foreach\x/\y in {#1}{\hiliA{\x}{\y}}
}

\newcommand{\FEhiliB}[1]{%
  \foreach\x/\y in {#1}{\hiliB{\x}{\y}}
}

\newcommand{\appsymb}{{\Large $\star$}}
\ifaaai{}
  \newcommand{\appref}[1]{{\appsymb}}
\else{}
  \newcommand{\appref}[1]{{\hyperref[proof:#1]{\appsymb}}}
  \newcommand{\apprefX}[1]{{\hyperref[#1]{\appsymb}}}
\fi{}

\newcommand{\appendixsection}[1]{%
  \gappto{\appendixProofText}{\section{Additional Material for Section~\ref{#1}}\label{app:#1}}
}

\newcommand{\toappendix}[1]{%
  \gappto{\appendixProofText}
  {{
    #1
  }}
}
\newcommand{\appendixproof}[2]{%
  \gappto{\appendixProofText}
  {
    \subsection{Proof of \cref{#1}}\label{proof:#1}
    #2
  }
}

\title{\thetitle}

\newcommand{\mgreen}{green circle}%
\newcommand{\mblue}{blue sphere}%

\newcommand{\respect}{satisfy}
\newcommand{\respects}{satisfies}
\newcommand{\respected}{satisfied}
\newcommand{\satisfy}{satisfy}
\newcommand{\satisfies}{satisfies}
\newcommand{\satisfied}{satisfied}
\newcommand{\reject}{violate}
\newcommand{\rejects}{violates}
\newcommand{\rejected}{violated}
\newcommand{\violate}{violate}
\newcommand{\violates}{violates}
\newcommand{\violated}{violated}
\NewDocumentCommand{\todor}{m}{\textcolor{red}{#1}}
\NewDocumentCommand{\changed}{m}{\textcolor{blue}{#1}}
\NewDocumentCommand{\proposal}{m}{\textcolor{violet}{#1}}

\newcommand{\theabstract}{%
  We study the task of electing egalitarian sequences of $\tau$ committees 
  given a set of agents 
  with additive utilities for candidates available on each of~$\tau$ levels.
  We introduce several rules for electing an egalitarian committee sequence 
  as well as properties for such rules.
  We settle the computational complexity of 
  finding a winning sequence
  for
  our rules
  and classify them against our properties.
  Additionally, we transform sequential election data from existing election data from the literature.
  Using this data set,
  we compare our rules empirically
  and
  test them experimentally against our properties.
}

\newcommand{\theacks}{%
 Till Fluschnik acknowledges support by Deutsche Forschungsgemeinschaft 
 (DFG, German Research Foundation), 
 projects \AFFA{} and \PACS{}.

 The emoji graphics are taken from \texttt{twemojis} and licensed under CC-BY 4.0: \url{https://creativecommons.org/licenses/by/4.0/}. Copyright 2019 Twitter, Inc and other contributors.
}

\begin{document}\tikzexternaldisable

\ifaaai{}
  \begin{frontmatter}

    \paperid{3862}

    \title{\thetitle{}}

    \author[A]{\fnms{Paula}~\snm{Böhm}\orcid{0000-0001-7499-8211}\thanks{Corresponding Author. Email: paula.boehm@tu-clausthal.de}}
    \author[A]{\fnms{Robert}~\snm{Bredereck}\orcid{0000-0002-6303-6276}}
    \author[B]{\fnms{Till}~\snm{Fluschnik}\orcid{0000-0003-2203-4386}}

    \address[A]{\TUCaffil}
    \address[B]{\HUBaffil}

    \begin{abstract}
    \theabstract{}
    \end{abstract}

  \end{frontmatter}

\else{}
  \maketitle

  \begin{abstract}
  \theabstract{}
  \end{abstract}
\fi{}

\section{Prologue}

Preferences can be more fine-grained than a single-committee election can represent.
For instance, instead of asking faculty members about their representative in an
appointment committee, one can ask which candidate they prefer to be elected for
each of several \emph{levels} (roles), e.g., (co-)heads, professors, researchers, students.
Consequently, people can vote in favor of candidates differing from their ‘overall favorite’
candidates, which permits respecting their opinions on a finer scale when selecting
a committee for each level. Moreover, such \emph{multilevel, sequential} elections allow
us to ask for additional fairness criteria fulfilled by the committee sequence.

In this work, we focus on \emph{egalitarian} committee sequences which maximize the satisfaction of the least satisfied agents.
Next to this primary fairness criterion, there are secondary goals one can aim for:
For example, a high overall satisfaction (i.e.~sum of satisfactions)
or maximizing the satisfaction of the next least satisfied agents lexicographically
(consistent with many other settings).
Furthermore, one could also exploit the multilevel setting by maximizing the minimum satisfaction per level.
We wonder: what is the impact of including each of these secondary goals,
evaluated both axiomatically and experimentally,
and how do they relate to each other?

As application example, consider creating a music playlist for an event
(to use the setting of one of our largest datasets, the labelled Spotify instances, see~\cref{subsec:expdata})
which shall contain a fixed number of songs from each genre of a predefined list.
The egalitarian approach ensures the representation of all guests’ preferences to a minimum degree,
including those with uncommon tastes.
As additional goal, one might aim to maximize the minimum total satisfaction of the guests
for a selection of a genre (maximin per genre)
or to maximize the total satisfaction over the whole playlist (utilitarian approach).
The former aims for high interest in the music over all genres
while the latter aims to maximize the overall appreciation of the music.

In addition to modelling scenarios in which there are natural categories (e.g.,
selecting the favorite movie or game in multiple genres),
sequential elections can be used for problems with a temporal component as well,
e.g.~which crop farmers prefer to grow in which month or which activity each member
of a traveling group prefers for which day.
Formally,
our model receives the following input:\footnote{We
could use unordered multisets instead of sequences,
as the order does not affect our rules or properties.
However, adhering to the literature maintains a consistent, simple syntax.
}
\newcommand{\ub}{\ensuremath{z}}
\begin{definition}
 A %
 (sequential, multilevel)
 election $\calE=(A,\CandSeq,U,\kSeq)$ consists of:
    A set~$A$ of $n$ agents, a sequence~$\CandSeq=(\Cand_1,\dots,\Cand_\tau)$ of subsets from a candidate set~$C$ of size~$m$,
    a sequence $U=(u_1,\dots,u_\tau)$ of 
    utilities~$u_t\colon A\times \Cand_t\to\Nzero$,
    and a sequence~$\kSeq=(k_1,\dots,k_\tau)$
    of nonnegative integers.
\end{definition}

\noindent
In applications, we would often assume normalized utilities 
for comparing satisfactions of different agents.
This is, however, not required for any of our results.

Since we are interested in egalitarian committee sequences,
we wish to find a 
sequence~$\ComSeq=(\Com_1,\dots,\Com_\tau)$
of committees~$\Com_t\subseteq C_t$ with~$|X_t|= k_t$ for every~$t\in \tauS\ceq\set{\tau}$
that maximizes
\begin{align*}
 \yscoremin(\Elec,\ComSeq) \ceq \min_{a\in A} &\yscore(\Elec,\ComSeq,a),
 \text{ where}\\
 \yscore(\Elec,\ComSeq,a) &\ceq \sum_{t\in\tauS} \sum_{c\in\Com_t} u_t(a,c)
  \ \ \text{(\emph{\yscrTxt} score)}.
\end{align*}

\NewDocumentCommand{\candF}{}{\twemoji{croissant}}%
\NewDocumentCommand{\candE}{}{\twemoji{sandwich}}%
\NewDocumentCommand{\candJ}{}{\twemoji{pizza}}%
\NewDocumentCommand{\candC}{}{\twemoji{spaghetti}}%
\NewDocumentCommand{\candS}{}{\twemoji{curry}}%
\NewDocumentCommand{\candI}{}{\twemoji{taco}}%
\begin{example}
 \label{ex:main}
  Four friends $(|A|=4)$ decide to spend a day together and want to plan their breakfast, lunch, and dinner ($\tau=3$).
  For each of their meals (levels), 
  they have two options ($|C_t|=2$)
  of which they want to take exactly one ($k_t=1$).
  They decide to distribute at most three points to each meal.
  Ben doesn't like any of the breakfast options and Fina needs to leave before dinner,
  so both give zero points to all corresponding options.
  Dora and Eric have identical preferences.
  The points (utilities) are displayed in \cref{tab:ex}.
  \begin{table}[h]
  \centering
  \caption{Tables for our~\cref{ex:main}.
  (Top) Our friend group's utilities.
  (Bottom) Winning committee sequences, each respecting a different optimization goal. 
  All omitted solutions have a $\yscoremin$ score smaller than~3.} %
  \label{tab:ex}
  \newcommand{\tabhl}[1]{#1}
  \setlength{\tabcolsep}{7pt}
  \ifaaai{}\def\fcwdth{5.5}\else{}\def\fcwdth{5.25}\fi{}
    \begin{tabular}{@{}rcccccc@{}}\toprule %
       & \multicolumn{2}{c}{Breakfast} 
       & \multicolumn{2}{c}{Lunch}
       & \multicolumn{2}{c}{Dinner}
       \\
       \cmidrule{2-7} 
       & \candF & \candE & \candJ & \candC & \candS & \candI
       \\\midrule
        Ben & 0 & 0 & 2 & 1 & 2 & 1 \\ 
        Dora & 3 & 0 & 3 & 0 & 1 & 2 \\ 
        Eric & 3 & 0 & 3 & 0 & 1 & 2 \\ 
        Fina & 0 & 3 & 0 & 3 & 0 & 0 \\ 
      \bottomrule
    \end{tabular}

    \smallskip
    \begin{tabular}{@{}rrrrr@{}}\toprule %
      & (a) \candF{} \candJ{} \candI{} & 
      (b) \candE{} \candJ{} \candI{} & 
      (c) \candF{} \candC{} \candS{} & 
      (d) \candE{} \candJ{} \candS{} \\\cmidrule{2-5}
      $\yscoremin$ & 0 & \tabhl{3} & \tabhl{3} & \tabhl{3} \\
      $\xscoremin$ & 5 & 3 & \tabhl{4} & 3 \\
      $\yscoresum$ & 19 & \tabhl{16} & 14 & 15 \\
      $\lexord$
      & 0,3,8,8 %
      & 3,3,5,5 %
      & 3,3,4,4 %
      & \tabhl{3,4,4,4} %
      \\
      \bottomrule
    \end{tabular}
 \end{table}
  The group's first approach is to
  (a) maximize the sum of points the selected options receive from each person.
  Consequently,
  \candF{} \candJ{} \candI{} would be selected.
  They note with discontent that Fina's satisfaction (measured by the sum of points she assigned to the selection options)
  is zero and wonder whether there is a solution where the least satisfied is better off.
  They decide to make sure as first criterion that the minimum agent satisfaction ($\yscoremin$) is maximal,
  and, only as second criterion, maximize the points received by the selected options.
  Since maximizing the sum of points as single criterion provided both,
  a maximal minimum points per level ($\xscoremin$) and a maximum sum of all points ($\yscoresum$),
  they wonder: Given all solutions where the minimum agent satisfaction is maximum,
  should they (b) first maximize $\yscoresum$ and then $\xscoremin$,
  or (c) the other way around?
  For (b),
  \candE{} \candJ{} \candI{} wins,
  while
  \candF{} \candC{} \candS{} wins for (c).
  Ben proposes as second optimization goal to 
  (d) maximize the satisfaction of the second least satisfied agent, then the third least satisfied, and so on ($\lexord$),
  which leads to
  \candE{} \candJ{} \candS{}
  as the winner.
  They think this is a good solution.
  Yet,
  they want to know more about the rules.
  ``If only there were a scientific paper on that.'', Dora says.
  \rqed
\end{example}

\paragraph{Our Contributions.}

\def\tlabA{$\ast$}
\def\tlabB{$\spadesuit$}
\def\tlabC{$\clubsuit$}
\def\tlabD{$\vardiamondsuit$}
\def\tlabE{$\varheartsuit$}
\def\tlabF{$\dagger$}
\def\tlabG{$\ddagger$}
\def\tlabH{$\P$}
\def\tlabI{$\circ$}
\def\tlabJ{$\times$}
\def\tlabK{$\diamond$}
\def\tlabL{$\triangle$}
\def\tlabM{$\blacksquare$}
\def\tlabN{\textleaf}
\def\tlabO{$\#$}
\def\tlabP{$\$$}
\def\tlabQ{$\%$}

\begin{table*}[t]
  \centering
  \caption{Summary of our results.
  The computational complexity refers to the problem of finding a winning committee sequence
  (\tsps{\tlabA}: easy to see).
  The (naturally rounded) proportion of the experimentally investigated (pairs of) instances for which a rule satisfies the conditions of the given property is given in braces: For \psref{propty:independentgroups}, the results are based only on instances with at least two independent groups, for \psref{propty:consistency}, on instance pairs with
  a common winning committee sequence.
  }
  \label{tab:results}
  \setlength{\tabcolsep}{6pt}
  \ifaaai{}\def\fcwdth{5.5}\else{}\def\fcwdth{5.25}\fi{}
  \begin{tabular}{@{}p{5cm}l|lllll@{}}
      \toprule
      Property \hfill Rule: & $\Ruleh$ & \rref{sec:lex}{$\Rulelex$} & \rref{sec:mmms}{$\Rulehsv$} & \rref{sec:mmms}{$\Rulehvs$} & \rref{sec:greedy}{$\Rulegreedy$} & \rref{sec:app}{$\Ruleapp$} \\
      \midrule[1pt]
      Computational Complexity & \NP-hard & \NP-hard & \NP-hard & \NP-hard & P & P \\
      \midrule
      \pref{propty:conchorzsuperadd} & \tyes\tsps{\tlabA} & \tyes\tsps{\tlabA} & \tyes\tsps{\tlabA} & \tyes\tsps{\tlabA} & \htno{obs:greedy:app:rejects:superadd} (\textcolor{GoogleGreen!98!GoogleYellow}{99.1}) & \tyes\tsps{\tlabA} \\
      \pref{propty:unitorzsuperadd} & \ttyes{} & \tyes\tsps{\tlabA} & \tyes\tsps{\tlabA} & \tyes\tsps{\tlabA} & \htno{obs:greedy:app:rejects:superadd} (\textcolor{GoogleGreen!99!GoogleYellow}{99.3}) & \htno{obs:greedy:app:rejects:superadd} (\textcolor{GoogleGreen!65!GoogleYellow}{82.7}) \\
      \pref{propty:consistency} & \htyes{obs:esum:lex:app:consistency} & \htyes{obs:esum:lex:app:consistency} & \htno{obs:mmmsmm:greedy:consist} (\textcolor{GoogleGreen!99!GoogleYellow}{99.3}) & \htno{obs:mmmsmm:greedy:consist} (\textcolor{GoogleGreen!98!GoogleYellow}{99.0}) & \htno{obs:mmmsmm:greedy:consist} (\textcolor{GoogleGreen!93!GoogleYellow}{96.7}) & \htyes{obs:esum:lex:app:consistency} \\
      \pref{propty:pareto} & \tno{}\tsps{\tlabA} & \tyes\tsps{\tlabA} & \htyes{obs:mmmsmm:hpareto} & \htno{obs:mmmsmm:hpareto} (\textcolor{GoogleGreen!92!GoogleYellow}{95.8}) & \htno{obs:greedy:rejects:pareto} (\textcolor{GoogleGreen!76!GoogleYellow}{88.2}) & \tyes\tsps{\tlabA} \\
      \pref{propty:independentgroups} & \htno{obs:mmmsmm-independentgroups} & \htyes{obs:lex:greedy:independentgroups} & \htno{obs:mmmsmm-independentgroups} (\textcolor{GoogleYellow!21!GoogleRed}{10.5}) & \htno{obs:mmmsmm-independentgroups} (\textcolor{GoogleYellow!19!GoogleRed}{9.7}) & \htyes{obs:lex:greedy:independentgroups} & \tyes\tsps{\tlabA} \\
      \bottomrule
  \end{tabular}
\end{table*}

We introduce several rules
for electing an egalitarian committee sequence.
Our five rules are
$\Rulelex$,
$\Rulehsv$,
$\Rulehvs$,
$\Rulegreedy$,
and~$\Ruleapp$ 
(see \cref{sec:centralrules}).
Intuitively,
committee sequences computed by
$\Rulelex$
firstly maximize the smallest agent score,
then the second smallest,
and so on;
$\Rulehsv$
firstly maximizes the smallest agent score,
secondly the sum of agent scores,
and finally the smallest level score;
$\Rulehvs$
firstly maximizes the smallest agent score,
then the smallest level score,
and finally the sum of agent scores;
$\Rulegreedy$ intuitively
mimics~$\Rulelex$
by iteratively selecting a candidate that improves $\Rulelex$'s objective function the most;
$\Ruleapp$ iteratively selects candidates
with the highest utility on each level.\footnote{$\Ruleapp$
is clearly non-egalitarian
as it ignores agent scores by definition.
Yet,
we consider~$\Ruleapp$ as a benchmark,
since it mimics the straightforward rule mostly used in practice when
customized rules are not available or known.}
We perform both a theoretical (\cref{sec:props}) and an experimental analysis (\cref{sec:experiments}).
On the theoretical side:
\begin{compactitem}
  \item We show that finding a winning committee sequence for~$\Rulelex$,
  $\Rulehvs$,
  or~$\Rulehsv$ is an~\NP-hard task,
  even in quite restricted settings.
  In contrast,
  a winning committee sequence for~$\Rulegreedy$ and~$\Ruleapp$ can be found in polynomial-time.
  \item We formulate five properties for rules electing egalitarian committee sequences
  (see~\cref{sec:props}).
  We show that~$\Rulelex$ is the only rule that \respects{} all of our properties
  (see~\cref{tab:results}).
  $\Rulegreedy$ \respects{} the least number of properties,
  $\Ruleapp$ the second-highest number.
   $\Rulehsv$ \respects{} one property more than~$\Rulehvs$.
\end{compactitem}
On the experimental side:
\begin{compactitem}
 \item We took 7888 PrefLib \cite{mattei2017apreflib} elections and assigned labels to the candidates based on
       real-world metadata, resulting in the
       largest preference dataset with labeled candidates, to the best of our knowledge.
       We used these labels to construct sequential elections
       by assigning candidates with the same label to the same level.
 \item We show that for each of our properties,
  $\Rulegreedy$ satisfies the property's conditions for at least $88\%$ of the instances.
 \item We see,
  e.g.,
  that $\Rulegreedy$ performs well compared to~$\Rulelex$
  and achieves a
  higher
  minimum agent score on average
  than~$\Ruleapp$
  (however, this is not always the case).
\end{compactitem}
Overall,
our results recommend~$\Rulelex$.
When runtime is central,
our results promote $\Rulegreedy$ as a good heuristic for $\Rulelex$.

Proof details to results marked with~\appsymb{} are deferred to
\ifapp%
  the appendix.
\else%
  a long version of the paper.
\fi%

\paragraph{Related Work.}
Selecting committees is an important topic from computational social choice with %
numerous applications %
following different goals (cf.\ \citet{EFSS17})
such as individual excellence, proportionality, and diversity.
The latter, our focus, is important whenever large parts of agents shall be covered or satisfied, but it is rather hard to formalize (see Section~4.2 in \citet{Drep17261}).
The classical way is to follow an egalitarian approach~\citep{AFGST18}, where 
the least satisfied agent defines the quality of a committee.

Our model falls into the category of temporal elections (see \citet{TV_ElkindAAAI} for a survey).
Computational aspects of finding sequences of committees have been considered in this context~\cite{TV_ElkindECAI},
including rules for the special setting with two committees that shall overlap as much as possible~\cite{TV_Zech}.
\citet{kellerhals2021parameterized} and \citet{BFK22} require a minimum satisfaction
in each time step with additional constraints on the difference between consecutive committees,
but without aiming for (a minimum) satisfaction of agents.
The model we use in this paper was essentially introduced by \citet{DeltlFB23}
(they consider~$\Cand_t=C$,
$\sum_{c\in \Cand_t} u_t(a,c) \leq 1$,
and~$k_t=k$ for some given~$k$
for every~$t\in\tauS,a\in A$).
\ifaaai%
They studied
  an immediate
\else%
\citet{DeltlFB23} studied
  the following
\fi%
computational problem with lower bounds
on both~$\yscoremin$ and~$\xscoremin$.
\ifaaai%
\else%
  \decprob{\smlaTsc\,(\smlaAcr{})\!\nolinebreak}{smla}
  {A set~$A$ of $n$~agents,
  a sequence $\calC=(C_1,\dots,C_\tau)$ of subsets of candidate set~$C$ with $m=|C|$,
  a sequence of nomination profiles~$U=(u_1,\dots,u_\tau)$ with~$u_t\colon A\to C_t\cup\{\emptyset\}$,
  a sequence~$\kSeq=(k_1,\dots,k_\tau)$, of non-negative integers,
  and two integers~$x,y\in\Nzero$.}
  {Is there a sequence~$\ComSeq=(\Com_1,\dots,\Com_\tau)$ with~$X_t\subseteq C_t$ and~$|X_t|\leq k_t$ for all~$t\in\tauS$ such that
  \begin{align}
  \xscoremin(\ComSeq) &\geq x
  \text{ and }
  \yscoremin(\ComSeq)  \geq y? \label{prob:smla:y}
  \end{align}
  }
  \noindent
\fi%
They focus on computational aspects,
in particular parameterized algorithmics.
While distinct formal properties of rules 
(the focus here) %
are well-studied
in the classical setting~\citep{AFGST18,BoehmerBFKN20,EFSS17,FSST19,LacknerS23},
research on (offline) sequential elections is very limited.
Only known to us is the work of \citet{CGP24},
where formal properties for proportional %
online, semi-online, and offline sequences are studied.

Categories (levels) are also used by \citet{BoehmerBFKN20},
but with the goal of allocating candidates to categories and not selecting from categories (as we do).
Other aspects of selecting multiple (sub)committees have been considered,
for example, by \citet{BKN20}, who %
also select a sequence of committees.
In that work, 
however,
agents have the same preferences for every time step.
Mostly focusing on single-winner decisions,
\citet{FZC17}, \citet{Lac20}, and \citet{PP13}
do allow evolving preferences, but in an online setting
with different measures of solution quality.
An offline setting is analyzed by \citet{BHPRT21}, but aiming for justified representation.

Related to our setting is participatory budgeting (PB), where a community votes on projects.
Each project comes with an individual price and there is an overall budget. %
\citet{LMR21} study a temporal PB setting where agents are partitioned into groups and the goal
is to have equal (temporal) fairness after some number of allocation rounds or to optimize
the Gini coefficient.
\citet{JSTZ21} study PB with project groups (levels), but aim for optimizing the utilitarian welfare.
\citet{SEH20} design a complex PB framework using judgment aggregation, allowing to model
dependencies between projects or quotas for project types, %
optimizing towards (super-)majorities.

\section{Preliminaries}
\label{sec:basics}

We denote by~$\Nzero$ and~$\N$ the set of natural numbers with and without zero,
respectively.
For two sequences~$a=(a_1,\dots,a_n)$ and~$b=(b_1,\dots,b_m)$,
$a\circ b$ denotes the sequence~$(a_1,\dots,a_n,b_1,\dots,b_m)$.
For two set sequences~$\calS=(S_1,\dots,S_n)$ and~$\calT=(T_1,\dots,T_n)$,
$S\cup T$ denotes the sequence~$(S_1\cup T_1,\dots,S_n \cup T_n)$.

Let 
$A$ be a set of $n$ agents, 
$\CandSeq=(\Cand_1,\dots,\Cand_\tau)$ be a sequence of subsets from a candidate set~$C$ of size~$m$,
and 
$U=(u_1,\dots,u_\tau)$ a sequence  of 
    utilities~$u_t\colon A\times \Cand_t\to\Nzero$.
For~$A'\subseteq A$,
we write~$u_t(A',\cdot) \ceq \sum_{a\in A'} u_t(a,\cdot)$ for short.
Similarly,
for~$X\subseteq C_t$,
we write~$u_t(\cdot,X) \ceq \sum_{c\in X} u_t(\cdot,c)$ for short.

Let~$\mathbf{E}$ denote the set of all elections.
For each election~$\calE=(A,\CandSeq,U,\kSeq)\in\mathbf{E}$,
let~$\ComSeqSet(\calE)=\{\ComSeq=(\Com_1,\dots,\Com_\tau) \mid \forall t:\:\Com_t\subseteq C_t\}$ denote the set of all committee sequences for~$\Elec$.
A committee $\Com\subseteq C$ is \emph{valid for level~$t$} if $\Com\subseteq C_t$ and $|\Com|= k_t$.
A committee sequence~$\ComSeq=(\Com_1,\dots,\Com_\tau)$ is \emph{valid} if for each~$t\in\set{\tau}$, committee~$\Com_t$ is valid for level~$t$.
A rule is a mapping
$\Rule(\Elec)\mapsto \ComSeqSet'$
with
$\ComSeqSet'\subseteq \ComSeqSet(\Elec)$.
For instance,
\ifaaai%
  $\Rulebase(\calE) \ceq \{\text{valid }\ComSeq\in\ComSeqSet{}(\calE)\} =\{(\Com_1,\dots,\Com_\tau)\in \ComSeqSet{}(\calE) \mid \forall t:\: |\Com_t|= k_t\}$
\else%
  \begin{align}
  \begin{aligned}
  \Rulebase(\calE) &\ceq \{\text{valid }\ComSeq\in\ComSeqSet{}(\calE)\}
  \\
  &=\{(\Com_1,\dots,\Com_\tau)\in \ComSeqSet{}(\calE) \mid \forall t:\: |\Com_t|= k_t\}.
  \label{eq:valid}
  \end{aligned}
  \end{align}
\fi%
is the rule that maps every election to its set of valid committee sequences.

\section{Central Rules}
\label{sec:centralrules}
\appendixsection{sec:centralrules}
Analogously to \cref{ex:main}, we first formalize the straightforward (but non-egalitarian) \emph{sum rule}~$\Ruleapp$, used ``as benchmark.''

\paragraph{Sum Rule.}
\label{sec:app}

Our first rule selects committees in an obvious way:
independently for each level,
select iteratively the candidate that receives the highest
utility.
For every~$t\in\set{\tau}$,
let
$\calP_t^*(\Elec) \ceq \argmaxlim_{\Com'\subseteq \Cand_t:\: |\Com'|= k_t} \sum_{a\in A} \sum_{c\in \Com'} u_t(a,c)$ %
be the set of all valid committees  %
that reach the highest
sum of utilities.
We define
\begin{align}
  \Ruleapp(\calE) \ceq \calP_1^*(\Elec)\times\dots\times\calP_\tau^*(\Elec).
  \label{eq:app}
\end{align}
\noindent \looseness -1
Greedy selection of candidates maximally increasing the sum of utilities
finds a winning committee sequence for~$\Ruleapp$ in polynomial time.

\paragraph{Egalitarian Sum Rules.}
\label{sec:mmms}

\newcommand{\mxlscr}{\theta}
\newcommand{\mxsscr}{\sigma}
\newcommand{\thevec}{\vec{\mathbf{v}}}
\newcommand{\whvs}{\vec{\mathbf{w}}_{\ymaxmin,\xmaxmin,\maxsum}}
\newcommand{\whsv}{\vec{\mathbf{w}}_{\ymaxmin,\maxsum,\xmaxmin}}

The most basic egalitarian rule
is
\begin{align}
\Ruleh(\Elec) &\ceq \argmax_{\ComSeq\in \Rulebase(\Elec)} \yscoremin(\Elec,\ComSeq).
\end{align}
We aim to study three rules that select committee sequences
from $\Ruleh$'s co-winning set
regarding further optimization goals.
These three rules are
$\Rulehsv$ and~$\Rulehvs$,
and
$\Rulelex$.
We will show that finding a winner for these rules is \NP-hard,
even in quite restricted settings.
\citet[Theorem 2]{DeltlFB23} proved that $\Ruleh$ when $u_t(a,C_t)\leq 1$ for every level~$t$ is \NP-hard even if~$\tau\geq 2$
or if~$k\geq 1$.
Their problem is solvable in polynomial-time if~$\tau=1$ and in~$f(n)\cdot |I|^{O(1)}$ time with instance~$I$ having~$n$ agents (and hence polynomial-time solvable for a constant number of agents).
We prove that this is quite different in our model
due to the generalized utilities
(for details about the Exponential Time Hypothesis (ETH),
see~\cite{ImpagliazzoP01}).

\begin{theorem}[\appref{thm:ruleh:nphard}]
 \label{thm:ruleh:nphard}
 The problem of finding a winning committee sequence for $\Ruleh$
 is 
 \begin{compactenum}[(a)]
  \item \NP-hard, when the utilities are encoded in binary,
 even if there are only two agents and~$k_t=1$ for every level~$t$;
  \item not solvable in~$f(n)\cdot |I|^{O(1)}$ time for any function~$f$ only depending on the number~$n$ of agents in instance~$I$, 
  even for unarily encoded utilities, 
  unless $W[1]= FPT$;
  \item \NP-hard
 even if $\tau=1$ and~$u_t(a,C)\leq 2$ for every level~$t$;
 This case cannot be solved in~$2^{o(m)}\cdot (n+m)^{O(1)}$ time
 unless the ETH breaks.
 \end{compactenum}
\end{theorem}

\begin{proof}
 We prove (b) and defer the proofs 
 of (a) and (c) due to the space constraints to 
 \ifapp%
  the appendix.
 \else%
  a long version of the paper.
 \fi%
 
 (b) We give a polynomial-time many-one reduction from \prob{Unary Bin Packing},
 where,
 given an instance~$I$ consisting of a multiset~$X=\{x_1,\dots,x_N\}$ of numbers and
 $k,B\in \N$,
 every number being unarily encoded,
 the question is whether there is a partition~$S_1\uplus \dots\uplus S_k=S\ceq \{1,\dots, N\}$ of the index set such that
 $\sum_{s\in S_i} x_s \leq B$ for all~$i\in \set{k}$.
 Let~$A=\{a_1,\dots,a_k\}$,
 $C=\{c_1,\dots,c_k\}$,
 and~$M=1+\sum_{s\in S} x_s$.
 For every~$s\in\set{N}$ construct a level and
 let~$C_s=C$,
 $k_s=1$,
 and the utilities be as follows:
 $u_s(a_i,c_j)\ceq M-x_s$ if~$i=j$,
 and 
 $u_s(a_i,c_j)\ceq M$ if~$i\neq j$.
 Note that the utilities can be unarily encoded in polynomial time.
 We claim that~$I$ is a \yes-instance
 if and only if
 $\yscoremin(\Elec,\ComSeq)\geq N\cdot M - B$.
 The following correspondence holds:
 $s\in S_i$ if and only if~$C_s=\{c_i\}$.
 It holds that~$\yscoremin(\Elec,\ComSeq,a_i)=\sum_{s\not\in S_i} M + \sum_{s\in S_i} (M - x_s) = N\cdot M - \sum_{s\in S_i} x_s$. 
 Hence,
 we have that $\yscoremin(\Elec,\ComSeq,a_i)\geq N\cdot M - B$ if and only if $\sum_{s\in S_i} x_s\leq B$. 
 Since this holds for each agent~$a_i$ and part~$S_i$ of the partition,
 the claim follows.
 
 \citet{JansenKMS13} proved that \prob{Unary Bin Packing}
 cannot be solved in~$f(k)\cdot |I|^{O(1)}$ time for any function~$f$ only depending on~$k$,
 unless~$W[1]=FPT$. 
 The reduction runs in polynomial time and
 constructs~$k$ agents in an equivalent instance.
 \appendixproof{thm:ruleh:nphard}
 {
 (a)
 We give a polynomial-time many-one reduction from \prob{Partition},
 where,
 given an instance~$I$ consisting of a multiset~$X=\{x_1,\dots,x_N\}$ of numbers,
 the question is whether there is a partition~$S_1\uplus S_2=S\ceq \{1,\dots, N\}$ of the index set such that
 $\sum_{s\in S_1} x_s = \sum_{s\in S_2} x_s$.
 Let~$A\ceq \{a_1,a_2\}$.
 Construct a level~$t$ for each number~$x_t$ with~$C_t\ceq \{c_1,c_2\}$ and
 $u_t(a_1,c_1)=u_t(a_2,c_2)=x_t$ and~$u_t(a_1,c_2)=u_t(a_2,c_1)=0$.
 We claim that~$I$ is a \yes-instance 
 if and only~$\yscoremin(\Elec,\ComSeq)\geq \frac{1}{2}\sum_{s\in S}x_s$.
 We have the following correspondence: 
 $s\in S_i \iff c_i\in \Com_s$.
 Given this,
 it then follows that~$\yscore(\Elec,(\Com_1,\dots,\Com_N),a_1)=\frac{1}{2}\sum_{s\in S}x_s$
 if and only if~$\sum_{s\in S_1} x_s=\frac{1}{2}\sum_{s\in S}x_s$.
 Thus, the claim follows.
 
 (c)
 We give a polynomial-time many-one reduction from \prob{Vertex Cover},
 where,
 given an instance~$I$ consisting of an undirected graph~$G=(V,E)$ and an integer~$k\in\N$,
 the question is whether there is a subset~$X\subseteq V$ with~$|X|\leq k$ such that each edge~$e\in E$ has an endpoint in~$X$. 
 Let~$A\ceq \{a_e\mid e\in E\}$, 
 with~$M\ceq|E|$,
 and~$C=V$.
 Let~$u(a_e,v)=1 \iff v\in e$.
 We claim that~$I$ is a \yes-instance if and only if
 there is winner~$\Com$ such that
 $\yscoremin(\Elec,\Com)\geq 1$,
 where~$\Elec\ceq(A,C,u,k)$.
 We have that for~$X\subseteq V$ with $|X|\leq k$,
 for every edge~$e\in E$ it holds true that
 $X\cap e\neq \emptyset$ if and only if~$\yscore(\Elec,X,a_e)\geq 1$,
 Hence,
 $X$ is a vertex cover
 if and only if~$\yscoremin(\Elec,X)\geq 1$.
 
 It is well-known that \prob{Vertex Cover} cannot be solved in~$2^{o(|V|+|E|)}\cdot (|V|+|E|)^{O(1)}$ time unless the ETH breaks.
 Since in our reduction we have~$n=|E|$ and~$m=|V|$,
 the claim follows.
 }
\end{proof}

We prove that every winner for
$\Rulehsv$, 
$\Rulehvs$, or
$\Rulelex$ is also winning for~$\Ruleh$,
and hence conclude \NP-hardness also for these rules.

Our first two rules' additional optimization goals are based on the
\emph{\xscrTxt} score~$\xscore(\Elec,\ComSeq,t) \ceq \sum_{a\in A}\sum_{c\in \Com_t} u_t(a,c)$
and directly take into account the minimum level score
$\xscoremin(\Elec,\ComSeq) \ceq \min_{t\in\tauS} \xscore(\Elec,\ComSeq,t)$ and
the sum score~$\yscoresum(\calE,\ComSeq) \ceq \sum_{a\in A} \yscore(\calE,\ComSeq,a)$.
While~$\Rulehvs$ maximizes the minimum level score before the sum score,
$\Rulehsv$ maximizes the sum score before the level score 
(recall that both maximize the minimum agent score first).
We formalize this by assigning suitable weights 
to the corresponding scores as follows.

Let 
$\mxlscr(\Elec) \ceq 1 + \max_{t\in T,\ComSeq\in\ComSeqSet(\Elec)} \xscore(\Elec,\ComSeq,t)$ be the maximum level score plus one
and
$\mxsscr(\Elec) \ceq 1 + \max_{\ComSeq\in\ComSeqSet(\Elec)} \yscoresum(\Elec,\ComSeq)$
the maximum sum score plus one.
Let~$\thevec(\Elec,\ComSeq)=(\yscoremin(\Elec,\ComSeq),\xscoremin(\Elec,\ComSeq),\yscoresum(\Elec,\ComSeq))$
be the vector consisting of the three scores~$\yscoremin$,
$\xscoremin$,
and 
$\yscoresum$.
Intuitively,
we define weights using $\mxlscr$ and~$\mxsscr$ 
to the scores in~$\thevec$
such that improving one score is better than one other,
and improving the third score is even better than improving both of the former.
We define
\begin{align*}
 \Rulehvs(\Elec) &\ceq \argmax_{\ComSeq\in\Rulebase(\Elec)} \thevec(\Elec,\ComSeq) \cdot
 \begin{pmatrix}
\mxlscr(\Elec)\cdot \mxsscr(\Elec) \\
\mxsscr(\Elec) \\
1
\end{pmatrix}
 \\
 \Rulehsv(\Elec) &\ceq \argmax_{\ComSeq\in\Rulebase(\Elec)} \thevec(\Elec,\ComSeq) \cdot
 \begin{pmatrix}
\mxlscr(\Elec)\cdot \mxsscr(\Elec) \\
1 \\
\mxlscr(\Elec)
\end{pmatrix}.
\end{align*}

\begin{lemma}
 \label{lem:supposedtodo}
 \begin{inparaenum}[(i)]
  \item $\ComSeq\in\Rulehvs(\Elec)$ if and only if~$\ComSeq\in\Ruleh(\Elec)$ has maximal sum score among all that have maximal level score.\label{lem:supposedtodo:hvs}
  \item $\ComSeq\in\Rulehsv(\Elec)$ if and only if~$\ComSeq\in\Ruleh(\Elec)$ has maximal level score among all that have maximal sum score.\label{lem:supposedtodo:hsv}
 \end{inparaenum}
\end{lemma}

\appendixproof{lem:supposedtodo}
{
\begin{proof}
 \eqref{lem:supposedtodo:hvs}:
 Let~$\whvs(\Elec) \ceq (\mxlscr(\Elec)\cdot \mxsscr(\Elec),\mxsscr(\Elec),1)^{\rm tr}$.
 
 \RD{} 
 Let $\ComSeq\in\Rulehvs(\Elec)$ and let~$\thevec(\Elec,\ComSeq)=(x,y,z)$.
 Suppose that~$\ComSeq\not\in\Ruleh(\Elec)$,
 that is,
 there is~$\ComSeq'\in\Ruleh(\Elec)$ with~$x'\ceq \yscoremin(\Elec,\ComSeq')>x$.
 Let~$\thevec(\Elec,\ComSeq')=(x',y',z')$.
 We have $(x',y',z')\cdot \whvs(\Elec) > x'\cdot \mxlscr(\Elec)\cdot \mxsscr(\Elec)
 \geq x\cdot \mxlscr(\Elec)\cdot \mxsscr(\Elec) + 1\cdot \mxlscr(\Elec)\cdot \mxsscr(\Elec)$.
 Now note that~$\mxlscr(\Elec)\cdot \mxsscr(\Elec) = (\mxlscr(\Elec)-1)\cdot \mxsscr(\Elec) + \mxsscr(\Elec) > y\cdot \mxsscr(\Elec) + z$.
 Thus,
 $(x',y',z')\cdot \whvs(\Elec)>(x,y,z)\cdot \whvs(\Elec)$
 and hence~$\ComSeq\not\in\Rulehvs(\Elec)$,
 a contradiction.
 
 Next,
 suppose that there is~$\ComSeq'\in \Ruleh(\Elec)$ with higher level score
 than~$\ComSeq$,
 that is,
 with
 $\thevec(\Elec,\ComSeq')=(x,y',z')$ and~$y'>y$.
 We have~$(x,y',z')\cdot \whvs(\Elec) - (x,y,z)\cdot \whvs(\Elec)
 = (y'-y)\cdot \mxsscr(\Elec) + (z' - z) \geq \mxsscr(\Elec) - z > 0$.
 Again,
 it follows that
 $\ComSeq\not\in\Rulehvs(\Elec)$,
 a contradiction.
 
 Finally,
 suppose that there is~$\ComSeq'\in \Ruleh(\Elec)$ with the same level score as~$\ComSeq$ but higher sum score,
 that is,
 with
 $\thevec(\Elec,\ComSeq')=(x,y,z')$ and~$z'>z$.
 We have~$(x,y',z')\cdot \whvs(\Elec) - (x,y,z)\cdot \whvs(\Elec)
 = z' - z > 0$.
 Again,
 it follows that
 $\ComSeq\not\in\Rulehvs(\Elec)$,
 a contradiction.
 
 \LD{}
 Let $\ComSeq\in\Ruleh(\Elec)$ and~$\ComSeq$ has maximal sum score among all that have maximal level score.
 Let~$\thevec(\Elec,\ComSeq)=(x,y,z)$.
 Suppose that~$\ComSeq\not\in\Rulehvs(\Elec)$,
 that is,
 there is~$\ComSeq'\in\Rulehvs(\Elec)$.
 Let~$\thevec(\Elec,\ComSeq')=(x',y',z')$.
 Since~$\ComSeq\in\Ruleh(\Elec)$,
 we know that~$x'\leq x$.
 Suppose that~$x'<x$.
 Then~$(x,y,z)\cdot \whvs(\Elec) > x\cdot \mxlscr(\Elec)\cdot \mxsscr(\Elec)
 \geq x'\cdot \mxlscr(\Elec)\cdot \mxsscr(\Elec) + 1\cdot \mxlscr(\Elec)\cdot \mxsscr(\Elec) = x'\cdot \mxlscr(\Elec)\cdot \mxsscr(\Elec) + (\mxlscr(\Elec)-1)\cdot \mxsscr(\Elec) + \mxsscr(\Elec) > x'\cdot \mxlscr(\Elec)\cdot \mxsscr(\Elec) + y'\cdot \mxsscr(\Elec) + z'$,
 a contradiction to the fact that~$\ComSeq'\in\Rulehvs(\Elec)$.
 Hence,
 $x'=x$.
 Thus,
 we know that~$y'\leq y$.
 We have~$(x,y,z)\cdot \whvs(\Elec) - (x,y',z')\cdot \whvs(\Elec)
 = (y-y')\cdot \mxsscr(\Elec) + (z - z')$.
 Suppose that~$y'<y$,
 then the right-hand side 
 is larger than~$\mxsscr(\Elec) - z' > 0$,
 a contradiction.
 It follows that~$y'=y$ and hence~$z'\leq z$.
 Now suppose that~$z'<z$.
 Then the right-hand side 
 equals~$z - z' > 0$,
 a contradiction.
 It follows that~$\ComSeq\in\Rulehvs(\Elec)$.
 
 \eqref{lem:supposedtodo:hsv}:
 Let $\whsv(\Elec) \ceq  (\mxlscr(\Elec)\cdot \mxsscr(\Elec),1,\mxlscr(\Elec))^{\rm tr}$.
 Let $\ComSeq\in\Rulehsv(\Elec)$ and let~$\thevec(\Elec,\ComSeq)=(x,y,z)$.
 
 Suppose that~$\ComSeq\not\in\Ruleh(\Elec)$,
 that is,
 there is~$\ComSeq'\in\Ruleh(\Elec)$ with~$x'\ceq \yscoremin(\Elec,\ComSeq')>x$.
 Let~$\thevec(\Elec,\ComSeq')=(x',y',z')$.
 We have $(x',y',z')\cdot \whsv(\Elec) > x'\cdot \mxlscr(\Elec)\cdot \mxsscr(\Elec)
 \geq x\cdot \mxlscr(\Elec)\cdot \mxsscr(\Elec) + 1\cdot \mxlscr(\Elec)\cdot \mxsscr(\Elec)$.
 Now note that~$\mxlscr(\Elec)\cdot \mxsscr(\Elec) = (\mxsscr(\Elec)-1)\cdot \mxlscr(\Elec) + \mxlscr(\Elec) > y + z\cdot \mxlscr(\Elec)$.
 Thus,
 $(x',y',z')\cdot \whsv(\Elec)>(x,y,z)\cdot \whsv(\Elec)$
 and hence~$\ComSeq\not\in\Rulehsv(\Elec)$,
 a contradiction.
 
 Next,
 suppose that there is~$\ComSeq'\in \Ruleh(\Elec)$ with higher sum score
 than~$\ComSeq$,
 that is,
 with
 $\thevec(\Elec,\ComSeq')=(x,y',z')$ and~$z'>z$.
 We have~$(x,y',z')\cdot \whsv(\Elec) - (x,y,z)\cdot \whsv(\Elec)
 = (y'-y) + (z' - z)\cdot \mxlscr(\Elec) \geq \mxlscr(\Elec) - y > 0$.
 Again,
 it follows that
 $\ComSeq\not\in\Rulehsv(\Elec)$,
 a contradiction.
 
 Finally,
 suppose that there is~$\ComSeq'\in \Ruleh(\Elec)$ with the same sum score as~$\ComSeq$ but higher level score,
 that is,
 with
 $\thevec(\Elec,\ComSeq')=(x,y',z)$ and~$y'>y$.
 We have~$(x,y',z)\cdot \whsv(\Elec) - (x,y,z)\cdot \whsv(\Elec)
 = y' - y > 0$.
 Again,
 it follows that
 $\ComSeq\not\in\Rulehsv(\Elec)$,
 a contradiction.
\end{proof}
}

Due to \cref{lem:supposedtodo}, it follows that 
$\Rulehvs(\Elec)\subseteq \Ruleh(\Elec)$ and $\Rulehsv(\Elec)\subseteq \Ruleh(\Elec)$
for every election~$\Elec$.
Thus,
due to~\cref{thm:ruleh:nphard},
finding a winner for each is \NP-hard.

\paragraph{Lex Rule.}
\label{sec:lex}

\newcommand{\satvec}{\operatorname{\overrightarrow{\sat}}}
\newcommand{\UB}{\ensuremath{Z}}

We rate our next rule~$\Rulelex$
as the ``most egalitarian''
of our rules:
It not only maximizes the minimum agent score,
but also the second smallest as the second goal,
and so on.
Let $\sat(\Elec,\ComSeq,i) \ceq |\{a\in A\mid \yscore(\Elec,\ComSeq,a)=i\}|$
be the number of agents with an agent score of~$i$.
$\Rulelex$ maximizes lexicographically
the agents' satisfaction histogram
$\satvec(\Elec,\ComSeq)=(\sat(\Elec,\ComSeq,0),\sat(\Elec,\ComSeq,1),\ldots,\sat(\Elec,\ComSeq,\UB(\Elec)))$,
where~$\UB(\Elec)\ceq \max_{a\in A} \sum_{t\in\ \tauS} u_t(a,\Cand_t)$.\footnote{%
Here,
we lexicographically order as usual:
For two number sequences~$x=(x_1,\dots,x_n)$ and~$y=(y_1,\dots,y_n)$,
we write~$x\prec y$ if for the smallest index~$i$ where 
$x$ and~$y$ differ we have $x_i<y_i$.
}
Note that this equivalent to maximizing lexicographically the vector~$\lexord$ from~\cref{ex:main}.
Again,
we can formalize~$\Rulelex$ by assigning suitable weights as follows.
\begin{align}
  \Rulelex(\calE) \ceq &\argmin_{\ComSeq\in\Rulebase(\Elec)} \ems(\Elec,\ComSeq)
  \text{ with}
  \label{eq:lex}
  \\
  \ems(&\Elec,\ComSeq) \ceq \sum_{i=0}^{\UB(\Elec)} \sat(\Elec,\ComSeq,i)\cdot (|A|+1)^{\UB(\Elec)-i} .
  \notag
\end{align}

The following connection is well-known.

\begin{lemma}[\appref{obs:lex:scoreandvec}]
 \label{obs:lex:scoreandvec}
 For every election~$\Elec$ and 
 committee sequences~$\ComSeq,\ComSeq'\in\ComSeqSet(\Elec)$ 
 it holds true that
 $\ems(\Elec,\ComSeq)<\ems(\Elec,\ComSeq') 
 \iff 
 \satvec(\Elec,\ComSeq) \prec \satvec(\Elec,\ComSeq')$.
\end{lemma}

\appendixproof{obs:lex:scoreandvec}
{
  \begin{proof}
  Let~$j\geq 0$ be the first index where $\satvec(\Elec,\ComSeq)$ and $\satvec(\Elec,\ComSeq')$ differ
  and $N\ceq |A|+1$.
  Then (with $\UB = \UB(\Elec)$):
  \begin{align*}
    &\ems(\Elec,\ComSeq)<\ems(\Elec,\ComSeq') 
    \\ &
    \iff 
    \sum_{i=j}^\UB (\sat(\Elec,\ComSeq,i)-\sat(\Elec,\ComSeq',i))\cdot N^{\UB-i} < 0
    \\
    &\iff 
    \sum_{i=j+1}^\UB (\sat(\Elec,\ComSeq,i)-\sat(\Elec,\ComSeq',i))\cdot N^{\UB-i} 
    \\ &\qquad\qquad 
    < (\sat(\Elec,\ComSeq',j)-\sat(\Elec,\ComSeq,j))\cdot N^{\UB-j}
      \\
    &\iff 
    \sum_{i=0}^{\UB-j-1} (\sat(\Elec,\ComSeq,\UB-i)-\sat(\Elec,\ComSeq',\UB-i))\cdot N^{i} 
    \\ &\qquad\qquad
    < (\sat(\Elec,\ComSeq',j)-\sat(\Elec,\ComSeq,j))\cdot N^{\UB-j}
    \\
    &\iff 
    \satvec(\Elec,\ComSeq) \prec \satvec(\Elec,\ComSeq').
  \end{align*}
  Note that
  if~$\satvec(\Elec,\ComSeq) \prec \satvec(\Elec,\ComSeq')$,
  then the right-hand side in the line before is at least~$N^{\UB-j}$.
  Since each coefficient is between $-|A|$ and~$|A|$,
  we have for the left-hand side:
  \begin{align*}
      &|\sum_{i=0}^{\UB-j-1} (\sat(\Elec,\ComSeq,\UB-i)-\sat(\Elec,\ComSeq',\UB-i))\cdot N^{i}|
      \\
      &\leq \sum_{i=0}^{\UB-j-1} |(\sat(\Elec,\ComSeq,\UB-i)-\sat(\Elec,\ComSeq',\UB-i))|\cdot N^{i}
      \\
      &\leq \sum_{i=0}^{\UB-j-1} |A|\cdot N^{i} = |A|\cdot \left(\frac{1-N^{\UB-j}}{1-N}\right)
      = N^{\UB-j}-1 
  \end{align*}
  Since~$j$ is chosen to be the first index where they differ,
  also the other direction follows.
  \end{proof}
}

Due to~\cref{obs:lex:scoreandvec}, it follows that
$\Rulelex(\Elec)\subseteq \Ruleh(\Elec)$ for every election~$\Elec$.
Hence, finding a winner for~$\Rulelex$ is \NP-hard due to~\cref{thm:ruleh:nphard}.

\paragraph{Greedy Rule.}
\label{sec:greedy}

Since finding a winning committee sequence for our three preceding egalitarian rules
is \NP-hard,
we present a polynomial-time greedy rule:
On each level $t$, the initial solution selects all candidates 
receiving a positive utility
from at least one agent if there are exactly $k_t$ many of them, and 
no agent
otherwise.
Then the following is repeated:
Choose a level $t$ with a
current committee smaller than~$k_t$ and a currently unselected candidate $c$ on level $t$ which leads to the largest
decrease of the current $\ems$ score and add this $c$ to level $t$.

\begin{definition}\label{def:gree}
 \label{def:greedy}
 Rule~$\Rulegreedy$ is formally defined as follows.
 \begin{compactenum}
  \item For each~$t\in\tauS$, let
  $C^+_t \ceq \{c\in C_t \mid u_t(A,c) > 0 \}$
  and~$k_t\ceq \min\{k_t,|C^+_t|\}$.
  Start with 
  $\ComSeq=(\Com_1,\dots,\Com_\tau)$, 
  where $\Com_t = C^+_t$ if $|C^+_t| = k_t$, 
  and $\Com_t = \emptyset$ otherwise, 
  for each $t\in\tauS$. 
  \item Let $T' \ceq \{ t\in \tauS \mid |\Com_t| < k_t \}$. 
    Stop if~$T'=\emptyset$.
  \item Pick some $(t', c') \in \argminlim_{(t, c)\in S} \ems(\Elec,\ComSeq \cup_t \{c\})$,
        where $S = \left\lbrace(t, c) \,\middle|\, t \in T', c \in C^+_t \setminus \Com_t \right\rbrace$ and
        $\ComSeq \cup_t \{c\} := (\Com_1, \dots, \Com_t \cup \{c\}, \dots, \Com_\tau)$.
        Add this $c'$ to $\Com_{t'}$.
  Go to step~2.
 \end{compactenum}
\end{definition}

\cref{def:gree}
also describes how
to find a winning committee sequence for~$\Rulegreedy$ in polynomial time.
Although~$\Rulegreedy$ treats the least-satisfied agent that still can improve in each step,
its outcome could be ``less'' egalitarian than one by our non-egalitarian rule~$\Ruleapp$.

\begin{observation}[\appref{lem:greedyvsapp}]
 \label{lem:greedyvsapp}
 There is an election~$\Elec$ and 
 two committee sequences~$\ComSeq\in\Ruleapp(\Elec)$
 and $\ComSeq'\in\Rulegreedy(\Elec)$ such that
 $\yscoremin(\Elec,\ComSeq)>\yscoremin(\Elec,\ComSeq')$.
\end{observation}

\appendixproof{lem:greedyvsapp}
{
\begin{proof}
 See~\cref{fig:greedyvsapp}.
 \begin{figure}[h!]
 \centering
 \begin{tikzpicture}
  \def\xr{1}
  \def\yr{1}
  \def\AIx{0.5}
  \def\AIy{0.5}
  \begin{scope}
    \agentInit{\AIx}{\AIy}
    \addagent{a,c,e}
    \addagent{a,c,e}
    \addagent{a,d,f}
    \addagent{b,c,-}
    \addagent{b,d,-}
    \addagent{b,d,-}
    \addagent{b,d,-}
    \addagent{-,d,-}
    
    \FEhiliA{4/1,5/1,6/1,7/1,3/2,5/2,6/2,7/2,8/2,1/3,2/3}
    \FEhiliB{1/1,2/1,3/1,3/2,5/2,6/2,7/2,8/2,1/3,2/3}
  \end{scope}
 \end{tikzpicture}
 \caption{%
 Example election~$\Elec$ for~\cref{lem:greedyvsapp} with~$\kSeq=(1,1,1)$.
 We have~$\{(\{b\},\{d\},\{e\})\}=\Ruleapp$ (\mblue{}) 
 with $\yscoremin(\Elec,(\{b\},\{d\},\{e\}))=1$
 and~$\{(\{a\},\{d\},\{e\})\}=\Rulegreedy$ (\mgreen{}) 
 with $\yscoremin(\Elec,(\{a\},\{d\},\{e\}))=0$.
 }
 \label{fig:greedyvsapp}
\end{figure}
\end{proof}
}

\section{Central Properties}
\label{sec:props}
\appendixsection{sec:centralrules}
In this section, 
we discuss the fingerprints of our rules 
on 
central properties 
(see \cref{tab:results} for an overview).
In our analysis,
the basic egalitarian rule~$\Ruleh$ serves as our baseline
in the following sense:
\begin{compactenum}[(a)]
  \item %
        We investigate $\Ruleh$ for some desirable property.
        \label{meth:props}
  \item We test our rules against this property and evaluate them as follows:
  The property \emph{must} be \respected{} if $\Ruleh$ \respects{} it,
  and \emph{should preferably} be \respected{} if $\Ruleh$ \rejects{} it.
  \label{meth:eval}
\end{compactenum}
Our obtained set of desirable properties
will allow each pair of rules to be distinguished (by one property). %
Through convention from~\eqref{meth:eval},
our analysis will show that~$\Rulelex$ is ``most egalitarian'' in the sense that
it is the only rule that not only \respects{} all properties which $\Ruleh$ \respects{},
but also \respects{} desirable properties that $\Ruleh$ \rejects{}.
We start with the desirable properties that $\Ruleh$ \respects{}.

\paragraph{Safeness.}
Since~$\Ruleh$ maximizes the minimum agent score,
$\Ruleh$ is safe against ``merging'' two elections in the following sense:
In a merged election,
the satisfaction of the least satisfied agent(s) shall not get worse.
This is desirable since,
conversely,
``splitting'' an election should not lead to better results.
We can merge in two different ways:
Firstly, 
along two different disjoint sets of levels with the same
agent set,
and secondly
along two disjoint agent sets
voting over the same set of levels, but possibly different candidates.
Accordingly,
we derive the following two
properties, which are slightly weaker, 
as they do \emph{not} demand that
the satisfaction of the least satisfied agent(s) does not get worse
in \emph{all} winning committee sequences
of the merged election,
but that there is \emph{at least one} such winning committee sequence.

\begin{property}[Safe Concatenation]
 \label{propty:conchorzsuperadd}
 Rule~$\Rule$
 fulfills \emph{safe concatenation}
 if for every two elections~$\Elec_1=(A,\calC,U,\kSeq)$ and~$\Elec_2=(A,\calC',U',\kSeq')$
 we have that
 for all~$\ComSeq_1\in\Rule(\Elec_1),\ComSeq_2\in\Rule(\Elec_2)$
 there is an~$\ComSeq\in\Rule(\Elec)$ with~$\Elec=(A,\calC\circ \calC',U\circ U',\kSeq\circ \kSeq')$
 so that~$\yscoremin(\Elec,\ComSeq)\geq \yscoremin(\Elec,\ComSeq_1\circ \ComSeq_2)$.
\end{property}

When merging along two disjoint agent sets over the same number of levels instead,
we first need to formalize how we merge utilities:
For utilities $U=(u_1,\dots,u_\tau)$ over $A,C$ and utilities $U'=(u'_1,\dots,u'_\tau)$ over $A',C'$ with $A\cap A'=\emptyset$,
$U\cup U'$ denotes
$\bar u_t(a,c) = u_t(a,c)$ if $a\in A$ and~$c\in \Cand_t$,
$\bar u_t(a,c) = u'_t(a,c)$ if $a\in A'$ and~$c\in \Cand_t'$,
and $\bar u_t(a,c)=0$ otherwise, for every~$t\in\tauS$.
Intuitively, each agent's utilities remain the same and zero utility is given to candidates only appearing in the other election.

\begin{property}[Safe Union]
 \label{propty:unitorzsuperadd}
 Rule~$\Rule$
 fulfills \emph{safe union}
 if for any two elections~$\Elec_1=(A_1,\CandSeq_1,U_1,\kSeq_1)$
 and~$\Elec_2=(A_2,\CandSeq_2,U_2,\kSeq_2)$
 over the same number of levels
 but disjoint agent sets,
 we have that for all~$\ComSeq_1\in\Rule(\Elec_1),\ComSeq_2\in\Rule(\Elec_2)$
 there is an~$\ComSeq\in\Rule(\Elec)$ with~$\Elec=(A_1\cup A_2,\CandSeq_1\cup \CandSeq_2,U_1\cup U_2,\kSeq_1+\kSeq_2)$
 such that~$\yscoremin(\Elec,\ComSeq)\geq \yscoremin(\Elec,\ComSeq_1\cup \ComSeq_2)$.
\end{property}

Since~$\Rulelex$,
$\Rulehvs$, 
and~$\Rulehsv$
maximize the minimum agent score,
they
\respect{} the two preceding properties.
$\Ruleapp$ also 
\respects{} \pref{propty:conchorzsuperadd}
since the concatenation of two winning committee sequences is again winning.
This is not true for~$\Rulegreedy$.

\begin{observation}%
  \label{obs:greedy:rejects:conc}
  $\Rulegreedy$ \rejects{}
  \pref{propty:conchorzsuperadd}.
\end{observation}
\begin{proof}
  See~\cref{fig:greedy:rejects:conc} with~$\kSeq=(1,\dots,1)$:
  \begin{figure}[t!]
   \centering
   \begin{tikzpicture}
    \def\xr{1}
    \def\yr{1}
    \def\AIx{0.6}
    \def\AIy{0.4}
    \def\dst{0.75}
    \begin{scope}[xshift=6*\dst*\xr cm]  
      \agentInit{\AIx}{\AIy}
      \addagent{a,c,e,g}
      \addagent{a,c,f,h}
      \addagent{b,d,e,-}
      \draw[dashed] (1.75*\Axsh,0*\Aysh) to (1.75*\Axsh,-2.75*\Aysh);
      
      \foreach\x/\y in {1/1,2/1,3/2}{\hiliA{\x}{\y}}
      \foreach\x/\y in {1/1,2/1,1/2,2/2}{\hiliB{\x}{\y}}
      \foreach\x/\y in {1/3,3/3,2/4}{\hiliA{\x}{\y}}
      \foreach\x/\y in {1/3,3/3,2/4}{\hiliB{\x}{\y}}
    \end{scope}
   \end{tikzpicture}
  \caption{%
  Counterexample to~\cref{obs:greedy:rejects:conc}.
  All our counterexamples use~$\{0,1\}$-utilities
  where, on each level $t\in\tauS$
  (here depicted column-wise),
  each agent gives a utility of one to at most one
  candidate and zero otherwise,
  (represented
  as $u_t\colon A\to \Cand_t\cup\{-\}$).
  }
  \label{fig:greedy:rejects:conc}
  \end{figure}
  After concatenation (along the dashed line), 
  $(\{a\},\{c\},\{e\},\{h\})$ (\mgreen{}) wins for~$\Rulegreedy$,
  leaving~$a_3$ with an agent score of 1.
  Yet,
  when concatenating individual winners (\mblue{})
  each agent gets an agent score of~2.
\end{proof}

\pref{propty:unitorzsuperadd} is \rejected{}
by both
$\Ruleapp$ and $\Rulegreedy$.

\begin{observation}[\appref{obs:greedy:app:rejects:superadd}]%
  \label{obs:greedy:app:rejects:superadd}
  $\Rulegreedy$ and $\Ruleapp$ \reject{}
   \pref{propty:unitorzsuperadd}
\end{observation}

\appendixproof{obs:greedy:app:rejects:superadd}
{
\begin{proof}
  For~$\Rulegreedy$,
  see~\cref{fig:unithorz}(i)
  with~$\Elec_1$ (top) and~$\Elec_2$ (bottom),
  separated by a dashed line,
  and~$\kSeq=(1,1)$ each:
  \begin{figure}[t]
   \centering
   \begin{tikzpicture}
    \def\xr{1}
    \def\yr{1}
    \def\AIx{0.5}
    \def\AIy{0.5}
    \def\dst{1}
    \newcommand{\mylabel}[1]{\node at (-0.9*\xr,0.0375*\yr)[anchor=south west]{(#1)};}
    \begin{scope}[xshift=9.5*\dst*\xr cm]
      \mylabel{i}
      \agentInit{\AIx}{\AIy}
      \addagent{a,d}
      \addagent{a,d}
      \addagent{a,e}
      \addagent{b,e}
      \foreach\x/\y in {1/1,2/1,3/1,3/2,4/2}{\hiliA{\x}{\y}}
      \foreach\x/\y in {1/1,2/1,3/1,4/1,1/2,2/2,3/2,4/2}{\hiliB{\x}{\y}}
    \end{scope}

    \begin{scope}[xshift=9.5*\dst*\xr cm,yshift=-4*\AIy*\yr cm]
      \agentInit{\AIx}{\AIy}
      \draw[dashed] (-1*\Axsh,0.15*\Aysh) to (1.75*\Axsh,0.15*\Aysh);
      \addagentp{b,f}
      \addagentp{c,-}
      \foreach\x/\y in {2/1,1/2}{\hiliA{\x}{\y}}
      \foreach\x/\y in {1/1}{\hiliB{\x}{\y}}
    \end{scope}

    \begin{scope}[xshift=12*\dst*\xr cm]
      \mylabel{ii}
      \agentInit{\AIx}{\AIy}
      \addagent{a,d}
      \addagent{a,d}
      \addagent{a,e}
      \addagent{b,e}
      \addagent{b,e}

      \foreach\x/\y in {1/1,2/1,3/1,3/2,4/2,5/2}{\hiliA{\x}{\y}}
      \foreach\x/\y in {1/1,2/1,3/1,4/1,5/1,1/2,2/2,3/2,4/2,5/2}{\hiliB{\x}{\y}}
    \end{scope}

    \begin{scope}[xshift=12*\dst*\xr cm,yshift=-5*\AIy*\yr cm]
      \agentInit{\AIx}{\AIy}
      \draw[dashed] (-1*\Axsh,0.15*\Aysh) to (1.75*\Axsh,0.15*\Aysh);
      \addagentp{c,f}
      \foreach\x/\y in {1/1,1/2}{\hiliA{\x}{\y}}
    \end{scope}
   \end{tikzpicture}
   \caption{Counterexamples to~\cref{obs:greedy:app:rejects:superadd}.}
   \label{fig:unithorz}
  \end{figure}
  Note that in the united election with~$\kSeq'=(2,2)$,
  $\Rulegreedy$ selects~$a$ first,
  then~$b$, and then~$d$ and~$e$,
  leaving agent~$a_2'$ with score zero (\mgreen{}).
  However,
  $(\{a\},\{e\})\in \Rulegreedy(\Elec_1)$
  together with $(\{c\},\{f\})\in\Rulegreedy(\Elec_2)$ satisfies every agent at least once
  (\mblue{}).

  For~$\Ruleapp$,
  see~\cref{fig:unithorz}(ii)
  with~$\Elec_1$ (top) and~$\Elec_2$ (bottom),
  separated by a dashed line,
  and $\kSeq=(1,1)$ each:
  Note that~$\Ruleapp(\Elec_1)=(\{a\},\{e\})$ and~$\Ruleapp(\Elec_2)=(\{c\},\{f\})$,
  leaving each agent satisfied at least once (\mblue{}).
  However,
  in the united election~$\Elec$ with~$\kSeq'=(2,2)$,
  $\Ruleapp(\Elec)=(\{a,b\},\{d,e\})$
  (\mgreen{}),
  leaving agent~$a_1'$ unsatisfied.
\end{proof}
}

\paragraph{Consistency.}
\pref{propty:unitorzsuperadd}
provides one way to unite two elections
(along two disjoint agent groups)
where the committee sizes are additive,
i.e.,
the committee sizes are summed up on each level.
Another scenario is where the committee sizes are the same in both elections and do not change in their union.
Note that
in this setting,
it makes no sense to
require that the satisfaction of the least satisfied agent(s) shall not get worse
in the united election.
Yet,
$\Ruleh$ provides the following:
If the two agent groups share a common solution,
then this is also a solution for the
elections' union.

\begin{property}[Sub-Consistency]
  \label{propty:consistency}
  Rule~$\Rule$
  \respects{}
  \emph{sub-consistency}
  if for every two elections~$\Elec_1=(A_1,\CandSeq,U_1,\kSeq)$
  and $\Elec_2=(A_2,\CandSeq,U_2,\kSeq)$
  over the same candidate sequence~$\CandSeq$,
  number of levels,
  and committee bounds,
  but disjoint agent sets~$A_1$ and~$A_2$,
  it holds that
  if~$M\ceq \Rule(\Elec_1)\cap \Rule(\Elec_2)\neq \emptyset$,
  then~$M\subseteq \Rule(A_1\cup A_2,\CandSeq,U_1\cup U_2,\kSeq)$.
\end{property}

Properties similar to~\pref{propty:consistency} are stated by~\citet{BoehmerBFKN20}
(for ``line-up elections'') and \citet{LacknerS23}.

$\Ruleh$
 \respects{} \pref{propty:consistency},
 and so does $\Rulelex$ and~$\Ruleapp$,
 as the only two of our five rules.

\begin{observation}[\appref{obs:esum:lex:app:consistency}]
 \label{obs:esum:lex:app:consistency}
 $\Ruleh$,
 $\Rulelex$
 and $\Ruleapp$
 \respect{}
 \pref{propty:consistency}.
\end{observation}

\appendixproof{obs:esum:lex:app:consistency}
{
\begin{proof}
 For $\Ruleh$:
 Let~$\ComSeq\in M\ceq \Ruleh(\Elec_1)\cap \Ruleh(\Elec_2)$.
 Suppose that~$\ComSeq\not\in\Ruleh(\Elec)$,
 that is,
 there is~$\ComSeq'\in \Rulebase(\Elec)$
 with~$\yscoremin(\Elec,\ComSeq')>\yscoremin(\Elec,\ComSeq)$.
 We have the following:
 \begin{claim}
  \label{lem:subscore}
  Let~$\Elec_i=(A_i,\CandSeq,U_i,\kSeq)$ for~$i\in\{1,2\}$
  with~$A_1\cap A_2=\emptyset$
  and let~$\Elec=(A_1\cup A_2,\CandSeq,U_1\cup U_2,\kSeq)$.
  Let~$\ComSeq\in \Rule(\Elec_1)\cap \Rule(\Elec_2)$.
  If there is~$\ComSeq'\in \Rule(\Elec)$ with~$\yscoremin(\Elec,\ComSeq')>\yscoremin(\Elec,\ComSeq)$,
  then there is an~$i\in\{1,2\}$ such that
  $\yscoremin(\Elec_i,\ComSeq')> \yscoremin(\Elec_i,\ComSeq)$.
  \end{claim}
  \begin{proof}
  Suppose not,
  that is,
  for every~$i\in\{1,2\}$ we have~$\yscoremin(\Elec_i,\ComSeq')\leq \yscoremin(\Elec_i,\ComSeq)$, which implies that
  \begin{align}
    \begin{aligned}
    && \yscoremin(\Elec_i,\ComSeq') &\leq \yscoremin(\Elec_i,\ComSeq) \\
    \implies && \min_{j\in\{1,2\}} \{\yscoremin(\Elec_j,\ComSeq')\} &\leq \min_{j\in\{1,2\}} \{\yscoremin(\Elec_j,\ComSeq)\}
    \label{eq:lem:subscore:min}
    \end{aligned}
  \end{align}
  It follows that
  \begin{align*}
    \yscoremin(\Elec,\ComSeq') &= \min_{j\in\{1,2\}} \{\yscoremin(\Elec_j,\ComSeq')\}
    \\
    &\stackrel{\eqref{eq:lem:subscore:min}}{\leq } \min_{j\in\{1,2\}} \{\yscoremin(\Elec_j,\ComSeq)\}
    \\
    &= \yscoremin(\Elec,\ComSeq),
  \end{align*}
  a contradiction.
  \end{proof}
 By~\cref{lem:subscore}
 it follow that
 $\ComSeq\not\in \Ruleh(\Elec_i)$ for some~$i\in\{1,2\}$,
 a contradiction.

\NewDocumentCommand{\emsb}{m d()}{\ems_{#1}\!\left( #2 \right)}
\NewDocumentCommand{\emsf}{d()}{\ems\!\left( #1 \right)}
 For~$\Rulelex$:
 Let $B\in\left\{ A_1, A_2, A \right\}$ with~$A=A_1\cup A_2$ and
  $\emsb{B}(\ComSeq) \ceq \sum_{i=0}^{\UB(\Elec)}
  \left|\left\{ a \in B \,\middle|\, \yscore(\Elec,\ComSeq,a) = i \right\}\right|
  \cdot \left(\left|A\right| + 1\right)^{\UB(\Elec)-i}$
 with~$\Elec=(A, \CandSeq, U_1 \cup U_2, \kSeq)$.
 It holds that
 $\emsb{A}(\ComSeq) = \ems(\Elec,\ComSeq)$.
 We use the following:
  $\ems(\calE_i, \ComSeq_1)
  \leq \ems(\calE_i, \ComSeq_2) \iff
  \emsb{A_i}(\ComSeq_1) \leq \emsb{A_i}(\ComSeq_2) \,\forall i \in \left\{ 1,2 \right\}$ $(*)$.
 We show that $M\subseteq \Rulelex(\Elec)$.
 Let~$\ComSeq_M \in M$. Then
 $\min_{\ComSeq} \emsb{A}(\ComSeq)
  = \min_{\ComSeq} \left( \emsb{A_1}(\ComSeq) + \emsb{A_2}(\ComSeq) \right)
  \geq \min_{\ComSeq} \emsb{A_1}(\ComSeq) + \min_{\ComSeq} \emsb{A_2}(\ComSeq)
  \oset{(*)}{=} \emsb{A_1}(\ComSeq_M) + \emsb{A_2}(\ComSeq_M)
  = \emsb{A}(\ComSeq_M)$
 and thus~$\ComSeq\in \Rulelex(\Elec)$.

 For~$\Ruleapp$:
 Let~$\Elec_i=(A_i,\CandSeq,U_i=(u_{i,1},\dots,u_{i,\tau}),\kSeq)$
 for each~$i\in\{1,2\}$
 and let~$\Elec=(A_1\cup A_2,\ComSeq,U_1\cup U_2=(u_1,\dots,u_\tau),\kSeq)$.
 Let~$\ComSeq=(\Com_1,\dots,\Com_\tau)\in M$.
 Assume that~$\ComSeq\not\in \Ruleapp(\Elec)$.
 Then there is a~$t\in\tauS$
 such that~$\Com_t\not\in P_t^*(\Elec)$.
 Let~$\Com_t'\in P_t^*(\Elec)$.
 We have that~$u_{1,t}(A,\Com_t)+u_{2,t}(A,\Com_t)<u_{t}(A,\Com_t') = u_{1,t}(A,\Com_t')+u_{2,t}(A,\Com_t')$,
 and thus~$u_{1,t}(A,\Com_t)<u_{1,t}(A,\Com_t')$ or~$u_{2,t}(A,\Com_t)<u_{2,t}(A,\Com_t')$,
 which contradicts the choice of~$\ComSeq$.
\end{proof}
}

Interestingly,
\pref{propty:consistency}
disqualifies not only $\Rulegreedy$,
but also $\Rulehsv$ and~$\Rulehvs$
as egalitarian rules in our sense.

\begin{observation}[\appref{obs:mmmsmm:greedy:consist}]
 \label{obs:mmmsmm:greedy:consist}
 Each of $\Rulehsv$,
 $\Rulehvs$,
 and $\Rulegreedy$
 \rejects{} \pref{propty:consistency}.
\end{observation}

\appendixproof{obs:mmmsmm:greedy:consist}
{
\begin{proof}
  For $\Rulehsv$, $\Rulehvs$: 
  See~\cref{fig:subconst}(i) with~$\kSeq=(1,1)$:
  \begin{figure}[h!]
   \centering
   \begin{tikzpicture}
    \def\xr{1}
    \def\yr{1}
    \def\AIx{0.6}
    \def\AIy{0.5}
    \def\dst{1.25}
    \newcommand{\mylabel}[1]{\node at (-0.9*\xr,0.0375*\yr)[anchor=south west]{(#1)};}
    \begin{scope}[xshift=18.5*\dst*\xr cm]
      \mylabel{i}
      \agentInit{\AIx}{\AIy}
      \addagent{a,c}
      \addagent{b,a}
      \addagent{b,b}
      \addagent{a,b}
      \FEhiliA{1/1,4/1,3/2,4/2}
      \FEhiliB{1/1,4/1,2/2}
    \end{scope}

    \begin{scope}[xshift=18.5*\dst*\xr cm,yshift=-4*\AIy*\yr cm]
      \agentInit{\AIx}{\AIy}
      \draw[dashed] (-1*\Axsh,0.125*\Aysh) -- (1.8*\Axsh,0.125*\Aysh);
      \addagentp{b,b}
      \addagentp{a,a}
      \addagentp{a,a}
      \addagentp{a,a}
      \FEhiliA{2/1,3/1,4/1,1/2}
      \FEhiliB{2/1,3/1,4/1,2/2,3/2,4/2}
    \end{scope}

    \begin{scope}[xshift=21*\dst*\xr cm]
      \mylabel{ii}
      \agentInit{\AIx}{\AIy}
      \addagent{a,c,e}
      \addagent{a,c,f}
      \FEhiliA{1/1,2/1,1/2,2/2,2/3}
      \FEhiliB{1/1,2/1,1/2,2/2,1/3}
    \end{scope}

    \begin{scope}[xshift=21*\dst*\xr cm,yshift=-2*\AIy*\yr cm]
      \agentInit{\AIx}{\AIy}
      \draw[dashed] (-1*\Axsh,0.125*\Aysh) -- (2.5*\Axsh,0.125*\Aysh);
      \addagentp{b,c,f}
      \addagentp{a,d,e}
      \addagentp{a,-,e}
      \FEhiliA{2/1,3/1,1/2,1/3}
      \FEhiliB{2/1,3/1,1/2,2/3,3/3}
    \end{scope}
   \end{tikzpicture}
   \caption{Counterexamples to~\cref{obs:mmmsmm:greedy:consist}}
   \label{fig:subconst}
  \end{figure}
  $(\{a\},\{b\})$ (\mblue{}) is the only common winner
  for~$\Elec_1$ (top) and~$\Elec_2$ (bottom), but
  $(\{a\},\{a\})$ (\mgreen{}) is the only winner for their union.
  
  For~$\Rulegreedy$:
  See~\cref{fig:subconst}(ii)
  with~$\kSeq=(1,1,1)$:
  A common winner is~$(\{a\},\{c\},\{f\})$ (\mblue{}):
  For~$\Elec_2$ (bottom),
  selecting first~$a$,
  then~$f$,
  and then~$c$ is possible.
  For the union,
  the only winner is $(\{a\},\{c\},\{e\})$
  (\mgreen{}).
\end{proof}
}

\paragraph{Pareto efficiency.}
One obvious weakness of~$\Ruleh$ 
is that there is no tie-breaking mechanism among all solutions that maximize~$\yscoremin$.
One very common mechanism is to output no solutions that are
\emph{dominated}:
For two committee sequences~$\ComSeq,\ComSeq'$ we say 
that $\ComSeq$ \emph{\hdominate{s}}
$\ComSeq'$
if 
$\forall\,a\in A:\: 
\yscore(\Elec,\ComSeq,a)\geq \yscore(\Elec,\ComSeq',a)$ 
and 
$\exists\,a'\in A:\: 
\yscore(\Elec,\ComSeq,a')> \yscore(\Elec,\ComSeq',a')$.
Consider~\cref{fig:super}(i).
\begin{figure}[h!]
\centering
  \begin{tikzpicture}
    \def\xr{1}
    \def\yr{1}
    \def\AIx{0.6}
    \def\AIy{0.4}
    \def\dst{1.25}
    \newcommand{\mylabel}[1]{\node at (-0.825*\xr,-0.375*\yr)[anchor=south east]{(#1)};}

    \begin{scope}[xshift=\xr*0.0 cm]
      \mylabel{i}
      \agentInit{\AIx}{\AIy}
      \addagent{a,c,-}
      \addagent{a,c,f}
      \addagent{-,c,e}
      \addagent{-,c,e}
      \addagent{b,d,-}
      \addagent{-,d,f}

      \foreach\x/\y in{1/1,2/1,5/2,6/2,3/3,4/3}{\hiliA{\x}{\y}}
      \foreach\x/\y in{1/2,2/2,3/2,4/2,5/1,2/3,6/3}{\hiliB{\x}{\y}}
    \end{scope}
    
    \begin{scope}[xshift=3*\dst*\xr cm]
      \mylabel{ii}
      \agentInit{\AIx}{\AIy}
      \addagent{a,c,e}
      \addagent{a,-,e}
      \addagent{b,-,f}
      \addagent{-,c,e}
      \addagent{-,d,f}
      
      \foreach\x/\y in {1/1,2/1,1/2,3/3,4/2,5/3}{\hiliA{\x}{\y}}
      \foreach\x/\y in {1/3,2/3,3/1,4/3,5/2}{\hiliB{\x}{\y}}
    \end{scope}
    
  \end{tikzpicture}
  \caption{%
  Counterexamples to~\cref{obs:mmmsmm:hpareto}. 
  }
  \label{fig:super}
\end{figure}
For the two highlighted committee sequences~$\ComSeq=(\{a\},\{d\},\{e\})$ (\mblue)
and~$\ComSeq'=(\{b\},\{c\},\{f\})$,
each agent is equally satisfied except for~$a_2$:
she is better off with~$\ComSeq'$, 
and hence~$\ComSeq'$ dominates~$\ComSeq$.
This is undesired and can be
formulated as follows:

\begin{property}[Pareto efficiency]
 \label{propty:pareto}
 A rule
 \respects{} 
 Pareto
 efficiency
 if it outputs no committee sequence
 which is
 \hdominate{d}
 by a valid committee sequence.
\end{property}

The property is our only property that 
distinguishes $\Rulehsv$ from~$\Rulehvs$ as follows.

\begin{observation}%
 \label{obs:mmmsmm:hpareto}
 $\Rulehsv$ and $\Ruleapp$
 \respect{} \pref{propty:pareto}.
 $\Ruleh$ and $\Rulehvs$
 \reject{}~\pref{propty:pareto}.
\end{observation}

\begin{proof}
  We have that
  if~$\ComSeq \text{ \hdominate{s} }\ComSeq'$,
  then
  $\yscoremin(\calE,\ComSeq)\geq \yscoremin(\calE,\ComSeq')$ and
  $\yscoresum(\calE,\ComSeq)>\yscoresum(\calE,\ComSeq')$.
  Hence,
  the statement for~$\Rulehsv$ and~$\Ruleapp$ follows.
 For~$\Rulehvs$ (and~$\Ruleh$),
 see~\cref{fig:super}(i)
 with~$\kSeq=(1,1,1)$:
  $\ComSeq=(\{a\},\{d\},\{e\})$ (\mblue{})
  and
  $\ComSeq'=(\{b\},\{c\},\{f\})$ (\mgreen{})
  are the only valid committee sequences with minimum agent score of one
  (clearly,
  $\yscoremin(\Elec,\ComSeq)\leq 1$ for every valid~$\ComSeq$).
  Since~$2=\xscoremin(\Elec,\ComSeq)>\xscoremin(\Elec,\ComSeq')=1$,
  $\Rulehvs$
  selects~$\ComSeq$.
  However,
  $\ComSeq'$ 
  \hdominate{s}~$\ComSeq$.
\end{proof}

By~definition,
$\Rulelex$ \respects{} \pref{propty:pareto}.
This is not the case for~$\Rulegreedy$.

\begin{observation}%
  \label{obs:greedy:rejects:pareto}
  $\Rulegreedy$ \rejects{}
  \pref{propty:pareto}.
\end{observation}

\begin{proof}
  See \cref{fig:super}(ii)
  with~$\kSeq=(1,1,1)$:
  $\Rulegreedy$ selects~$(\{b\},\{d\},\{e\})$ (\mgreen{}),
  which is \hdominate{d} by~$(\{a\},\{c\},\{f\})$ (\mblue{}).
\end{proof}

\paragraph{Independent Groups.}
When different agent groups 
have disjoint interests
(i.e., positive utilities only on disjoint level sets),
then the outcome should be the same as if each
agent group attended a separate election.
This is not true for $\Ruleh$,
as shown in~\cref{fig:mmmsmm-independentgroups}:
\begin{figure}[h!]
  \centering
  \begin{tikzpicture}
    \def\xr{1}
    \def\yr{1}
    \def\AIx{0.6}
    \def\AIy{0.4}
    \def\dst{0.75}
    \def\teps{0.055}
    \newcommand{\mylabel}[1]{\node at (-0.9*\xr,0.0375*\yr)[anchor=south west]{(#1)};}
    
    \begin{scope}[xshift=14.5*\dst*\xr cm]
      \agentInit{\AIx}{\AIy}
      \addagent{x,-,-,-,-} %
      \addagent{-,a,a,a,a}
      \addagent{-,b,b,b,b}
      \addagent{-,b,b,b,b}
      \draw[-,dashed,lightgray] ($(nx22.north west)+(-\teps,\teps)$) to ($(nx25.north east)+(\teps,\teps)$) to ($(nx45.south east)+(\teps,-\teps)$) to ($(nx42.south west)+(-\teps,-\teps)$) to cycle;
      \draw[-,dashed,lightgray] ($(nx11.north west)+(-\teps,\teps)$) to ($(nx11.north east)+(\teps,\teps)$) to ($(nx11.south east)+(\teps,-\teps)$) to ($(nx11.south west)+(-\teps,-\teps)$) to cycle;
      \foreach\x/\y in {2/2,2/3,3/4,3/5,4/4,4/5}{\hiliA{\x}{\y}}
      \foreach\x/\y in {2/2,3/3,3/4,3/5,4/3,4/4,4/5}{\hiliB{\x}{\y}}
    \end{scope}
    
    \end{tikzpicture}
  \caption{%
  Counterexample to~\cref{obs:mmmsmm-independentgroups}.
  }
  \label{fig:mmmsmm-independentgroups}
\end{figure}
On the separate election formed by~$a_2,a_3,a_4$ on the levels 2 to 5,
committee sequence~$(\{a\},\{a\},\{b\},\{b\})$ (\mblue) witnesses that
$(\{a\},\{b\},\{b\},\{b\})$ (\mgreen) is not winning for~$\Ruleh$.
Yet,
$(\{x\},\{a\},\{b\},\{b\},\{b\})$ wins for~$\Ruleh$ on the complete election,
implausibly leaving~$a_2$ less satisfied than possible.
Before we can state our corresponding property,
we formalize elections like the one in~\cref{fig:mmmsmm-independentgroups}.
We say that an election $\Elec=(A,\CandSeq=(\Cand_1,\dots,\Cand_\tau),U=(u_1,\dots,u_\tau),\kSeq=(k_1,\dots,k_\tau))$
is \emph{groupable} (into~$r$ groups)
if there is a partition~$A=A_1\uplus\dots\uplus A_r$ of the agent set and
levels~$1\leq t_1<t_2<\dots<t_{r-1}<\tau=t_r$ 
such that
for every~$s\in\set{r}$,
among all levels~$t\in\set[t_{s-1}+1]{t_{s}}\eqqcolon T_s$,
where~$t_0=0$,
exactly the group~$A_s$ has positive utilities.
For example,
the election given in~\cref{fig:mmmsmm-independentgroups} is groupable into $r=2$ groups with $A_1=\{a_1\}, A_2=\{a_2,a_3,a_4\}, T_1=\{1\}$, and $T_2=\{2,3,4,5\}$.

\begin{property}[Independent Groups]
 \label{propty:independentgroups}
 Rule~$\Rule$ \respects{} 
 \emph{independent groups} 
 if for every election~$\Elec=(A,\CandSeq=(\Cand_1,\dots,\Cand_\tau),U=(u_1,\dots,u_\tau),\kSeq=(k_1,\dots,k_\tau))$ groupable into~$r$ groups,
 it holds true that
 $\Rule(\Elec)=\Rule(\Elec_1)\times\dots\times\Rule(\Elec_r)$,
 where~$\Elec_s=(A_s,(\Cand_t)_{t\in T_s},(u_t)_{t\in T_s},(k_t)_{t\in T_s})$,
 $s\in\set{r}$,
 are the corresponding elections.
\end{property}

In addition to~$\Ruleh$, $\Rulehsv$ and~$\Rulehvs$ also \reject{} \pref{propty:independentgroups}:
A group with small maximal minimum agent score
can lead to other scores being optimized for another group
even at cost of the maximal minimum agent score.

\begin{observation}[\appref{obs:mmmsmm-independentgroups}]
 \label{obs:mmmsmm-independentgroups}
 Each of $\Ruleh$, 
 $\Rulehsv$, and~$\Rulehvs$ 
 \rejects{}~\pref{propty:independentgroups}.
\end{observation}

\appendixproof{obs:mmmsmm-independentgroups}
{
\begin{proof}
  See~\cref{fig:mmmsmm-independentgroups}
  with~$\kSeq=(1,\dots,1)$
  and elections~$\Elec_1$ and~$\Elec_2$ indicated by dashed boxes.
  We have that~$\Rulehsv(\Elec_1)=\{x\}$ and 
  $\Rulehsv(\Elec_2)$ selects each of candidate~$a$ and~$b$ exactly twice (\mblue{}).
  $\Rulehsv(\Elec)$
  selects~$a$ only once (\mgreen{}).
\end{proof}
}

$\Ruleapp$ trivially \respects{} \pref{propty:independentgroups} 
since it selects committees for each level independently.
Intuitively,
$\ems$ also optimizes each group independently.
Indeed,
we have the following.

\begin{proposition}%
 \label{obs:lex:greedy:independentgroups}
 $\Rulelex$
 and $\Rulegreedy$
 \respect{}~\pref{propty:independentgroups}.
\end{proposition}

\begin{proof}
  \newcommand{\ccut}[1]{\bar#1}
  For~$\Rulelex$:
  Let~$\ComSeq\in\Rulelex(\Elec)$.
  Let~$\ComSeq=\bigcup_{s=1}^r \ComSeq_s$, 
  where~$\ComSeq_s$ has only non-empty committees on~$T_s$,
  and let~$\ccut{\ComSeq_s}$ be~$\ComSeq_s$ restricted to the levels in~$T_s$.
  For every~$A'\subseteq A$ and committee sequence~$\ComSeq'\in\Rulebase(\Elec)$,
  let~$\sat(\ComSeq',A',i) \ceq |\{a\in A'\mid \yscore(\Elec,\ComSeq',a)=i\}|$
  and 
  $\satvec(\ComSeq',A') \ceq (\sat(\ComSeq',A',0),\dots,\sat(\ComSeq',A',\UB(\Elec)))$.
  We have
    $\satvec(\ComSeq,A)
    =\sum_{s=1}^r \satvec(\ComSeq_s,A_s).$
  Suppose there is an~$\ccut{\ComSeq_s'}\in\Rulelex(\Elec_s)$ with $\ems(\Elec_s,\ccut{\ComSeq_s'})<\ems(\Elec_s,\ccut{\ComSeq_s})$,
  which is equivalent to~$\satvec(\Elec_s,\ccut{\ComSeq_s'})\prec \satvec(\Elec_s,\ccut{\ComSeq_s})$ (see~\cref{obs:lex:scoreandvec}).
  The latter is equivalent to~$\satvec(\ComSeq_s',A_s)\prec \satvec(\ComSeq_s,A_s)$,
  where~$\ComSeq_s'$ is~$\ccut{\ComSeq_s'}$ with~$\sum_{q=1}^{s-1} |T_q|$ preceding and~$\sum_{q=s+1}^{r} |T_q|$ succeeding empty committees.
  Thus,
  $\ComSeq'$,
  where we replace~$\ComSeq_s$ with~$\ComSeq_s'$,
  has~$\satvec(\ComSeq',A)\prec \satvec(\ComSeq,A)$ and 
  hence~$\ems(\Elec,\ComSeq')<\ems(\Elec,\ComSeq)$
  (see~\cref{obs:lex:scoreandvec}),
  a contradiction to~$\ComSeq\in\Rulelex(\Elec)$.
  The other direction goes analogously.
  
 The proof for~$\Rulegreedy$ is deferred to
 \ifapp%
  the appendix~(\appref{obs:lex:greedy:independentgroups}). %
 \else%
  a paper's long version.
 \fi%
 \appendixproof{obs:lex:greedy:independentgroups}
 {
 For~$\Rulegreedy$:
 Let $\Elec=(A,\calC,U=(u_1,\dots,u_\tau),\kSeq)$
 be an election with~$A=A_1\uplus\dots\uplus A_r$ 
 and~$1\leq t_1<t_2<\dots<t_{r-1}<\tau=t_r$ such that
 among all~$t\in\set[t_{s-1}+1]{t_{s}}\eqqcolon T_s$,
 where~$t_0=0$,
 exactly the group~$A_s$ has positive utilities.
 Let~$\Elec_s=(A_s,(\Cand_t)_{t\in T_s},(u_t)_{t\in T_s},(k_t)_{t\in T_s})$,
 $s\in\set{r}$,
 be the corresponding elections.
 We need to show that 
 $\ComSeq\in\Rulegreedy(\Elec)$ if and only if~$\ComSeq\in\Rulegreedy(\Elec_1)\times\dots\times\Rulegreedy(\Elec_r)$.
 
 \RD{}
 Let~$\ComSeq=\ComSeq_1\circ\dots\circ\ComSeq_r\in\Rulegreedy(\Elec)$.
 Since~$\Rulegreedy$ is algorithmically defined,
 there is a sequence~$((\lambda_i,\chi_i))_{i=1}^q$ 
 of actions performed by~$\Rulegreedy$ that lead to~$\ComSeq$,
 where~$\lambda_i\in\tauS$ is the chosen level,
 and~$\chi_i\in C_{\lambda_i}$ is the selected candidate.
 For every~$s\in\set{r}$,
 let~$1\leq i_{s,1}<\dots<i_{s,q_s}\leq q$ be the maximal index sequence
 such that~$\lambda_{s,j}\in T_s$ for all~$j\in\set{q_s}$.
 We claim that~$((\lambda_{i_j},\chi_{i_j}))_{j=1}^{q_s}$
 is a sequence of actions performed by~$\Rulegreedy$ on~$\Elec_s$ leading to~$\ComSeq_s$.
 Suppose not,
 that is,
 there is a smallest index~$j^*$ with level~$\lambda_{i_{j^*}}'$ and a candidate~$\chi_{i_{j^*}}'$
 such that~$\ems(\Elec_s,\ComSeq_s' \cup_{\lambda_{i_{j^*}}'} \{\chi_{i_{j^*}}'\})<\ems(\Elec_s,\ComSeq_s' \cup_{\lambda_{i_{j^*}}} \{\chi_{i_{j^*}}\})$,
 where~$\ComSeq_s'$ is the committee sequence build from~$((\lambda_{i_j},\chi_{i_j}))_{j=1}^{j^*-1}$.
 We claim towards a contradiction that $\ems(\Elec,\ComSeq' \cup_{\lambda_{i_{j^*}}'} \{\chi_{i_{j^*}}'\})<\ems(\Elec,\ComSeq' \cup_{\lambda_{i_{j^*}}} \{\chi_{i_{j^*}}\})$,
 where~$\ComSeq'$ is the committee sequence build from~$((\lambda_{i},\chi_{i}))_{i=1}^{i_{j^*}-1}$.
 Let~$\ComSeq'=\ComSeq''\uplus \ComSeq_s'$.
 With~\cref{obs:lex:scoreandvec} 
 we have the following:
 \begin{align}
 \begin{aligned}
  & \satvec(\Elec_s,\ComSeq_s' \cup_{\lambda_{i_{j^*}}'} \{\chi_{i_{j^*}}'\}) 
  \\ &\qquad
  \prec 
  \satvec(\Elec_s,\ComSeq_s' \cup_{\lambda_{i_{j^*}}} \{\chi_{i_{j^*}}\})
  \\
  \iff & \satvec(\Elec,\ComSeq_s' \cup_{\lambda_{i_{j^*}}'} \{\chi_{i_{j^*}}'\}) 
  \\ &\qquad\prec 
  \satvec(\Elec,\ComSeq_s' \cup_{\lambda_{i_{j^*}}} \{\chi_{i_{j^*}}\})
  \\
  \iff & \satvec(\Elec,\ComSeq'') + \satvec(\Elec,\ComSeq_s' \cup_{\lambda_{i_{j^*}}'} \{\chi_{i_{j^*}}'\}) 
  \\&\qquad\prec 
  \satvec(\Elec,\ComSeq'') + \satvec(\Elec,\ComSeq_s' \cup_{\lambda_{i_{j^*}}} \{\chi_{i_{j^*}}\})
  \\
  \iff & \satvec(\Elec,\ComSeq'\cup_{\lambda_{i_{j^*}}'} \{\chi_{i_{j^*}}'\}) 
  \\&\qquad\prec 
  \satvec(\Elec,\ComSeq'\cup_{\lambda_{i_{j^*}}} \{\chi_{i_{j^*}}\}),
  \label{greddy:ig:equivs}
  \end{aligned}
 \end{align}
 a contradiction,
 where the first equivalence holds by~\pref{propty:independentgroups},
 the second equivalence by the definition of~$\prec$,
 and the last equivalence by the definition of~$\ComSeq'$.

 \LD{}
 Let~$\ComSeq_s\in\Rulegreedy(\Elec_s)$ for every~$s\in\set{r}$.
 Let $((\lambda_{i_j^s},\chi_{i_j^s}))_{j=1}^{q_s}$
 be a sequence of actions performed by~$\Rulegreedy$ on~$\Elec_s$ leading to~$\ComSeq_s$,
 where~$\lambda_{i_j^s}\in\tauS$ is the chosen level,
 and~$\chi_{i_j^s}\in C_{\lambda_{i_j^s}}$ is the selected candidate.
 We show that there is a sequence~$((\lambda_i,\chi_i))_{i=1}^q$
 of actions of $\Rulegreedy$ on~$\Elec$ leading to~$\ComSeq\ceq \ComSeq_1\circ\dots\circ\ComSeq_r$.
 To this end,
 we claim that each action can be taken
 in increasing order from some of the~$r$ action sequences above.
 Let~$(j_1,\dots,j_r)$ be the sequence of action indices which are to be considered next.
 We only have to prove that there is one valid action 
 (unless they all passed their respective last index).
 Towards a contradiction,
 assume this is not the case.
 That is,
 $\Rulegreedy$ performs action~$(\theta,\chi)$ that is better
 (in the sense of minimizing the current $\ems$-score) 
 than any of the actions~$(\lambda_{i_{j_s}^s},\chi_{i_{j_s}^s})$, 
 $s\in\set{r}$.
 Let~$s'\in \{1,\dots,r\}$ be such that~$\theta \in T_{s'}$.
 Then, 
 following the lines in~\eqref{greddy:ig:equivs},
 $(\theta,\chi)$ would be considered and performed by~$\Rulegreedy$ on~$\Elec_{s'}$
 instead of $(\lambda_{i_{j_{s'}}^{s'}},\chi_{i_{j_{s'}}^{s'}})$---a contradiction.
 }
\end{proof}

\section{Experiments}
\label{sec:experiments}
\def\mytodoinline#1{\textcolor{GoogleRed}{#1}}
\def\mytodo#1{\textcolor{GoogleRed}{#1}}
\def\lex{\ensuremath{\Rulelex}}
\def\app{\ensuremath{\Ruleapp}}
\def\gree{\ensuremath{\Rulegreedy}}
\def\mms{\ensuremath{\Rulehvs}}
\def\msm{\ensuremath{\Rulehsv}}

\def\hm{\ensuremath{\yscoremin}}
\def\vm{\ensuremath{\xscoremin}}
\def\hs{\ensuremath{\yscoresum}}
\def\lexo{\ensuremath{\ems}}
\def\maxa#1{$\leq #1$-Approval}
\def\kSeql#1{\kSeq^{(#1)}}

\NewDocumentCommand\paren{d()}{\left( #1 \right)}

\NewDocumentCommand\len{d()}{\mathit{len}\!\left(#1\right)}
\NewDocumentCommand\maxF{d()}{\max\!\left\{#1\right\}}
\NewDocumentCommand\minF{d()}{\min\!\left\{#1\right\}}
\NewDocumentCommand\logF{d()}{\log\!\left(#1\right)}
\NewDocumentCommand\ISym{d[]}{I\!\left[#1\right]}

\colorlet{ApprovalColour}{MaterialPink300}
\colorlet{GreedyColour}{GoogleGreen}
\colorlet{LexColour}{MaterialOrange}
\colorlet{MmsColour}{MaterialPurple}
\colorlet{MsmColour}{MaterialCyan}

\colorlet{IdentityColour}{GoogleBlue}
\NewDocumentCommand{\plotExp}{}{
  \begin{figure*}[ht!]\tikzexternalenable\centering%
    \makebox{ 

}
    \caption{A comparison of our rules. In the scatter plots, each point represents one instance from the experimental data, the $x$ value the corresponding score (see title of plot) achieved by the rule specified on the $x$ axis, and the $y$ value the score achieved by the rule specified on the $y$ axis or in the legend. If a point is on the blue line, both rules reach the same score.}
    \label{fig:expplots}%
  \end{figure*}%
}
We analyze the behavior of our rules when applied to experimental data discussed in \cref{subsec:expdata} using the experimental setup discussed in \cref{subsec:expsetup}.
We then discuss multiple aspects of the rules' behavior in \cref{subsec:ones}:
The values of different scores of the winning committee sequences determined by different rules
and the proportion of the experimental data for which a given rule satisfies the conditions of a given property from Section~\ref{sec:props}---even if a rule violates a given property in general,
its conditions can be satisfied for a specific instance.

\ifarxiv{The code for creating the instances and running the experiments is published in \cite{Our_Repo}.}

\subsection{Experimental Data}\label{subsec:expdata}
To the best of our knowledge, 
there are no (real-life) datasets for electing sequences of committees.
Thus,
we transformed preference datasets from PrefLib \cite{mattei2017apreflib},
where preferences are represented as strict (possibly incomplete) linear orders,
into ``\maxa{2} instances'' in which each agent can approve at most two candidates per level with an approved candidate getting a utility of one\footnote{
  We conducted analogous experiments for instances where each agent 
  (a) approves at most one candidate per level and 
  (b) distributes 10 points among its two most preferred candidates per 
  \ifapp%
    level (see \cref{app:exp:data}).
  \else%
    level.
  \fi%
  The results support the main conclusions made here (see 
  \ifapp%
    \cref{app:exp:res}).
  \else%
    a long version of the paper).
  \fi%
  }:
Based on the respective metadata, we chose meaningful %
levels and assigned %
every candidate %
accordingly (possibly to multiple levels).
On each level, each agent approves its two most preferred candidates if two candidates on this level are in the agent's preference order, otherwise all present ones.

Our new multilevel elections do not describe a temporal setting in which different levels correspond to different time steps.
This is because
there are no datasets available in which the same agent can be identified in elections carried out at various points in time,
to the best of our knowledge.
Yet,
our new multilevel elections describe a multimodal setting.
For example, we used the “Spotify Daily Chart” dataset \cite{BoSc22}
by defining levels as genres.
Such a transformed election describes, for instance, a scenario where a
playlist is created in which each genre should be represented
twice, and, for each genre (i.e.~level), an agent approves the two songs belonging to this genre that are placed highest in the agent's preference order.
For other datasets, levels describe continents, countries, U.S.~states, age groups, or board game types; 
\ifapp%
  see \cref{tab:data} in the appendix for a more detailed overview.
\else%
  see a long version of the paper for a more detailed overview.
\fi%

Using this approach, we have created a total of 7888 sequential elections %
with the following additional steps:
We removed candidates not approved by any agent and levels on which all agents approve the same candidate.
After this, we removed instances with at most one level or three candidates and those with more than 160 candidates.
\cref{fig:dims} gives an overview of the dimensions of these instances.

In the following, we discuss results for $\kSeq = \kSeql{2}$ with $\kSeql{l} \ceq \paren(\maxF(\minF(l, \abs{C_i}-1), 1))_{i=1}^{\tau}$.
\ifapp%
 In the appendix,
\else%
 In a long version of the paper,
\fi%
we show analogous results for $\kSeql{1}$ and $\kSeql{3}$,
for which the qualitative results only change a little 
\ifapp%
 overall (see \cref{app:exp:res}).
\else%
 overall.
\fi%

\begin{figure}[t!]\tikzexternalenable\centering%
  \makebox{ 

}%
  \caption{The dimensions of the experimental data, where the color of each point represents the average $\tau$ of all instances with the given $m$ and $n$.}
  \label{fig:dims}
\end{figure}
\plotExp{}

\subsection{Experimental Setup}\label{subsec:expsetup}
We implemented our rules in Python to analyze their behavior.
Additionally, \gree{} is implemented in C++ to compute all winning committee sequences for investigating properties.
While the implementations of \gree{} and \app{} follow directly from their definitions in \cref{sec:centralrules}, the other rules are defined as constraint optimization problems and implemented as well as solved using the CP-SAT solver from Google OR-Tools (with default parameters).
All experiments were conducted on an Ubuntu 22.04.4 LTS server (Intel\textregistered{} Xeon\textregistered{} Silver 4310 CPU with 12 physical cores at \SI{2.1}{\giga\hertz}, \SI{128}{\giga\byte} RAM).

When investigating the performance of the rules regarding the scores and runtimes,
we consider each of the above-mentioned 7888 sequential elections, but 
only one winning committee sequence for each rule and election.
For \lex{}, \mms{}, and \msm{}, 
the first winning committee sequence found by the CP-SAT solver is considered.
For \gree{} and \app{}, the first possible candidate in the candidate order of the underlying PrefLib dataset is chosen.

We need to compute all winning committee sequences when investigating the properties.
To avoid excessive runtimes,
for each property, we discard  a specific instance,
its variant,
or a pair of instances if
(1) determining at most 30 winning committee sequences of \gree{} takes more than $\SI{10}{\second}$ or 
(2) one of the rules violating the property
determines at least 30 winning committee sequences.
For some other properties, we have further particularities or restrictions
(e.g., upper bounds on~$n$ and~$m$) 
to reduce runtimes 
\ifapp%
  (see appendix).
\else%
  (see a long version).
\fi%
Nevertheless, each property is tested for at least 6129 (pairs of) instances.

\subsection{Experimental Results}\label{subsec:ones}
We want to highlight the following takeaways: %

\app{} is the fastest (see \cref{fig:expplots}, 1st plot) and reaches around $88\%$ of the optimal \hm{} score on average, but reaches the optimal \hm{} score for only $20\%$ of the data (see \cref{tab:overviewexp}) with severe outliers (see \cref{fig:expplots}, 2nd plot).
This reflects the fact that \app{} is not egalitarian.

\lex{}, on the other hand, has the worst runtimes (see \cref{fig:expplots}, 1st plot), which shows that a good heuristic is important for larger instances or when computing time is limited.

\gree{} appears to be such a heuristic for \lexo{} and \hm{} and a far better choice than \app{}: It performs well regarding the runtimes (see \cref{fig:expplots}, 1st plot) and achieves $99\%$ of the optimal \hm{} score on average, the optimal \hm{} score for $89\%$ of the data (see \cref{tab:overviewexp}), and has no severe outliers (see \cref{fig:expplots}, 2nd plot).
It also outperforms \app{} regarding the average \lexo{} score:
It achieves an \lexo{} score that is only about $1.2$ times as large as the optimum on average, as opposed to five times in the case of \app{}, and reaches the optimum for about $50\%$ of the instances, while all other rules (except \lex{}) achieve it for less than $43\%$ of the data (see \cref{tab:overviewexp}).
Additionally, \gree{} violates many properties in general, but, for each property, it fulfills its condition for at least $88\%$ of the instances (see \cref{tab:results}). 

If, however, one prioritizes a higher \hs{} or a higher \vm{} score over the \lexo{} score,
\msm{} and \mms{} can be good choices.
Interestingly, both achieve only a slightly worse \lexo{} score than \gree{} on average: The optimal \lexo{} score is $80\%$ and $82\%$ of the score reached by \mms{} and \msm{} on average (see \cref{tab:overviewexp}), respectively, despite neither rule optimizing $\lexo{}$.
While both rules behave similarly on average regarding the considered scores (see \cref{tab:overviewexp}), these two rules achieve different \hs{} and \vm{} scores for around $18\%$ of the instances and can differ significantly (see \cref{fig:expplots}, 3rd and 4th plot):
While \mms{} reaches very similar \hs{} scores to \msm{} with very few outliers, there are significant outliers regarding the \vm{} score.
Lastly, both rules satisfy each property in general or its condition for at least $95\%$ of data with one exception: They violate the conditions of \pref{propty:independentgroups} for around $90\%$ of the instances with more than one independent group (see \cref{tab:results}).

\begin{table}[t!]\centering
  \caption{The first value is the percentage of the experimental data for which a rule achieves the optimal 
  score. 
  The value in parentheses is the percentage of the optimal score that a rule achieves on average, with one exception: 
  For \lexo{}, the percentage of the reciprocal proportion is shown, as \lexo{} is the only score to be minimized. 
  If a rule is optimal 
  with respect to a score, 
  then the corresponding cell contains~100.
  The percentages are rounded naturally.
  }\label{tab:overviewexp}
  \begin{tblr}{
    colspec={lrrrr},
  }\toprule
    Rule     & \lexo{}                                                                                    & \hm{}                                                                                      & \vm{}                                                                                   & \hs{}                                                                                   \\\midrule
  \lex{}   & \textcolor{GoogleGreen!100!GoogleYellow}{100}                                            & \textcolor{GoogleGreen!100!GoogleYellow}{100}                                             & \textcolor{GoogleYellow!78!GoogleRed}{39} (\textcolor{GoogleGreen!77!GoogleYellow}{88})  & \textcolor{GoogleYellow!23!GoogleRed}{11} (\textcolor{GoogleGreen!93!GoogleYellow}{97}) \\
  \gree{}  & \textcolor{GoogleYellow!99!GoogleRed}{50} (\textcolor{GoogleGreen!67!GoogleYellow}{84})  & \textcolor{GoogleGreen!78!GoogleYellow}{89} (\textcolor{GoogleGreen!99!GoogleYellow}{99}) & \textcolor{GoogleYellow!78!GoogleRed}{39} (\textcolor{GoogleGreen!77!GoogleYellow}{88})  & \textcolor{GoogleYellow!23!GoogleRed}{12} (\textcolor{GoogleGreen!93!GoogleYellow}{97}) \\
  \msm{}   & \textcolor{GoogleYellow!84!GoogleRed}{42} (\textcolor{GoogleGreen!64!GoogleYellow}{82})  & \textcolor{GoogleGreen!100!GoogleYellow}{100}                                             & \textcolor{GoogleYellow!99!GoogleRed}{50} (\textcolor{GoogleGreen!83!GoogleYellow}{91})  & \textcolor{GoogleYellow!47!GoogleRed}{23} (\textcolor{GoogleGreen!95!GoogleYellow}{98}) \\
  \mms{}   & \textcolor{GoogleYellow!76!GoogleRed}{38} (\textcolor{GoogleGreen!61!GoogleYellow}{80})  & \textcolor{GoogleGreen!100!GoogleYellow}{100}                                             & \textcolor{GoogleGreen!11!GoogleYellow}{56} (\textcolor{GoogleGreen!87!GoogleYellow}{94})& \textcolor{GoogleYellow!47!GoogleRed}{23} (\textcolor{GoogleGreen!95!GoogleYellow}{97}) \\
  \app{}   & \textcolor{GoogleYellow!17!GoogleRed}{9} (\textcolor{GoogleYellow!41!GoogleRed}{20})     & \textcolor{GoogleYellow!39!GoogleRed}{20} (\textcolor{GoogleGreen!77!GoogleYellow}{88})   & \textcolor{GoogleGreen!100!GoogleYellow}{100}                                            & \textcolor{GoogleGreen!100!GoogleYellow}{100}                                           \\\bottomrule
  \end{tblr}
\end{table}

\section{Epilogue}
\label{sec:epilogue}

We introduced rules and properties for %
egalitarian committee sequences and tested them against each other,
both theoretically and experimentally.
We conclude with the following.
\begin{compactitem}
 \item Our work promotes~$\Rulelex$ for egalitarian committee sequences.
  $\Rulelex$ \respects{} many preferable properties.
  Our experiments indicate that while computationally demanding,
  no other rule achieves $\ems$ scores close to $\Rulelex$'s.
 \item No rule \rejects{} as many of our preferable properties as $\Rulegreedy$.
  However,
  $\Rulegreedy$ performs well in our experiments---regarding runtime,
  solution quality,
  and fraction of instances where a generally \rejected{} property is \satisfied.
  Thus,
  $\Rulegreedy$ qualifies
  itself (on the available data) as a good heuristic for~$\Rulelex$,
  in particular on larger instances or when (computing) time is limited.
  When two elections are merged, we recommend comparing a winning committee sequence of the merged election with the merged winners of the separate elections.
 \item If one accepts a smaller $\ems$ score for a higher sum score
  or a higher minimum level score,
  then $\Rulehsv$ and~$\Rulehvs$ can be good choices.
  Our analysis indicates that $\Rulehsv$ is slightly preferable over $\Rulehvs$.
  For both,
  we recommend computing and combining winning committee sequences for disjoint groups with disjoint interests,
  if they exist in the election at hand.
 \item In general,
  we advise against~$\Ruleapp$ for finding egalitarian committee sequences:
  Experimentally,
  while being very fast,
  $\Ruleapp$ performs poorly for~$\yscoremin$ (and~$\ems$).
\end{compactitem}

Our work constitutes the first axiomatic and experimental study on egalitarian committee sequences 
and hence paves the way for future work in many directions. 
Additional research can add further properties and rules
not only for committee sequences that are egalitarian,
but also for, e.g., equitable committee sequences.
On the experimental side,
comparing winners
between non-sequential elections and their ``sequentialized'' ones
may be of interest
and more real-world data could be collected.
Nevertheless, we created the largest preference dataset with labeled candidates, to the best of our knowledge.
On the one hand, this could be used in the context of multilevel elections
to experimentally investigate other goals and properties.
On the other hand, the data can also be valuable in the classical, non-sequential setting whenever the labels (i.e.~attributes) of the candidates should be taken into account, e.g.~when aiming for a diverse committee.

\ifaaai%
  \begin{ack}\theacks{}\end{ack}
\else%
  \section*{Acknowledges}
  \theacks{}
\fi%

\ifaaai%

\else%

\fi%

\ifapp%
  \clearpage
  \appendix
  \section*{\section*{Appendix}}
  \appendixProofText

\subsection{Additional Material for \cref{subsec:expdata}}\label{app:exp:data}

Next to the \maxa{2} instances which are introduced in \cref{subsec:expdata}, we also consider instances with different, normalized utilities, i.e. different \textit{instance classes}.
Regardless of the choice of utilities, we chose the same set of
levels and assigned the same subset of candidates to each level for each instance from PrefLib that we consider,
based on the respective metadata.
In the following, let $m_t(a)$ be the number of candidates on level $t$ that are present in the preference order of agent $a$ (in the original PrefLib file),
and, for each candidate $c$ in the preference order of agent $a$, let $p(a,c)$ be the position of $c$ in $a$'s preference order (starting at zero).
Furthermore, let $m_o$ be the number of candidates in the original PrefLib instance and
$b(a,c)=m_o-p(a,c)$ be the Borda score of candidate $c$ based on $a$'s preference order.

One class of instances we consider consists of \maxa{1} instances: 
Instead of allowing each agent to approve at most two candidates per level by assigning to them a utility of one (\maxa{2} instances), each agent can approve at most one candidate per level with an approved candidate always getting a utility of one.
Thus, on each level $t$, $a$ approves the $\min(1, m_t(a))$ most preferred candidates on level $t$, and it holds that $\forall a\in A, t\in\tauS: \sum_{c\in \Cand_t} u_t(a,c) \leq 1 \land \forall c\in \Cand_t: u_t(a,c)\leq 1$.

Furthermore, we constructed \textit{point instances} based on the idea that an agent gets $10$ points which it can distribute over its two most preferred candidates on each level $t$.
More precisely, if $m_t(a) = 0$, $u_t(a,c)=0$ for each candidate $c$ on level $t$.
If $m_t(a)=1$, the only candidate on level $t$ present in $a$'s preference order gets a utility of $10$.
Otherwise, only the highest ranked candidate $c_1$ and the second-highest ranked candidate $c_2$ of level $t$ get a utility larger than zero from agent $a$ on level $t$
with $u_t(a,c_1) = \mathrm{round}\!\paren(10\cdot \frac{b(a,c_1)}{b(a,c_1)+b(a,c_2)})$ (rounding half up) and $u_t(a,c_2) = 10-u_t(a,c_1)$.
Thus, the utilities are based on the Borda scores for the agent's preference order
and scaled so that they sum up to $10$ ($\forall a\in A, t\in\tauS: \sum_{c\in \Cand_t} u_t(a,c) \leq 10$).

Based on these instances, we applied the following (clean-up) steps on all instances:
We removed candidates that did not get a utility larger than zero from any agent and levels on which all agents give the same utility to the same candidate(s).
After this, we removed instances with at most one level or at most three candidates.
In addition, we removed point instances with $n>80$ or $m>80$ to avoid excessive runtimes.

This approach leads to 8870 \maxa{1} instances and 4160 point instances
whose dimensions are visualized in \cref{fig:dims4}.
An overview of the sequential elections constituting the experimental data can be seen in \cref{tab:data},
in which the elections are grouped by the dataset names used in PrefLib for the underlying preferences.
For each instance class, the same experiments were conducted using $\kSeq \in\{\kSeql{1},\kSeql{2},\kSeql{3}\}$.

\begin{figure*}\tikzexternalenable
  \centering%
  \parbox{.49\textwidth}{%
		\centering

\par
		{(c) Point Instances}
	}%
  \caption{The dimensions of the experimental data for each of our instance classes,
  where the color of each point represents the average $\tau$ of all instances with the given $m$ and $n$.}
  \label{fig:dims4}
\end{figure*}

\begin{table*}\tikzexternalenable\centering
  \caption{Overview of the sequential elections used in the experiments. The elections are grouped by the dataset names used in PrefLib for the underlying preferences.
  ($^\dagger$\,\protect\cite{BoSc22}; $^\ddagger$\,\protect\cite{Cities})}\label{tab:data}\medskip
  %
\end{table*}

\subsection{Additional Material for \cref{subsec:expsetup}}
\subsubsection{Implementations of \app{} and \gree{}}
As mentioned in Section~\ref{subsec:expsetup}, the implementations of \app{} and \gree{} for determining a single winning committee sequence follow directly from their definitions.
This is also the case for computing all winning committee sequences of \app{}.
However, we want to note that we have two different implementations of \gree{} for determining one winning committee sequence in Python: one for \maxa{l} instances (each agent can approve at most $l$ candidates per level) and one for general utilities (including point instances).
The implementation for \maxa{l} instances is more efficient and differs when determining the candidate to be added, which leads to the highest decrease of the current $\lex{}$ score, as adding a candidate approved by an agent (i.e. a candidate assigned a utility larger than zero) will always increase its satisfaction (i.e. the agent score) by one.
This assumption cannot be made for general utilities where two candidates could, for example, increase the agent score of the same agents, but by different amounts, leading to different $\lex{}$ scores (while the amounts would be the same in case of \maxa{l} instances).

To compute all winning committee sequences of \gree{}, the following search tree is traversed depth-first:
The root of the search tree represents the committee sequence resulting from the first step of \gree{} defined in Definition~\ref{def:gree}.
The inner nodes represent “incomplete” committee sequences---those containing at least one level $t$ with less than $k_t$ chosen candidates---while the leaves represent complete winning committee sequences.
In both cases, only nodes that \gree{} can reach are considered:
An edge from one committee sequence $\ComSeq$ to another $\ComSeq'$ exists if $\ComSeq'$ can be the result of applying the third step of Definition~\ref{def:gree} to $\ComSeq$.
Incomplete committee sequences that have already been visited are stored in a cache to avoid visiting them again, i.e. to prune the search tree.

\subsubsection{Implementations of \mms{}, \msm{}, and \lex{}}
Here, we define the constraint optimization problems which we use to find a winning committee sequence of \mms{}, \msm{}, and \lex{}.

\def\BbbB{\mathbb{B}}
\def\solx{x_{\left( t,c \right)}}
The two constraint optimization problems that are used to find a winning committee sequence of \mms{} and \msm{} share the same variables and constraints, but use different objective functions.
The following variables are used:
\begin{itemize}
  \item $\solx\in\BbbB$ with $t\in\tauS, c\in\Cand_t$ is the solution, i.e.~$\solx$ is 1 if and only if $c\in\ComSeq_t$.
  \item $s_a \in \left\{ 0,\dots,\sigma_a \right\}$ with $a\in A$ corresponds to $\yscore(\Elec,\ComSeq,a)$
  with $\sigma_a=\tau$ if used for \maxa{1} instances,
  $\sigma_a = \sum_{t\in T} \min\left(\kSeq_t, \abs{\left\{c\in \Cand_t: u_t(a, c)=1\right\}}\right)$ for \maxa{2} instances,
  and $\sigma_a = \sum_{t\in T} \sum_{v\in V} v$ with $V$ containing the utilities of $\kSeq_t$ many candidates with the highest utilities received from $a$ otherwise.
  \item $s'_t \in \left\{ 0,\dots,\lambda_t\right\}$ with $t\in \tauS$ corresponds to $\xscore(\Elec,\ComSeq,t)$
  with $\lambda_t=\abs{A}$ if used for \maxa{1} instances,
  $\lambda_t = \sum_{a\in A} \min\left(\kSeq_t, \abs{\left\{c\in \Cand_t: u_t(a, c)=1\right\}}\right)$ for \maxa{2} instances,
  and $\lambda_t = \sum_{a\in A} \sum_{v\in V} v$ with $V$ containing the utilities of $\kSeq_t$ many candidates with the highest utilities received from $a$ otherwise.
  \item $h_m \in \left\{ 0,\dots,\max_{a\in A} \sigma_a \right\}$ corresponds to $\yscoremin(\Elec,\ComSeq)$
  \item $v_m \in \left\{ 0,\dots,\max_{t\in T} \lambda_t \right\}$ corresponds to $\xscoremin(\Elec,\ComSeq)$
  \item $s \in \left\{ 0,\dots, \sum_{a\in A} \sigma_a \right\}$ corresponds to $\yscoresum(\calE,\ComSeq)$
\end{itemize}
The constraints are:
\begin{itemize}
  \item The solution has to be valid:
  \begin{align}
    \forall t \in \tauS: \sum_{c\in\Cand_t} \solx = \kappa_t\label{eq:lexvalid}
  \end{align}
  \item Define $s_a$:
  \begin{align}
    \forall a \in A: s_a = \sum_{t\in\tauS} \sum_{c\in\Cand_t} \solx\cdot u_t(a,c)\label{eq:sa}
  \end{align}
  \item Define $s'_t$:
  \begin{align*}
    \forall t \in \tauS: s'_t = \sum_{c\in\Cand_t}\sum_{a\in A} \solx\cdot u_t(a,c)
  \end{align*}
  \item Define $h_m$:
  \begin{align*}
    h_m = \min_{a\in A} s_a
  \end{align*}
  \item Define $v_m$:
  \begin{align*}
    v_m = \min_{t\in \tauS} s'_t
  \end{align*}
  \item Define $s$:
  \begin{align*}
    s = \sum_{a\in A} s_a
  \end{align*}
\end{itemize}
In the case of \mms{}, the objective function to be maximized is:
\begin{align*}
  s +& \left( \sum_{a\in A} \sigma_a + 1 \right) v_m
  \\ +& \left( \sum_{a\in A} \sigma_a + 1 \right) \left( \max_{t\in T} \lambda_t + 1 \right) h_m
\end{align*}
In the case of \msm{}, the objective function to be maximized is:
\begin{align*}
  v_m +& \left( \max_{t\in T} \lambda_t + 1 \right) s\\
  +& \left( \max_{t\in T} \lambda_t + 1 \right)  \left( \sum_{a\in A} \sigma_a + 1\right)  h_m 
\end{align*}

\def\soldi{d_{\left(p,t,c \right)}}
To compute a committee sequence that differs from $w>1$ previously computed committee sequences $\ComSeq^{(p)}=(\Com^{(p)}_1, \dots, \Com^{(p)}_\tau)$ with $p\in \{1,\dots, w\}$, the following variables and constraints are added to the above optimization problem:
For each $\ComSeq^{(p)}$, $t\in \tauS$, and $c\in C_t$, a Boolean variable $\soldi\in\BbbB$ is added to indicate whether a solution $\solx$ differs from $\ComSeq^{(p)}$ with respect to choosing candidate $c$ on level $t$.
This is ensured by adding the following constraint:
\begin{itemize}
  \item If $c\not\in \Com^{(p)}_t$, add the constraint $\soldi = \solx$.
  \item If $c\in \Com^{(p)}_t$, add the constraint $\soldi = \neg\solx$.
\end{itemize}
To ensure that the currently searched solution differs from $\ComSeq^{(p)}$ by at least one candidate on any level, the constraint $\sum_{t\in\tauS} \sum_{c\in C_t} \soldi \geq 1$ is added for each $p\in\{ 1,\dots,w\}$.

Using this approach, all winning committee sequences can be found for a given election.

Our approach to determine a winning committee sequence of \lex{} is based on a constraint optimization problem that consists of the variables $\solx$ and $s_a$.
Additionally, there is a variable $s_{\left( a,i \right)}^= \in\BbbB$ with $a\in A, i\in \sigma_m$, and $\sigma_m\ceq \max_{a\in A} \sigma_a$, which is 1 if and only if $s_a = i$.

The constraints include constraints~\eqref{eq:lexvalid} and \eqref{eq:sa} and the following two constraints that define $s_{\left( a,i \right)}^=$:
\begin{align*}
  \forall a\in A:{}& \sum_{i=0}^{\sigma_m} s_{\left( a,i \right)}^= = 1 \\
  \forall a\in A:{}& \sum_{i=0}^{\sigma_m} i\cdot s_{\left( a,i \right)}^= = s_a
\end{align*}

The objective function to be minimized is:
\begin{equation*}
  \ems(\Elec,\ComSeq)
  = \sum_{i=0}^{\sigma_m} \sum_{a \in A} s_{\left( a,i \right)}^= \cdot \left(\left|A\right|+1\right)^{\sigma_m-i}
\end{equation*}

As maximizing the \lexo{} score directly leads to overflows for part of the experimental data when using the CP-Sat solver, we determined $10^{17}$ experimentally as an approximation of the maximum possible objective value that does not cause an overflow.
Based on this, we approximate the number $\zeta$ of consecutive $i$ that can be considered in one optimization problem by first determining an upper bound of $\ems(\Elec,\ComSeq)$:
\begin{align*}
  &\sum_{i=0}^{\sigma_m} \left|\left\{a\in A\mid \yscore(\ComSeq,a)=i\right\}\right|\cdot \left(\left|A\right|+1\right)^{\sigma_m-i}\\
  \leq{} & \sum_{i=0}^{\sigma_m} \left|A\right|\cdot \left(\left|A\right|+1\right)^i\\
  ={} & \left(\left|A\right|+1\right)^{\sigma_m+1}-1
  \leq \left(\left|A\right|+1\right)^{\sigma_m+1}
  \mathop{\overset{!}{\leq}} 10^{17} \\
  \implies & \sigma_m + 1 \leq \frac{\logF(10^{17})}{\logF(\left|A\right|+1)} \\
  \implies & \zeta = \left\lfloor \frac{\log\!\left(10^{17}\right)}{\log\!\left(\left|A\right|+1\right)}\right\rfloor - 1
\end{align*}
The optimization problem is split into $\left\lceil \left( \sigma_m+1 \right) / \zeta \right\rceil$ optimization problems $O_d$ with $d \in \{ 0,\dots, \left\lceil \left(\sigma_m+1\right) / \zeta \right\rceil -1 \}$, where the $d$-th optimization problem maximizes
\begin{equation*}
  \sum_{i=0}^{\zeta'} |\{a\in A\mid \yscore(\ComSeq,a)=d\cdot \zeta + i\}|\cdot \left(\left|A\right|+1\right)^{\zeta'-i}
\end{equation*}
with $\zeta' = \min\!\left\{ \zeta-1, \sigma_m -d \cdot \zeta \right\}$.
The optimization problem $O_d$ is executed after the optimization problems $O_{d'}$ with $d'\in \left\{ 0,\dots, d-1 \right\}$ and preserves the results of the previous optimization problems regarding the number of agents with an agent score of $\yscore(\ComSeq,a)=s$ for all $s\in \left\{ 0,\dots,d\cdot \zeta - 1 \right\}$. 

This optimization problem is always used for \maxa{1} and \maxa{2} instances.
For point instances, this approach is only used if $\frac{\sigma_m}{\zeta}\leq 16$ (i.e. if the original optimization problem is split into at most 16 optimization problems).
Otherwise, a different optimization problems is executed multiple times with different inputs:
In the first round, the primary goal is to maximize the minimum agent score and the secondary goal is to minimize the number of agents that have this agent score.
In the second round, constraints ensure that the minimum agent score as well as the number of agents that have this score remain the same, and the goal is to maximize the next largest agent score appearing in the solution (primary goal) as well as minimizing the number of agents that have this score (secondary goal), and so on.
This is based on the possibility that there are large distances between an agent score of an agent and the next largest agent score in the solution.
While the $\ems$ score sums over all possible agent scores from zero to $\sigma_m $, the following optimization problem will skip the agent scores that no agent has in the final solution:
Let $P$ contain the results from previous rounds of executing the optimization problem, i.e. tuples $(s', n')$ where $s'$ is an agent score and $n'$ is the number of agents that have this score.
The variables include the variable $\solx$ and the following variables:
\begin{itemize}
  \item $\beta_{\left( a,i \right)} \in\BbbB$ with $a\in A, i\in P$, which is 1 if and only if agent $a$ has an agent score of $i_1$ (which was considered in the objective function of a previous round).
  \item $\beta^*_a \in\BbbB$ with $a\in A$, which is 1 if and only if agent $a$ has an agent score which was considered in the objective function of a previous round.
  \item $\omega^* \in \left\{ 0,\dots,\sigma_m \right\}$, representing the minimum agent score out of the scores not considered in a previous round.
  \item $\omega_a \in\BbbB$ with $a\in A$, which is 1 if and only if agent $a$ has an agent score of $\omega^*$.
\end{itemize}

The constraints include constraint~\eqref{eq:lexvalid} and the following constraints:
\begin{itemize}
  \item Define $\beta_{\left( a,i \right)}$ and $\beta^*_a \in\BbbB$ in case that $\abs{P} > 0$:
  \begin{gather*}
    \forall a\in A, i\in P:{}\\
    \beta_{\left( a,i \right)} \implies i_1 = \sum_{t\in T}\sum_{c\in\Cand_t} \solx\cdot u_t(a,c) \\
    \neg\beta_{\left( a,i \right)} \implies i_1 \neq \sum_{t\in T}\sum_{c\in\Cand_t} \solx\cdot u_t(a,c) \\
    \forall i\in P:{} \sum_{a\in A} \beta_{\left( a,i \right)} = i_2\\
    \beta^*_a  = \max_{i\in P} \beta_{\left( a,i \right)}
  \end{gather*}
  \item Define $\beta^*_a \in\BbbB$ in the case that $\abs{P} = 0$: $\forall a \in A: \beta^*_a = \false$
  \item Define $\omega^*$:
  \begin{align*}
    \forall a \in A:{}& \omega^* \leq \sum_{t\in T}\sum_{c\in\Cand_t} \solx\cdot u_t(a,c) + \sigma_m \cdot \beta^*_a\\
    \text{If } \abs{P} > 0:{}& \omega^* \geq \max_{i\in P} i_1 +1
  \end{align*}
  \item Define $\omega_a$:
  \begin{align*}
    \forall a\in A:{}& \omega_a \implies \omega^* = \sum_{t\in T}\sum_{c\in\Cand_t} \solx\cdot u_t(a,c) \\
    \forall a\in A:{}& \neg\omega_a \implies \omega^* \neq \sum_{t\in T}\sum_{c\in\Cand_t} \solx\cdot u_t(a,c) \\
  \end{align*}
\end{itemize}
The objective function to be maximized is:
\begin{equation*}
  \left(\abs{A}+1\right) \cdot \omega^* - \sum_{a\in A} \omega_a
\end{equation*}

To compute all winning committee sequences of \lex{}, an analogous approach as described for \mms{} and \msm{} is used.

\subsection{Additional Material for \cref{subsec:ones}}\label{app:exp:res}
\subsubsection{The Scores}
For each instance class and $\kSeq$ considered, we show
\begin{compactitem}
  \item a table analogous to \cref{tab:overviewexp} in \cref{tab:kdiff4},
  \item the runtimes of the rules for finding one winning committee sequence in \cref{fig:timesl} (for each rule and \maxa{2} or point instance, the runtime of one computation is considered, while for each \maxa{1} instance, the runtimes of three computations were averaged using the hardware mentioned in \cref{subsec:expsetup}),
  \item a comparison of \lex{}, \gree{}, and \app{} regarding the \vm{} score in \cref{fig:threevl} and regarding the \hm{} score in \cref{fig:threehl},
  \item a comparison of \msm{} and \mms{} regarding the \hs{} score in \cref{fig:mmsuml} and regarding the \vm{} score in \cref{fig:mmsvl}, and
  \item the percentage of the optimal \lexo{} score relative to the \lexo{} score determined by each rule that does not minimize \lexo{} in Figure~\ref{fig:lexoboxl}.
\end{compactitem}
While all main conclusions form the main body of the paper also hold for instance classes other than \maxa{2} instances, we want to highlight the following differences between the results for different instance classes: 

The difference between \gree{} and \app{} is bigger for \maxa{1} instances than for \maxa{2} and point instances, both regarding the proportion of the optimal \hm{} and the proportion of the optimal \vm{} score achieved on average.
Overall, for each $\kSeq$ and score, each rule achieves at least as good scores for \maxa{2} and point instances as for \maxa{1} instances relative to the optimal one on average, with \gree{} and \app{} being the only exception for the point instances when considering the \lexo{} score.

Another noteworthy difference is that for most instance classes and $\kSeq$s considered, $\msm{}$ and $\mms{}$ are preferable to \gree{} regarding the \lexo{} score (e.g. point instances with $\kSeq = \kSeql{1}$), but not for all (e.g.~\maxa{1} instances with $\kSeq = \kSeql{1}$).

Furthermore, \mms{} and \msm{} achieve different \hs{} and \vm{} scores
for around $3\%$ ($13\%$) of the \maxa{2} (\maxa{1}) instances using $\kSeq=\kSeql{1}$,
for around $18\%$ ($6\%$) using $\kSeq=\kSeql{2}$,
and for around $12\%$ ($5\%$) using $\kSeq=\kSeql{3}$.
For points instances, they achieve different \hs{} and \vm{} scores for around
$3\%$ of instances using $\kSeq=\kSeql{1}$,
for around $8\%$ using $\kSeq=\kSeql{2}$,
and for around $6\%$ using $\kSeq=\kSeql{3}$.
When they differ, they can differ significantly, independently of the instance class and $\kSeq$ considered (see \cref{fig:mmsvl}).

Lastly, when comparing the results for the different $\kSeq=\kSeql{l}$ we consider, we want to note that, for each instance class, \gree{} performs worse overall regarding the \lexo{} score when increasing $l$.

\begin{table*}[t]\tikzexternalenable\small\centering
  \caption{For each score and each $\kSeql{l}$ considered, the first value is the percentage of instances with $\kSeql{l}$ for which a rule achieves the optimal value of the respective score.
  The value in parentheses is the percentage of the optimal score that a rule achieves on average for instances with $\kSeql{l}$, with one exception: 
  For \lexo{}, the percentage of the reciprocal proportion is shown, as \lexo{} is the only score to be minimized. 
  If a rule is optimal 
  with respect to a score, 
  the corresponding cell contains~100.
  The percentages are rounded naturally.
  }\label{tab:kdiff4}\medskip
  \parbox{\textwidth}{\centering
    {(a) \maxa{1} Instances}\par\medskip

\par
		{(i) Point Instances, $\kSeql{3}$}
	}\bigskip
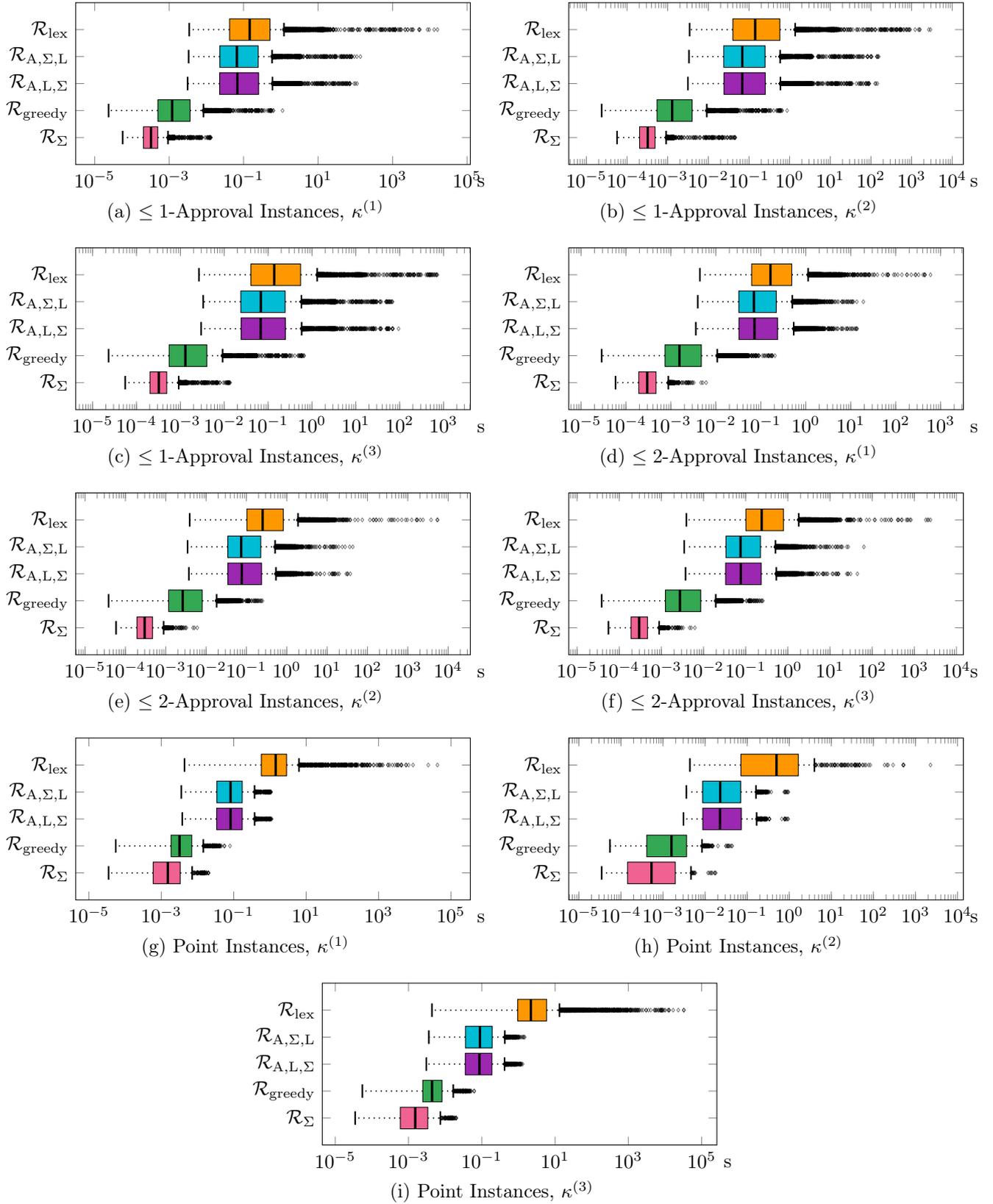
  \caption{The runtimes of the rules for finding a winning committee sequence for each of our instance classes and $\kSeq$.}
  \label{fig:timesl}%
\end{figure*}%

\begin{figure*}\tikzexternalenable\centering%
  \parbox{.3\textwidth}{%
		\centering
%
\\[-1ex]
		{(i) Point Instances, $\kSeql{3}$}
	}\bigskip
  \caption{A comparison of \lex{}, \gree{}, and \app{} for each $\kSeq$ considered: The $x$ value represents the \vm{} score achieved by \app{}, the $y$ value represents the \vm{} score achieved by another rule. If a point is on the blue line, the scores reached by both rules are the same. A point represents one instance from the corresponding instance class.}
  \label{fig:threevl}%
\end{figure*}%

\begin{figure*}
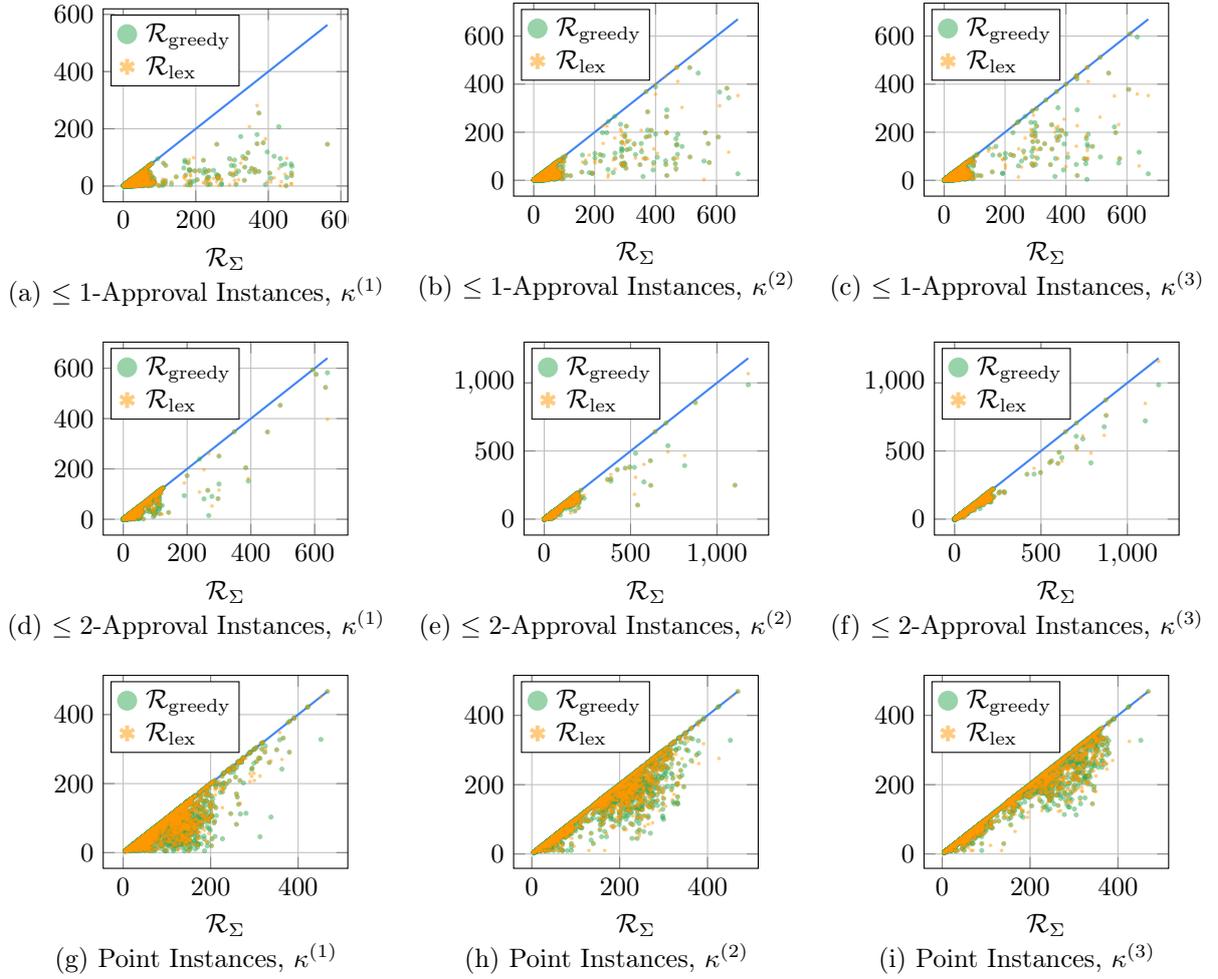
\tikzexternalenable\centering%
  \parbox{.3\textwidth}{%
		\centering
%
\\[-1ex]
		{(i) Point Instances, $\kSeql{3}$}
	}\bigskip
  \caption{A comparison of \lex{}, \gree{}, and \app{} for each $\kSeq$ considered: The $x$ value represents the \hm{} score achieved by \lex{}, the $y$ value represents the \hm{} score achieved by another rule. If a point is on the blue line, the scores reached by both rules are the same. A point represents one instance from the corresponding instance class.}
  \label{fig:threehl}%
\end{figure*}%

\begin{figure*}\tikzexternalenable\centering%
  \parbox{.3\textwidth}{%
		\centering
%
\\[-1ex]
		{(i) Point Instances, $\kSeql{3}$}
	}\bigskip
  \caption{A comparison of \mms{} and \msm{} for each $\kSeq$ considered: The $x$ value represents the \hs{} score achieved by \msm{}, the $y$ value represents the same score achieved by \mms{}. If a point is on the blue line, the scores reached by both rules are the same. A point represents one instance from the corresponding instance class.}
  \label{fig:mmsuml}%
\end{figure*}%

\begin{figure*}
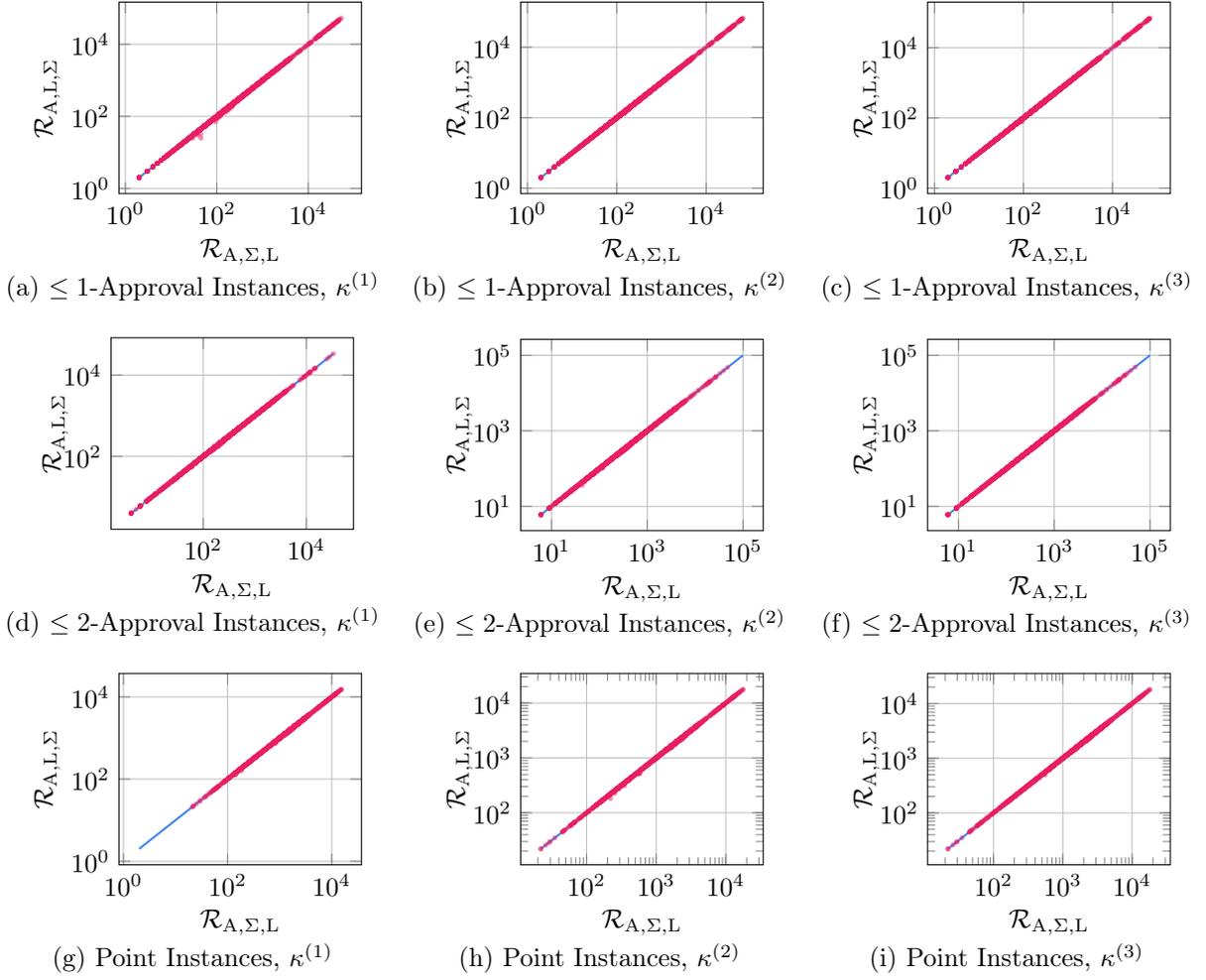
\tikzexternalenable\centering%
  \parbox{.3\textwidth}{%
		\centering
%
\\[-1ex]
		{(i) Point Instances, $\kSeql{3}$}
	}\bigskip
  \caption{A comparison of \mms{} and \msm{} for each $\kSeq$ considered: The $x$ value represents the \vm{} score achieved by \mms{}, the $y$ value represents the same score achieved by \msm{}. If a point is on the blue line, the scores reached by both rules are the same. A point represents one instance from the corresponding instance class.
  }
  \label{fig:mmsvl}%
\end{figure*}%

\begin{figure*}
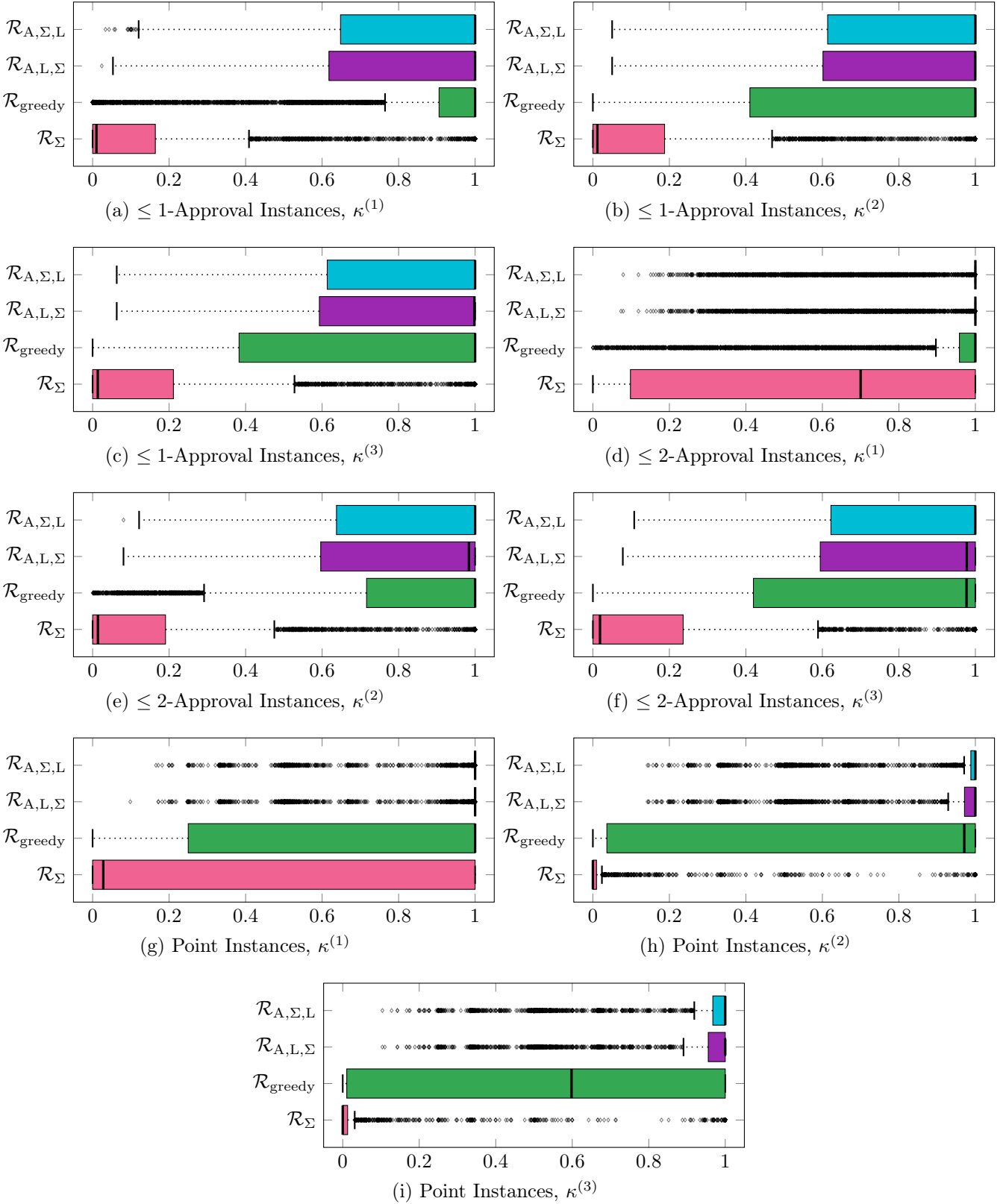
\tikzexternalenable
  \centering%
  \parbox{.49\textwidth}{%
		\centering
%

		{(i) Point Instances, $\kSeql{3}$}
	}\bigskip
  \caption{The proportions of the optimal \lexo{} score to the \lexo{} score determined by each rule that does not minimize this score for the instances from the corresponding instance class.}
  \label{fig:lexoboxl}%
\end{figure*}%

\subsubsection{Investigating the Properties}
In this section, we describe the approach we used to investigate the properties in more detail.
In the following, we assume that the agents are ordered, i.e.\ $A=\left\{ a_1,\dots,a_n \right\}$, reflecting the order of the agents in the underlying preference dataset from PrefLib, and make an analogous assumption with regard to the candidates on each level.

As mentioned in \cref{subsec:expsetup}, we discard an instance when testing the properties (to avoid excessive runtimes) if it meets one of the following two conditions, which are later referred to as \textit{Main Discard Conditions}:
\begin{enumerate}
  \item \gree{} violates the property and determining at most 30 winning committee sequences of \gree{} takes more than $\SI{10}{\second}$ using the C++ program and hardware mentioned in \cref{subsec:expsetup}.
  \item The first condition is not met, but at least one rule among all rules that violate the property determines at least 30 winning committee sequences.
\end{enumerate}
For each of the investigated properties, such instances from the experimental data are discarded.

For \pref{propty:pareto}---for which the exact results are shown in Tables~\ref{tab:prop_inst_disc4} and \ref{tab:prop_inst_res4}---these instances are the only instances discarded.

\pref{propty:unitorzsuperadd} is examined by partitioning the instances which do not meet the \textit{Main Discard Conditions} based on the number of levels (discarding every partition with only one instance), as the condition applies to two instances with the same $\tau$.
In each partition, the instances are sorted according to their file name (which matches the file name of the underlying preference dataset from PrefLib).
The condition of the property is then tested for each instance in combination with each of the $10$ instances following it in the ordering (if there are that many and if the merged version of the instances does not meet the \textit{Main Discard Conditions}).

To investigate \pref{propty:consistency} we first discard instances with $n>120$ or $m > 120$, then instances that meet one of the \textit{Main Discard Conditions}.
The condition of the property is then checked by splitting each of the remaining instances into two instances with $A_1 = \left\{a_1,\dots,a_j \right\}$ and $A_2 = \left\{a_{j+1},\dots,a_n \right\}$ for all $j\in\left\{1,\dots,n-1\right\}$, as $\CandSeq$, $\tau$, and $\kSeq$ have to be the same for both instances.
The resulting instance pair is discarded if at least one of the instances contains at least one level with no candidate getting a utility of at least one or meats one of the \textit{Main Discard Conditions}.
We use an analogous approach for~\pref{propty:conchorzsuperadd} by splitting the levels instead of the agents, but we do not discard instances with $n>120$ or $m > 120$ here.

As there are no instances with independent groups in all of our instances, we investigate the property using the following approach:
First, the instances with $n\leq 120$ and $m\leq 120$ which do not meet the \textit{Main Discard Conditions} are sorted according to their file name.
For each instance $\Elec=(A,\CandSeq,U=(u_1,\dots,u_\tau),\kSeq)$ and its successor $\Elec'=(A',\CandSeq',U'=(u_1',\dots,u_\tau'), \kSeq')$ in the ordering (if there is one), the property is evaluated for $\Elec_1=(A \cup A', \CandSeq \circ \CandSeq', (u_1'',\dots,u_{\tau + \tau'}''),\kSeq\circ \kSeq')$ (if $n\leq 120$ and $m\leq 120$ and if it does not meet the \textit{Main Discard Conditions}) with
$u_i''(a) = u_i(a)$ if $i\in\set{\tau}$ and $a\in A$,
$u_i''(a) = 0$ if $i\in\{\tau+1, \tau'\}$ and $a\in A$,
$u_i''(a) = u_i'(a)$ if $i\in\{\tau+1, \tau'\}$ and $a\in A'$,
and $u_i''(a) = 0$ if $i\in\set{\tau}$ and $a\in A'$, which leads to two independent groups.
If $\Elec'$ is not discarded and has a successor $\Elec''$, the same approach is used to create an instance $\Elec_2$ using $\Elec_1$ instead of $\Elec$ and $\Elec''$ instead of $\Elec'$---which leads to three independent groups---and the property is evaluated for $\Elec_2$ as well (if $n\leq 120$ and $m\leq 120$ and if it does not meet the \textit{Main Discard Conditions}).
 
The exact results regarding \pref{propty:conchorzsuperadd}, \pref{propty:unitorzsuperadd}, \pref{propty:consistency}, and \pref{propty:independentgroups} can be seen in Tables~\ref{tab:prop_pair_disc4_nom} and \ref{tab:prop_pair_res4_nom} for \maxa{1} instances, in Tables~\ref{tab:prop_pair_disc4_app} and \ref{tab:prop_pair_res4_app} for \maxa{2} instances, and in Tables~\ref{tab:prop_pair_disc4_point} and \ref{tab:prop_pair_res4_point} for point instances.

For each property and instance class investigated, the proportion of the investigated (and non-discarded pairs of) instances for which a rule satisfies the conditions of the given property can be seen in Table~\ref{tab:propsexpall4}:
For $\msm$ and $\mms$ it holds that, for each instance class and $\kSeq$, they satisfy the conditions of \pref{propty:independentgroups} for only at most $37.3\%$ of the instances, but satisfy, for each of the remaining properties, its conditions for at least $95.8\%$ of the instances.
\gree{} is, apart from \lex{}, the only rule that, for each instance class, $\kSeq$, and property, satisfies its conditions for at least $88\%$ of the instances.
For \app{}, on the other hand, the results can vary significantly between different instance types or $\kSeq$: For example, \app{} satisfies the conditions of \pref{propty:unitorzsuperadd}---the only property \app{} does not satisfy in general---for around $85\%$ of the \maxa{1} instances and $\kSeq=\kSeql{1}$, but only for around $50\%$ of the point instances for each $\kSeq$ considered.

\begin{table*}[t]\tikzexternalenable\centering
  \caption{The number of instances considered and discarded from the instances of the corresponding instance class for \pref{propty:pareto}. ``C1'' (``C2'') refers to the number of instances discarded due to the first (second) \textit{Main Discard Conditions}.}\label{tab:prop_inst_disc4}\medskip
  \begin{tikzpicture}\node[inner sep=0pt]{%
  \begin{tblr}{
    colspec={lrrrrrrrrr}
    }\toprule
    & \SetCell[c=3]{c}{$\kSeql{1}$}&& &\SetCell[c=3]{c}{$\kSeql{2}$}&& &\SetCell[c=3]{c}{$\kSeql{3}$}&&   \\%
    \cmidrule[lr]{2-4}\cmidrule[lr]{5-7}\cmidrule[lr]{8-10}
  Instance Class & Considered & {C1} & {C2}& Considered & {C1} & {C2}& Considered & {C1} & {C2}\\ \midrule
  \maxa{1} & 8727 & 13 & 130   & 8304 & 14 & 552    & 8227 & 14 & 629 \\
  \maxa{2} & 6256 & 646 & 986    & 6153 & 842 & 893    & 6057 & 651 & 1180\\
  Point & 4157 & 0 & 3   & 4145 & 0 & 15  & 4121 & 0 & 39\\
  \bottomrule
  \end{tblr}};
    \end{tikzpicture}
\end{table*}

\begin{table*}[t]\tikzexternalenable\centering
  \caption{The number of non-discarded instances from the experimental data for which a rule satisfies the conditions of \pref{propty:pareto} (``Sat.'') and the number of those for which they are violated (``Viol.'') are shown for each of our instance classes. If a rule satisfies \pref{propty:pareto} in general, the corresponding cells contain a \checkmark.
  }\label{tab:prop_inst_res4}\medskip
  \parbox{0.48\textwidth}{\centering
    {(a) \maxa{1} Instances}\par\medskip
    \begin{tikzpicture}\node[inner sep=0pt]{%
    \begin{tblr}{
      colspec={lrrrrrr},
      column{1} = {rightsep=0pt},
      column{2} = {rightsep=3.5pt},
      column{3} = {leftsep=3.5pt},
      column{4} = {rightsep=3.5pt},
      column{4} = {leftsep=3.5pt},
      column{6} = {rightsep=3.5pt},
      column{7} = {leftsep=3.5pt},
    }\toprule
    & \SetCell[c=2]{c}{$\kSeql{1}$}& &\SetCell[c=2]{c}{$\kSeql{2}$}& &\SetCell[c=2]{c}{$\kSeql{3}$}&   \\%
    \cmidrule[lr]{2-3}\cmidrule[lr]{4-5}\cmidrule[lr]{6-7}
         & Sat. & Viol. & Sat. & Viol. & Sat. & Viol. \\\midrule
    \lex{}   & \checkmark &       & \checkmark &        & \checkmark &      \\
    \msm{}   & \checkmark &       & \checkmark &        & \checkmark &      \\
    \mms{}   & 8713       & 14    & 8261       & 43     & 8190       & 37   \\
    \gree{}  & 7957       & 770   & 7585       & 719    & 7649       & 578  \\
    \app{}   & \checkmark &       & \checkmark &        & \checkmark &      \\
    \end{tblr}};%
    \end{tikzpicture}%
  }
  \parbox{0.48\textwidth}{\centering
    {(b) \maxa{2} Instances}\par\medskip
    \begin{tikzpicture}\node[inner sep=0pt]{%
    \begin{tblr}{
      colspec={lrrrrrr},
      column{1} = {rightsep=0pt},
      column{2} = {rightsep=3.5pt},
      column{3} = {leftsep=3.5pt},
      column{4} = {rightsep=3.5pt},
      column{4} = {leftsep=3.5pt},
      column{6} = {rightsep=3.5pt},
      column{7} = {leftsep=3.5pt},
    }\toprule
    & \SetCell[c=2]{c}{$\kSeql{1}$}& &\SetCell[c=2]{c}{$\kSeql{2}$}& &\SetCell[c=2]{c}{$\kSeql{3}$}&   \\%
    \cmidrule[lr]{2-3}\cmidrule[lr]{4-5}\cmidrule[lr]{6-7}
         & Sat. & Viol. & Sat. & Viol. & Sat. & Viol. \\\midrule
    
    \lex{}   & \checkmark &        & \checkmark &       & \checkmark &        \\
    \msm{}   & \checkmark &        & \checkmark &       & \checkmark &        \\
    \mms{}   & 6249       & 7      & 5893       & 260   & 5971       & 86     \\
    \gree{}  & 5780       & 476    & 5427       & 726   & 5320       & 737    \\
    \app{}   & \checkmark &        & \checkmark &       & \checkmark &        \\
  
    \end{tblr}};%
    \end{tikzpicture}%
  }\bigskip

  \parbox{\textwidth}{\centering
    {(c) Point Instances}\par\medskip
    \begin{tikzpicture}\node[inner sep=0pt]{%
    \begin{tblr}{
      colspec={lrrrrrr},
      column{1} = {rightsep=0pt},
      column{2} = {rightsep=3.5pt},
      column{3} = {leftsep=3.5pt},
      column{4} = {rightsep=3.5pt},
      column{4} = {leftsep=3.5pt},
      column{6} = {rightsep=3.5pt},
      column{7} = {leftsep=3.5pt},
    }\toprule
    & \SetCell[c=2]{c}{$\kSeql{1}$}& &\SetCell[c=2]{c}{$\kSeql{2}$}& &\SetCell[c=2]{c}{$\kSeql{3}$}&   \\%
    \cmidrule[lr]{2-3}\cmidrule[lr]{4-5}\cmidrule[lr]{6-7}
         & Sat. & Viol. & Sat. & Viol. & Sat. & Viol. \\\midrule
    
    \lex{}   & \checkmark &         & \checkmark &         & \checkmark &     \\
    \msm{}   & \checkmark &         & \checkmark &         & \checkmark &     \\
    \mms{}   & 4151       & 6       & 4114       & 31      & 4111       & 10  \\
    \gree{}  & 4047       & 110     & 3833       & 312     & 3846       & 275 \\
    \app{}   & \checkmark &         & \checkmark &         & \checkmark &     \\
  
    \end{tblr}};%
    \end{tikzpicture}%
  }
\end{table*}

\begin{table*}[t]\tikzexternalenable\centering
  \caption{The number of instances and instance pairs considered and discarded from the \maxa{1} instances for \pref{propty:conchorzsuperadd}, \pref{propty:unitorzsuperadd}, \pref{propty:consistency}, and \pref{propty:independentgroups}.}\label{tab:prop_pair_disc4_nom}\medskip
  \parbox{\textwidth}{\centering
    {(a) $\kSeql{1}$}\par\medskip
    \begin{tikzpicture}\node[inner sep=0pt]{%
    \begin{tblr}{
      colspec={lccccccc},
      column{1} = {rightsep=0pt},
      column{2} = {rightsep=3.5pt},
      column{3} = {rightsep=3.5pt},
      column{4} = {rightsep=3.5pt},
      column{5} = {rightsep=3.5pt},
      column{6} = {rightsep=3.5pt},
      column{7} = {rightsep=3.5pt},
      column{8} = {rightsep=3.5pt},
    }\toprule
    Property & {Considered\\Instance\\Pairs} & {Discarded\\Instances:\\Condition 1} & {Discarded\\Instances:\\Condition 2} & {Discarded\\Instances:\\$n>120$ or\\$m > 120$} & {Discarded\\Pairs:\\Condition 1} & {Discarded\\ Pairs:\\Condition 2} & {Discarded Pairs:\\Other reason\\(see text)}\\ \midrule
    P1 & 162628 & 13 & 121 & 0 & 1 & 783 & 0 \\
    P2 & 74804 & 13 & 170 & 0 & 1554 & 7938 & 0 \\
    P3 & 143782 & 13 & 101 & 464 & 895 & 11751 & 40963 \\
    P5 & 14343 & 0 & 53 & 464 & 0 & 459 & 1901 \\
    \bottomrule
    \end{tblr}};%
    \end{tikzpicture}%
  }\bigskip

  \parbox{\textwidth}{\centering
    {(b) $\kSeql{2}$}\par\medskip
    \begin{tikzpicture}\node[inner sep=0pt]{%
    \begin{tblr}{
      colspec={lccccccc},
      column{1} = {rightsep=0pt},
      column{2} = {rightsep=3.5pt},
      column{3} = {rightsep=3.5pt},
      column{4} = {rightsep=3.5pt},
      column{5} = {rightsep=3.5pt},
      column{6} = {rightsep=3.5pt},
      column{7} = {rightsep=3.5pt},
      column{8} = {rightsep=3.5pt},
    }\toprule
    Property & {Considered\\Instance\\Pairs} & {Discarded\\Instances:\\Condition 1} & {Discarded\\Instances:\\Condition 2} & {Discarded\\Instances:\\$n>120$ or\\$m > 120$} & {Discarded\\Pairs:\\Condition 1} & {Discarded\\ Pairs:\\Condition 2} & {Discarded Pairs:\\Other reason\\(see text)}\\ \midrule
    P1 & 152525 & 14 & 502 & 0 & 1 & 1416 & 0 \\
    P2 & 65271 & 14 & 756 & 0 & 186 & 12988 & 0 \\
    P3 & 125902 & 14 & 437 & 464 & 106 & 28674 & 29553 \\
    P5 & 12152 & 0 & 161 & 464 & 0 & 2507 & 1828 \\
    \bottomrule
    \end{tblr}};%
    \end{tikzpicture}%
  }\bigskip

  \parbox{\textwidth}{\centering
    {(c) $\kSeql{3}$}\par\medskip
    \begin{tikzpicture}\node[inner sep=0pt]{%
    \begin{tblr}{
      colspec={lccccccc},
      column{1} = {rightsep=0pt},
      column{2} = {rightsep=3.5pt},
      column{3} = {rightsep=3.5pt},
      column{4} = {rightsep=3.5pt},
      column{5} = {rightsep=3.5pt},
      column{6} = {rightsep=3.5pt},
      column{7} = {rightsep=3.5pt},
      column{8} = {rightsep=3.5pt},
    }\toprule
    Property & {Considered\\Instance\\Pairs} & {Discarded\\Instances:\\Condition 1} & {Discarded\\Instances:\\Condition 2} & {Discarded\\Instances:\\$n>120$ or\\$m > 120$} & {Discarded\\Pairs:\\Condition 1} & {Discarded\\ Pairs:\\Condition 2} & {Discarded Pairs:\\Other reason\\(see text)}\\ \midrule
    P1 & 152132 & 14 & 567 & 0 & 1 & 1137 & 0 \\
    P2 & 61877 & 14 & 930 & 0 & 149 & 14679 & 0 \\
    P3 & 119532 & 14 & 488 & 464 & 109 & 34626 & 27605 \\
    P5 & 11228 & 0 & 321 & 464 & 0 & 3253 & 1686 \\
    \bottomrule
    \end{tblr}};%
    \end{tikzpicture}%
  }
\end{table*}

\begin{table*}[t]\tikzexternalenable\centering
  \caption{The number of instances and instance pairs considered and discarded from the \maxa{2} instances for \pref{propty:conchorzsuperadd}, \pref{propty:unitorzsuperadd}, \pref{propty:consistency}, and \pref{propty:independentgroups}.}\label{tab:prop_pair_disc4_app}\medskip
  \parbox{\textwidth}{\centering
    {(a) $\kSeql{1}$}\par\medskip
    \begin{tikzpicture}\node[inner sep=0pt]{%
    \begin{tblr}{
      colspec={lccccccc},
      column{1} = {rightsep=0pt},
      column{2} = {rightsep=3.5pt},
      column{3} = {rightsep=3.5pt},
      column{4} = {rightsep=3.5pt},
      column{5} = {rightsep=3.5pt},
      column{6} = {rightsep=3.5pt},
      column{7} = {rightsep=3.5pt},
      column{8} = {rightsep=3.5pt},
    }\toprule
    Property & {Considered\\Instance\\Pairs} & {Discarded\\Instances:\\Condition 1} & {Discarded\\Instances:\\Condition 2} & {Discarded\\Instances:\\$n>120$ or\\$m > 120$} & {Discarded\\Pairs:\\Condition 1} & {Discarded\\ Pairs:\\Condition 2} & {Discarded Pairs:\\Other reason\\(see text)}\\ \midrule
    P1 & 79118 & 646 & 980 & 0 & 0 & 8 & 0 \\
    P2 & 34315 & 646 & 999 & 0 & 13900 & 12275 & 0 \\
    P3 & 61452 & 614 & 355 & 779 & 1877 & 50788 & 18296 \\
    P5 & 12178 & 0 & 344 & 779 & 0 & 282 & 1067 \\
    \bottomrule
    \end{tblr}};%
    \end{tikzpicture}%
  }\bigskip

  \parbox{\textwidth}{\centering
    {(b) $\kSeql{2}$}\par\medskip
    \begin{tikzpicture}\node[inner sep=0pt]{%
    \begin{tblr}{
      colspec={lccccccc},
      column{1} = {rightsep=0pt},
      column{2} = {rightsep=3.5pt},
      column{3} = {rightsep=3.5pt},
      column{4} = {rightsep=3.5pt},
      column{5} = {rightsep=3.5pt},
      column{6} = {rightsep=3.5pt},
      column{7} = {rightsep=3.5pt},
      column{8} = {rightsep=3.5pt},
    }\toprule
    Property & {Considered\\Instance\\Pairs} & {Discarded\\Instances:\\Condition 1} & {Discarded\\Instances:\\Condition 2} & {Discarded\\Instances:\\$n>120$ or\\$m > 120$} & {Discarded\\Pairs:\\Condition 1} & {Discarded\\ Pairs:\\Condition 2} & {Discarded Pairs:\\Other reason\\(see text)}\\ \midrule
    P1 & 77191 & 842 & 883 & 0 & 1 & 95 & 0 \\
    P2 & 35398 & 842 & 917 & 0 & 18743 & 5158 & 0 \\
    P3 & 102834 & 647 & 426 & 779 & 1102 & 9326 & 17830 \\
    P5 & 11880 & 0 & 368 & 779 & 0 & 528 & 1071 \\
    \bottomrule
    \end{tblr}};%
    \end{tikzpicture}%
  }\bigskip

  \parbox{\textwidth}{\centering
    {(c) $\kSeql{3}$}\par\medskip
    \begin{tikzpicture}\node[inner sep=0pt]{%
    \begin{tblr}{
      colspec={lccccccc},
      column{1} = {rightsep=0pt},
      column{2} = {rightsep=3.5pt},
      column{3} = {rightsep=3.5pt},
      column{4} = {rightsep=3.5pt},
      column{5} = {rightsep=3.5pt},
      column{6} = {rightsep=3.5pt},
      column{7} = {rightsep=3.5pt},
      column{8} = {rightsep=3.5pt},
    }\toprule
    Property & {Considered\\Instance\\Pairs} & {Discarded\\Instances:\\Condition 1} & {Discarded\\Instances:\\Condition 2} & {Discarded\\Instances:\\$n>120$ or\\$m > 120$} & {Discarded\\Pairs:\\Condition 1} & {Discarded\\ Pairs:\\Condition 2} & {Discarded Pairs:\\Other reason\\(see text)}\\ \midrule
    P1 & 75278 & 651 & 1158 & 0 & 4 & 243 & 0\\
    P2 & 33849 & 651 & 1277 & 0 & 14561 & 9251 & 0 \\
    P3 & 92336 & 620 & 541 & 779 & 321 & 19700 & 16979 \\
    P5 & 10471 & 0 & 389 & 779 & 0 & 1899 & 1067 \\
    \bottomrule
    \end{tblr}};%
    \end{tikzpicture}%
  }
\end{table*}

\begin{table*}[t]\tikzexternalenable\centering
  \caption{The number of instances and instance pairs considered and discarded from the point instances for \pref{propty:conchorzsuperadd}, \pref{propty:unitorzsuperadd}, \pref{propty:consistency}, and \pref{propty:independentgroups}.}\label{tab:prop_pair_disc4_point}\medskip
  \parbox{\textwidth}{\centering
    {(a) $\kSeql{1}$}\par\medskip
    \begin{tikzpicture}\node[inner sep=0pt]{%
    \begin{tblr}{
      colspec={lccccccc},
      column{1} = {rightsep=0pt},
      column{2} = {rightsep=3.5pt},
      column{3} = {rightsep=3.5pt},
      column{4} = {rightsep=3.5pt},
      column{5} = {rightsep=3.5pt},
      column{6} = {rightsep=3.5pt},
      column{7} = {rightsep=3.5pt},
      column{8} = {rightsep=3.5pt},
    }\toprule
    Property & {Considered\\Instance\\Pairs} & {Discarded\\Instances:\\Condition 1} & {Discarded\\Instances:\\Condition 2} & {Discarded\\Instances:\\$n>120$ or\\$m > 120$} & {Discarded\\Pairs:\\Condition 1} & {Discarded\\ Pairs:\\Condition 2} & {Discarded Pairs:\\Other reason\\(see text)}\\ \midrule
    P1 & 68529 & 0 & 1 & 0 & 0 & 1 & 0 \\
    P2 & 39310 & 0 & 3 & 0 & 2 & 314 & 0 \\
    P3 & 53210 & 0 & 3 & 0 & 0 & 15100 & 18258 \\
    P5 & 7784 & 0 & 3 & 0 & 0 & 32 & 495 \\
    \bottomrule
    \end{tblr}};%
    \end{tikzpicture}%
  }\bigskip

  \parbox{\textwidth}{\centering
    {(b) $\kSeql{2}$}\par\medskip
    \begin{tikzpicture}\node[inner sep=0pt]{%
    \begin{tblr}{
      colspec={lccccccc},
      column{1} = {rightsep=0pt},
      column{2} = {rightsep=3.5pt},
      column{3} = {rightsep=3.5pt},
      column{4} = {rightsep=3.5pt},
      column{5} = {rightsep=3.5pt},
      column{6} = {rightsep=3.5pt},
      column{7} = {rightsep=3.5pt},
      column{8} = {rightsep=3.5pt},
    }\toprule
    Property & {Considered\\Instance\\Pairs} & {Discarded\\Instances:\\Condition 1} & {Discarded\\Instances:\\Condition 2} & {Discarded\\Instances:\\$n>120$ or\\$m > 120$} & {Discarded\\Pairs:\\Condition 1} & {Discarded\\ Pairs:\\Condition 2} & {Discarded Pairs:\\Other reason\\(see text)}\\ \midrule
    P1 & 68258 & 0 & 14 & 0 & 0 & 23 & 0 \\
    P2 & 38868 & 0 & 22 & 0 & 1 & 567 & 0 \\
    P3 & 62322 & 0 & 15 & 0 & 0 & 5954 & 18167 \\
    P5 & 7728 & 0 & 7 & 0 & 0 & 80 & 495 \\
    \bottomrule
    \end{tblr}};%
    \end{tikzpicture}%
  }\bigskip

  \parbox{\textwidth}{\centering
    {(c) $\kSeql{3}$}\par\medskip
    \begin{tikzpicture}\node[inner sep=0pt]{%
    \begin{tblr}{
      colspec={lccccccc},
      column{1} = {rightsep=0pt},
      column{2} = {rightsep=3.5pt},
      column{3} = {rightsep=3.5pt},
      column{4} = {rightsep=3.5pt},
      column{5} = {rightsep=3.5pt},
      column{6} = {rightsep=3.5pt},
      column{7} = {rightsep=3.5pt},
      column{8} = {rightsep=3.5pt},
    }\toprule
    Property & {Considered\\Instance\\Pairs} & {Discarded\\Instances:\\Condition 1} & {Discarded\\Instances:\\Condition 2} & {Discarded\\Instances:\\$n>120$ or\\$m > 120$} & {Discarded\\Pairs:\\Condition 1} & {Discarded\\ Pairs:\\Condition 2} & {Discarded Pairs:\\Other reason\\(see text)}\\ \midrule
    P1 & 67966 & 0 & 33 & 0 & 0 & 21 & 0 \\
    P2 & 38196 & 0 & 61 & 0 & 0 & 850 & 0 \\
    P3 & 57774 & 0 & 39 & 0 & 0 & 10288 & 17741 \\
    P5 & 7570 & 0 & 15 & 0 & 0 & 222 & 495 \\
    \bottomrule
    \end{tblr}};%
    \end{tikzpicture}%
  }
\end{table*}

\begin{table*}[t]\tikzexternalenable\centering
  \caption{For each of the properties \pref{propty:conchorzsuperadd}, \pref{propty:unitorzsuperadd}, \pref{propty:consistency}, and \pref{propty:independentgroups}, the number of non-discarded \maxa{1} instances from the experimental data for which a rule satisfies the conditions of the property and the number for which they are violated is shown. If a rule satisfies a property in general, the corresponding cells contain a \checkmark.
  }\label{tab:prop_pair_res4_nom}\medskip
  \parbox{\textwidth}{\centering
    {(a) $\kSeql{1}$}\par\medskip
    \begin{tikzpicture}\node[inner sep=0pt]{%
    \begin{tblr}{
      colspec={lrrrrrrrrr},
      row{2} = {c},
      column{1} = {rightsep=0pt},
      column{2} = {rightsep=3.5pt},
      column{3} = {leftsep=3.5pt},
      column{4} = {rightsep=3.5pt},
      column{5} = {leftsep=3.5pt},
      column{6} = {rightsep=3.5pt},
      column{7} = {colsep=3.5pt},
      column{8} = {leftsep=3.5pt},
      column{9} = {rightsep=3.5pt},
      column{10} = {leftsep=3.5pt},
    }\toprule
    & \SetCell[c=2]{c}{P1}  &                  & \SetCell[c=2]{c}{ P2}        &                  & \SetCell[c=3]{c}{ P3} & & &\SetCell[c=2]{c}{P5}          & \\%
    \cmidrule[lr]{2-3}\cmidrule[lr]{4-5}\cmidrule[lr]{6-8}\cmidrule[lr]{9-10}
         & Satisfied            & Violated            & Satisfied             & Violated   & {Satisfied\\$M\neq\emptyset$}                  & {Satisfied\\$M=\emptyset$}           & Violated     & Satisfied             & Violated                                 \\\midrule
    
    \msm{}    & \checkmark &       & \checkmark &        & 9610       & 134075        & 97   &  1801 &  12542  \\
    \mms{}    & \checkmark &       & \checkmark &        & 9396       & 134274        & 112  &  1750  &  12593 \\
    \gree{}   & 159294     &  3334 & 73731      & 1073   & 8233       & 135446        & 103  &  \checkmark &   \\
    \app{}    & \checkmark &       & 63347      & 11457  & \checkmark &  \checkmark   &      &  \checkmark &   \\
    \bottomrule
  \end{tblr}};%
    \end{tikzpicture}%
  }\bigskip

  \parbox{\textwidth}{\centering
    {(b) $\kSeql{2}$}\par\medskip
    \begin{tikzpicture}\node[inner sep=0pt]{%
    \begin{tblr}{
      colspec={lrrrrrrrrr},
      row{2} = {c},
      column{1} = {rightsep=0pt},
      column{2} = {rightsep=3.5pt},
      column{3} = {leftsep=3.5pt},
      column{4} = {rightsep=3.5pt},
      column{5} = {leftsep=3.5pt},
      column{6} = {rightsep=3.5pt},
      column{7} = {colsep=3.5pt},
      column{8} = {leftsep=3.5pt},
      column{9} = {rightsep=3.5pt},
      column{10} = {leftsep=3.5pt},
    }\toprule
    & \SetCell[c=2]{c}{P1}  &                  & \SetCell[c=2]{c}{ P2}        &                  & \SetCell[c=3]{c}{ P3} & & &\SetCell[c=2]{c}{P5}          & \\%
    \cmidrule[lr]{2-3}\cmidrule[lr]{4-5}\cmidrule[lr]{6-8}\cmidrule[lr]{9-10}
         & Satisfied            & Violated            & Satisfied             & Violated   & {Satisfied\\$M\neq\emptyset$}                  & {Satisfied\\$M=\emptyset$}           & Violated     & Satisfied             & Violated                                 \\\midrule
    
    \msm{}    & \checkmark &       & \checkmark &        & 8223       & 117586        & 93   &  1598  &  10554  \\
    \mms{}    & \checkmark &       & \checkmark &        & 8124       & 117692        & 86   &  1554  &  10598 \\
    \gree{}   & 145832     &  6693 & 64271      & 1000   & 4711       & 121049        & 142  &  \checkmark &   \\
    \app{}    & \checkmark &       & 50600      & 14671  & \checkmark &  \checkmark   &      &  \checkmark &   \\
    \bottomrule  
  \end{tblr}};%
    \end{tikzpicture}%
  }\bigskip

  \parbox{\textwidth}{\centering
    {(c) $\kSeql{3}$}\par\medskip
    \begin{tikzpicture}\node[inner sep=0pt]{%
    \begin{tblr}{
      colspec={lrrrrrrrrr},
      row{2} = {c},
      column{1} = {rightsep=0pt},
      column{2} = {rightsep=3.5pt},
      column{3} = {leftsep=3.5pt},
      column{4} = {rightsep=3.5pt},
      column{5} = {leftsep=3.5pt},
      column{6} = {rightsep=3.5pt},
      column{7} = {colsep=3.5pt},
      column{8} = {leftsep=3.5pt},
      column{9} = {rightsep=3.5pt},
      column{10} = {leftsep=3.5pt},
    }\toprule
    & \SetCell[c=2]{c}{P1}  &                  & \SetCell[c=2]{c}{ P2}        &                  & \SetCell[c=3]{c}{ P3} & & &\SetCell[c=2]{c}{P5}          & \\%
    \cmidrule[lr]{2-3}\cmidrule[lr]{4-5}\cmidrule[lr]{6-8}\cmidrule[lr]{9-10}
         & Satisfied            & Violated            & Satisfied             & Violated   & {Satisfied\\$M\neq\emptyset$}                  & {Satisfied\\$M=\emptyset$}           & Violated     & Satisfied             & Violated                                 \\\midrule
    
    \msm{}    & \checkmark &       & \checkmark &        & 8476       & 110979        & 77   &  1646       &  9582  \\
    \mms{}    & \checkmark &       & \checkmark &        & 8372       & 111087        & 73   &  1616       &  9612 \\
    \gree{}   & 144578     &  7554 & 60955      & 922    & 4364       & 115094        & 74   &  \checkmark &   \\
    \app{}    & \checkmark &       & 48669      & 13208  & \checkmark &  \checkmark   &      &  \checkmark &   \\
    \bottomrule  
  \end{tblr}};%
    \end{tikzpicture}%
  }
\end{table*}

\begin{table*}[t]\tikzexternalenable\centering
  \caption{For each of the properties \pref{propty:conchorzsuperadd}, \pref{propty:unitorzsuperadd}, \pref{propty:consistency}, and \pref{propty:independentgroups}, the number of non-discarded \maxa{2} instances from the experimental data for which a rule satisfies the conditions of the property and the number for which they are violated is shown. If a rule satisfies a property in general, the corresponding cells contain a \checkmark.
  }\label{tab:prop_pair_res4_app}\medskip
  \parbox{\textwidth}{\centering
    {(a) $\kSeql{1}$}\par\medskip
    \begin{tikzpicture}\node[inner sep=0pt]{%
    \begin{tblr}{
      colspec={lrrrrrrrrr},
      row{2} = {c},
      column{1} = {rightsep=0pt},
      column{2} = {rightsep=3.5pt},
      column{3} = {leftsep=3.5pt},
      column{4} = {rightsep=3.5pt},
      column{5} = {leftsep=3.5pt},
      column{6} = {rightsep=3.5pt},
      column{7} = {colsep=3.5pt},
      column{8} = {leftsep=3.5pt},
      column{9} = {rightsep=3.5pt},
      column{10} = {leftsep=3.5pt},
    }\toprule
    & \SetCell[c=2]{c}{P1}  &                  & \SetCell[c=2]{c}{ P2}        &                  & \SetCell[c=3]{c}{ P3} & & & \SetCell[c=2]{c}{P5}          & \\%
    \cmidrule[lr]{2-3}\cmidrule[lr]{4-5}\cmidrule[lr]{6-8}\cmidrule[lr]{9-10}
         & Satisfied            & Violated            & Satisfied             & Violated   & {Satisfied\\$M\neq\emptyset$}                  & {Satisfied\\$M=\emptyset$}           & Violated     & Satisfied             & Violated                                 \\\midrule
    
    \msm{}  & \checkmark &      & \checkmark &            & 22932      & 38496         & 24   & 4539       &  7639 \\
    \mms{}  & \checkmark &      & \checkmark &            & 22840      & 38571         & 41   & 4494       &  7684 \\
    \gree{} & 78422      &  696 & 34216      & 99         & 22148      & 39104         & 200  & \checkmark &       \\
    \app{}  & \checkmark &      & 16949      & 17366      & \checkmark & \checkmark    &      & \checkmark &       \\
    \bottomrule
  \end{tblr}};%
    \end{tikzpicture}%
  }\bigskip

  \parbox{\textwidth}{\centering
    {(b) $\kSeql{2}$}\par\medskip
    \begin{tikzpicture}\node[inner sep=0pt]{%
    \begin{tblr}{
      colspec={lrrrrrrrrr},
      row{2} = {c},
      column{1} = {rightsep=0pt},
      column{2} = {rightsep=3.5pt},
      column{3} = {leftsep=3.5pt},
      column{4} = {rightsep=3.5pt},
      column{5} = {leftsep=3.5pt},
      column{6} = {rightsep=3.5pt},
      column{7} = {colsep=3.5pt},
      column{8} = {leftsep=3.5pt},
      column{9} = {rightsep=3.5pt},
      column{10} = {leftsep=3.5pt},
    }\toprule
    & \SetCell[c=2]{c}{P1}  &                  & \SetCell[c=2]{c}{ P2}        &                  & \SetCell[c=3]{c}{ P3} & & & \SetCell[c=2]{c}{P5}          & \\%
    \cmidrule[lr]{2-3}\cmidrule[lr]{4-5}\cmidrule[lr]{6-8}\cmidrule[lr]{9-10}
         & Satisfied            & Violated            & Satisfied             & Violated   & {Satisfied\\$M\neq\emptyset$}                  & {Satisfied\\$M=\emptyset$}           & Violated     & Satisfied             & Violated                                 \\\midrule
    
    \msm{}  & \checkmark &      & \checkmark &            & 6691      & 96099         & 44   & 1245       &  10635 \\
    \mms{}  & \checkmark &      & \checkmark &            & 6628      & 96142         & 64   & 1148       &  10732 \\
    \gree{} & 76481      &  710 & 35146      & 252        & 4817      & 97852         & 165  & \checkmark &       \\
    \app{}  & \checkmark &      & 29268      & 6130       & \checkmark & \checkmark    &      & \checkmark &       \\
    \bottomrule  
  \end{tblr}};%
    \end{tikzpicture}%
  }\bigskip

  \parbox{\textwidth}{\centering
    {(c) $\kSeql{3}$}\par\medskip
    \begin{tikzpicture}\node[inner sep=0pt]{%
    \begin{tblr}{
      colspec={lrrrrrrrrr},
      row{2} = {c},
      column{1} = {rightsep=0pt},
      column{2} = {rightsep=3.5pt},
      column{3} = {leftsep=3.5pt},
      column{4} = {rightsep=3.5pt},
      column{5} = {leftsep=3.5pt},
      column{6} = {rightsep=3.5pt},
      column{7} = {colsep=3.5pt},
      column{8} = {leftsep=3.5pt},
      column{9} = {rightsep=3.5pt},
      column{10} = {leftsep=3.5pt},
    }\toprule
    & \SetCell[c=2]{c}{P1}  &                  & \SetCell[c=2]{c}{ P2}        &                  & \SetCell[c=3]{c}{ P3} & & & \SetCell[c=2]{c}{P5}          & \\%
    \cmidrule[lr]{2-3}\cmidrule[lr]{4-5}\cmidrule[lr]{6-8}\cmidrule[lr]{9-10}
         & Satisfied            & Violated            & Satisfied             & Violated   & {Satisfied\\$M\neq\emptyset$}                  & {Satisfied\\$M=\emptyset$}           & Violated     & Satisfied             & Violated                                 \\\midrule
    
    \msm{}  & \checkmark &      & \checkmark &            & 5920      & 86374         & 42   & 1191       &  9280 \\
    \mms{}  & \checkmark &      & \checkmark &            & 5904      & 86373         & 59   & 1095       &  9376 \\
    \gree{} & 73159      & 2119 & 33394      & 455        & 3639      & 88576         & 121  & \checkmark &       \\
    \app{}  & \checkmark &      & 26027      & 7822       & \checkmark & \checkmark    &      & \checkmark &       \\
    \bottomrule  
  \end{tblr}};%
    \end{tikzpicture}%
  }
\end{table*}

\begin{table*}[t]\tikzexternalenable\centering
  \caption{For each of the properties \pref{propty:conchorzsuperadd}, \pref{propty:unitorzsuperadd}, \pref{propty:consistency}, and \pref{propty:independentgroups}, the number of non-discarded point instances from the experimental data for which a rule satisfies the conditions of the property and the number for which they are violated is shown. If a rule satisfies a property in general, the corresponding cells contain a \checkmark.
  }\label{tab:prop_pair_res4_point}\medskip
  \parbox{\textwidth}{\centering
    {(a) $\kSeql{1}$}\par\medskip
    \begin{tikzpicture}\node[inner sep=0pt]{%
    \begin{tblr}{
      colspec={lrrrrrrrrr},
      row{2} = {c},
      column{1} = {rightsep=0pt},
      column{2} = {rightsep=3.5pt},
      column{3} = {leftsep=3.5pt},
      column{4} = {rightsep=3.5pt},
      column{5} = {leftsep=3.5pt},
      column{6} = {rightsep=3.5pt},
      column{7} = {colsep=3.5pt},
      column{8} = {leftsep=3.5pt},
      column{9} = {rightsep=3.5pt},
      column{10} = {leftsep=3.5pt},
    }\toprule
    & \SetCell[c=2]{c}{P1}  &                  & \SetCell[c=2]{c}{ P2}        &                  & \SetCell[c=3]{c}{ P3} & & & \SetCell[c=2]{c}{P5}          & \\%
    \cmidrule[lr]{2-3}\cmidrule[lr]{4-5}\cmidrule[lr]{6-8}\cmidrule[lr]{9-10}
         & Satisfied            & Violated            & Satisfied             & Violated   & {Satisfied\\$M\neq\emptyset$}                  & {Satisfied\\$M=\emptyset$}           & Violated     & Satisfied             & Violated                                 \\\midrule
    
    \msm{}  & \checkmark &       & \checkmark &            & 11580       & 41566        & 64  & 2274       &  5510 \\
    \mms{}  & \checkmark &       & \checkmark &            & 11483       & 41659        & 68  & 2244       &  5540 \\
    \gree{} & 64040      &  4489 & 37366      & 1944       & 10692       & 42367        & 151 & \checkmark &        \\
    \app{}  & \checkmark &       & 18470      & 20840      & \checkmark &  \checkmark   &     & \checkmark &        \\
    \bottomrule
  \end{tblr}};%
    \end{tikzpicture}%
  }\bigskip

  \parbox{\textwidth}{\centering
    {(b) $\kSeql{2}$}\par\medskip
    \begin{tikzpicture}\node[inner sep=0pt]{%
    \begin{tblr}{
      colspec={lrrrrrrrrr},
      row{2} = {c},
      column{1} = {rightsep=0pt},
      column{2} = {rightsep=3.5pt},
      column{3} = {leftsep=3.5pt},
      column{4} = {rightsep=3.5pt},
      column{5} = {leftsep=3.5pt},
      column{6} = {rightsep=3.5pt},
      column{7} = {colsep=3.5pt},
      column{8} = {leftsep=3.5pt},
      column{9} = {rightsep=3.5pt},
      column{10} = {leftsep=3.5pt},
    }\toprule
    & \SetCell[c=2]{c}{P1}  &                  & \SetCell[c=2]{c}{ P2}        &                  & \SetCell[c=3]{c}{ P3} & & & \SetCell[c=2]{c}{P5}          & \\%
    \cmidrule[lr]{2-3}\cmidrule[lr]{4-5}\cmidrule[lr]{6-8}\cmidrule[lr]{9-10}
         & Satisfied            & Violated            & Satisfied             & Violated   & {Satisfied\\$M\neq\emptyset$}                  & {Satisfied\\$M=\emptyset$}           & Violated     & Satisfied             & Violated                                 \\\midrule
    
    \msm{}  & \checkmark &       & \checkmark &            & 3527       & 58755         & 40  & 703        &  7025 \\
    \mms{}  & \checkmark &       & \checkmark &            & 3479        & 58803        & 40  & 651        &  7077 \\
    \gree{} & 63378      &  4880 & 37436      & 1432       & 3004        & 59213        & 105 & \checkmark &        \\
    \app{}  & \checkmark &       & 21041      & 17827      & \checkmark &  \checkmark   &     & \checkmark &        \\
    \bottomrule  
  \end{tblr}};%
    \end{tikzpicture}%
  }\bigskip

  \parbox{\textwidth}{\centering
    {(c) $\kSeql{3}$}\par\medskip
    \begin{tikzpicture}\node[inner sep=0pt]{%
    \begin{tblr}{
      colspec={lrrrrrrrrr},
      row{2} = {c},
      column{1} = {rightsep=0pt},
      column{2} = {rightsep=3.5pt},
      column{3} = {leftsep=3.5pt},
      column{4} = {rightsep=3.5pt},
      column{5} = {leftsep=3.5pt},
      column{6} = {rightsep=3.5pt},
      column{7} = {colsep=3.5pt},
      column{8} = {leftsep=3.5pt},
      column{9} = {rightsep=3.5pt},
      column{10} = {leftsep=3.5pt},
    }\toprule
    & \SetCell[c=2]{c}{P1}  &                  & \SetCell[c=2]{c}{ P2}        &                  & \SetCell[c=3]{c}{ P3} & & & \SetCell[c=2]{c}{P5}          & \\%
    \cmidrule[lr]{2-3}\cmidrule[lr]{4-5}\cmidrule[lr]{6-8}\cmidrule[lr]{9-10}
         & Satisfied            & Violated            & Satisfied             & Violated   & {Satisfied\\$M\neq\emptyset$}                  & {Satisfied\\$M=\emptyset$}           & Violated     & Satisfied             & Violated                                 \\\midrule
    
    \msm{}  & \checkmark &       & \checkmark &            & 3509        & 54247        & 18  & 637        &  6933 \\
    \mms{}  & \checkmark &       & \checkmark &            & 3493        & 54263        & 18  & 617        &  6953 \\
    \gree{} & 61763      &  6203 & 36775      & 1421       & 2743        & 54950        & 81 & \checkmark  &        \\
    \app{}  & \checkmark &       & 20342      & 17854      & \checkmark &  \checkmark   &     & \checkmark &        \\
    \bottomrule  
  \end{tblr}};%
    \end{tikzpicture}%
  }
\end{table*}

\begin{table*}[t]\tikzexternalenable\centering
  \caption{The (naturally rounded) percentages of the investigated (pairs of) instances for which a rule satisfies the conditions of the given property, ignoring discarded (pairs of) instances, for each instance class investigated.
  If a rule satisfies a property in general, the corresponding cell contains a \checkmark{}.}\label{tab:propsexpall4}\medskip
  \parbox{0.48\textwidth}{\centering
    {(a) \maxa{1} Instances}\par\medskip
    \begin{tikzpicture}\node[inner sep=0pt]{%
    \begin{tblr}{
      colspec={llccccc},
      column{1} = {rightsep=0pt},
      column{2} = {rightsep=0pt},
      column{3} = {rightsep=3.5pt},
      column{4} = {rightsep=3.5pt},
      column{5} = {rightsep=3.5pt},
      column{6} = {rightsep=3.5pt},
      column{7} = {rightsep=3.5pt},
    }\toprule
    $\kappa$ & Property & \lex{} & \msm{} & \mms{} & \gree{} & \app{}\\ \midrule
                                     & $\kSeql{1}$ & \checkmark & \checkmark & \checkmark & \textcolor{GoogleGreen!96!GoogleYellow}{97.9} & \checkmark \\ %
    P1  & $\kSeql{2}$ & \checkmark & \checkmark & \checkmark & \textcolor{GoogleGreen!91!GoogleYellow}{95.6} & \checkmark \\ %
                                     & $\kSeql{3}$ & \checkmark & \checkmark & \checkmark & \textcolor{GoogleGreen!90!GoogleYellow}{95.0} & \checkmark \\\midrule %
                                     & $\kSeql{1}$ & \checkmark & \checkmark & \checkmark & \textcolor{GoogleGreen!97!GoogleYellow}{98.6} & \textcolor{GoogleGreen!69!GoogleYellow}{84.7} \\ %
    P2   & $\kSeql{2}$ & \checkmark & \checkmark & \checkmark & \textcolor{GoogleGreen!97!GoogleYellow}{98.5} & \textcolor{GoogleGreen!55!GoogleYellow}{77.5} \\ %
                                     & $\kSeql{3}$ & \checkmark & \checkmark & \checkmark & \textcolor{GoogleGreen!97!GoogleYellow}{98.5} & \textcolor{GoogleGreen!57!GoogleYellow}{78.7} \\\midrule %
                                     & $\kSeql{1}$ & \checkmark & \textcolor{GoogleGreen!98!GoogleYellow}{99.0} & \textcolor{GoogleGreen!98!GoogleYellow}{98.8} & \textcolor{GoogleGreen!98!GoogleYellow}{98.8} & \checkmark \\ %
    P3       & $\kSeql{2}$ & \checkmark & \textcolor{GoogleGreen!98!GoogleYellow}{98.9} & \textcolor{GoogleGreen!98!GoogleYellow}{99.0} & \textcolor{GoogleGreen!94!GoogleYellow}{97.1} & \checkmark \\ %
                                     & $\kSeql{3}$ & \checkmark & \textcolor{GoogleGreen!98!GoogleYellow}{99.1} & \textcolor{GoogleGreen!98!GoogleYellow}{99.1} & \textcolor{GoogleGreen!97!GoogleYellow}{98.3} & \checkmark \\\midrule %
                                     & $\kSeql{1}$ & \checkmark & \checkmark & \textcolor{GoogleGreen!100!GoogleYellow}{99.8} & \textcolor{GoogleGreen!82!GoogleYellow}{91.2} & \checkmark \\ 
    P4            & $\kSeql{2}$ & \checkmark & \checkmark & \textcolor{GoogleGreen!99!GoogleYellow}{99.5} & \textcolor{GoogleGreen!83!GoogleYellow}{91.3} & \checkmark \\ 
                                     & $\kSeql{3}$ & \checkmark & \checkmark & \textcolor{GoogleGreen!99!GoogleYellow}{99.6}  & \textcolor{GoogleGreen!86!GoogleYellow}{93.0} & \checkmark \\\midrule 
                                     & $\kSeql{1}$ & \checkmark & \textcolor{GoogleYellow!25!GoogleRed}{12.6} & \textcolor{GoogleYellow!24!GoogleRed}{12.2} & \checkmark & \checkmark \\
    P5 & $\kSeql{2}$ & \checkmark & \textcolor{GoogleYellow!26!GoogleRed}{13.2} & \textcolor{GoogleYellow!26!GoogleRed}{12.8} & \checkmark & \checkmark \\
                                     & $\kSeql{3}$ & \checkmark & \textcolor{GoogleYellow!29!GoogleRed}{14.7} & \textcolor{GoogleYellow!29!GoogleRed}{14.4} & \checkmark & \checkmark \\
    \bottomrule
    \end{tblr}};%
    \end{tikzpicture}%
  }
  \parbox{0.48\textwidth}{\centering
    {(b) \maxa{2} Instances}\par\medskip
    \begin{tikzpicture}\node[inner sep=0pt]{%
    \begin{tblr}{
      colspec={llccccc},
      column{1} = {rightsep=0pt},
      column{2} = {rightsep=0pt},
      column{3} = {rightsep=3.5pt},
      column{4} = {rightsep=3.5pt},
      column{5} = {rightsep=3.5pt},
      column{6} = {rightsep=3.5pt},
      column{7} = {rightsep=3.5pt},
    }\toprule
    $\kappa$ & Property & \lex{} & \msm{} & \mms{} & \gree{} & \app{}\\ \midrule
                                     & $\kSeql{1}$ & \checkmark & \checkmark & \checkmark & \textcolor{GoogleGreen!98!GoogleYellow}{99.1} & \checkmark \\ %
    P1  & $\kSeql{2}$ & \checkmark & \checkmark & \checkmark & \textcolor{GoogleGreen!98!GoogleYellow}{99.1} & \checkmark \\ %
                                     & $\kSeql{3}$ & \checkmark & \checkmark & \checkmark & \textcolor{GoogleGreen!94!GoogleYellow}{97.2} & \checkmark \\\midrule %
                                     & $\kSeql{1}$ & \checkmark & \checkmark & \checkmark & \textcolor{GoogleGreen!99!GoogleYellow}{99.7} & \textcolor{GoogleYellow!99!GoogleRed}{49.4}   \\ %
    P2   & $\kSeql{2}$ & \checkmark & \checkmark & \checkmark & \textcolor{GoogleGreen!99!GoogleYellow}{99.3} & \textcolor{GoogleGreen!65!GoogleYellow}{82.7} \\ %
                                     & $\kSeql{3}$ & \checkmark & \checkmark & \checkmark & \textcolor{GoogleGreen!97!GoogleYellow}{98.7} & \textcolor{GoogleGreen!54!GoogleYellow}{76.9} \\\midrule %
                                     & $\kSeql{1}$ & \checkmark & \textcolor{GoogleGreen!100!GoogleYellow}{99.9} & \textcolor{GoogleGreen!100!GoogleYellow}{99.8} & \textcolor{GoogleGreen!98!GoogleYellow}{99.1} & \checkmark \\ %
    P3       & $\kSeql{2}$ & \checkmark & \textcolor{GoogleGreen!99!GoogleYellow}{99.3}  & \textcolor{GoogleGreen!98!GoogleYellow}{99.0}  & \textcolor{GoogleGreen!93!GoogleYellow}{96.7} & \checkmark \\ %
                                     & $\kSeql{3}$ & \checkmark & \textcolor{GoogleGreen!99!GoogleYellow}{99.3}  & \textcolor{GoogleGreen!98!GoogleYellow}{99.0}  & \textcolor{GoogleGreen!94!GoogleYellow}{96.8} & \checkmark \\\midrule %
                                     & $\kSeql{1}$ & \checkmark & \checkmark & \textcolor{GoogleGreen!100!GoogleYellow}{99.9} & \textcolor{GoogleGreen!85!GoogleYellow}{92.4}& \checkmark \\
    P4            & $\kSeql{2}$ & \checkmark & \checkmark & \textcolor{GoogleGreen!92!GoogleYellow}{95.8}  & \textcolor{GoogleGreen!76!GoogleYellow}{88.2}& \checkmark \\ 
                                     & $\kSeql{3}$ & \checkmark & \checkmark & \textcolor{GoogleGreen!97!GoogleYellow}{98.6}  & \textcolor{GoogleGreen!76!GoogleYellow}{87.8}& \checkmark \\\midrule 
                                     & $\kSeql{1}$ & \checkmark & \textcolor{GoogleYellow!75!GoogleRed}{37.3} & \textcolor{GoogleYellow!74!GoogleRed}{36.9} & \checkmark & \checkmark \\
    P5 & $\kSeql{2}$ & \checkmark & \textcolor{GoogleYellow!21!GoogleRed}{10.5} & \textcolor{GoogleYellow!19!GoogleRed}{9.7} & \checkmark & \checkmark \\
                                     & $\kSeql{3}$ & \checkmark & \textcolor{GoogleYellow!23!GoogleRed}{11.4} & \textcolor{GoogleYellow!21!GoogleRed}{10.5} & \checkmark & \checkmark \\
    \bottomrule
    \end{tblr}};%
    \end{tikzpicture}%
  }\bigskip
  
  \parbox{0.48\textwidth}{\centering
    {(c) Point Instances}\par\medskip
    \begin{tikzpicture}\node[inner sep=0pt]{%
    \begin{tblr}{
      colspec={llccccc},
      column{1} = {rightsep=0pt},
      column{2} = {rightsep=0pt},
      column{3} = {rightsep=3.5pt},
      column{4} = {rightsep=3.5pt},
      column{5} = {rightsep=3.5pt},
      column{6} = {rightsep=3.5pt},
      column{7} = {rightsep=3.5pt},
    }\toprule
    $\kappa$ & Property & \lex{} & \msm{} & \mms{} & \gree{} & \app{}\\ \midrule
                                       & $\kSeql{1}$ & \checkmark & \checkmark & \checkmark & \textcolor{GoogleGreen!87!GoogleYellow}{93.4} & \checkmark \\ %
      P1  & $\kSeql{2}$ & \checkmark & \checkmark & \checkmark & \textcolor{GoogleGreen!86!GoogleYellow}{92.9} & \checkmark \\ %
                                       & $\kSeql{3}$ & \checkmark & \checkmark & \checkmark & \textcolor{GoogleGreen!82!GoogleYellow}{90.9} & \checkmark \\\midrule %
                                       & $\kSeql{1}$ & \checkmark & \checkmark & \checkmark & \textcolor{GoogleGreen!90!GoogleYellow}{95.1} & \textcolor{GoogleYellow!94!GoogleRed}{47.0}  \\ %
      P2   & $\kSeql{2}$ & \checkmark & \checkmark & \checkmark & \textcolor{GoogleGreen!93!GoogleYellow}{96.3} & \textcolor{GoogleGreen!8!GoogleYellow}{54.1} \\ %
                                       & $\kSeql{3}$ & \checkmark & \checkmark & \checkmark & \textcolor{GoogleGreen!93!GoogleYellow}{96.3} & \textcolor{GoogleGreen!7!GoogleYellow}{53.3} \\\midrule %
                                       & $\kSeql{1}$ & \checkmark & \textcolor{GoogleGreen!99!GoogleYellow}{99.5} & \textcolor{GoogleGreen!99!GoogleYellow}{99.4} & \textcolor{GoogleGreen!97!GoogleYellow}{98.6} & \checkmark \\ %
      P3       & $\kSeql{2}$ & \checkmark & \textcolor{GoogleGreen!98!GoogleYellow}{98.9} & \textcolor{GoogleGreen!98!GoogleYellow}{98.9} & \textcolor{GoogleGreen!93!GoogleYellow}{96.6} & \checkmark \\ %
                                       & $\kSeql{3}$ & \checkmark & \textcolor{GoogleGreen!99!GoogleYellow}{99.5} & \textcolor{GoogleGreen!99!GoogleYellow}{99.5} & \textcolor{GoogleGreen!94!GoogleYellow}{97.1} & \checkmark \\\midrule %
                                       & $\kSeql{1}$ & \checkmark & \checkmark & \textcolor{GoogleGreen!100!GoogleYellow}{99.9} & \textcolor{GoogleGreen!95!GoogleYellow}{97.4}  & \checkmark \\ 
      P4            & $\kSeql{2}$ & \checkmark & \checkmark & \textcolor{GoogleGreen!99!GoogleYellow}{99.3}  & \textcolor{GoogleGreen!85!GoogleYellow}{92.5}  & \checkmark \\ 
                                       & $\kSeql{3}$ & \checkmark & \checkmark & \textcolor{GoogleGreen!100!GoogleYellow}{99.8} & \textcolor{GoogleGreen!87!GoogleYellow}{93.3}  & \checkmark \\\midrule 
                                       & $\kSeql{1}$ & \checkmark & \textcolor{GoogleYellow!58!GoogleRed}{29.2} & \textcolor{GoogleYellow!58!GoogleRed}{28.8} & \checkmark & \checkmark \\
      P5 & $\kSeql{2}$ & \checkmark & \textcolor{GoogleYellow!18!GoogleRed}{9.1}  & \textcolor{GoogleYellow!17!GoogleRed}{8.4}  & \checkmark & \checkmark \\
                                       & $\kSeql{3}$ & \checkmark & \textcolor{GoogleYellow!17!GoogleRed}{8.4}  & \textcolor{GoogleYellow!16!GoogleRed}{8.2}  & \checkmark & \checkmark \\
    \bottomrule
    \end{tblr}};%
    \end{tikzpicture}%
  }
\end{table*}

\fi%

\end{document}